\documentclass[traditabstract,longauth]{aa}
\usepackage{graphicx,amsmath,amssymb, natbib, wasysym}
\usepackage{multirow}
\usepackage{booktabs}
\title{GRB 120422A/SN 2012bz: Bridging the Gap between Low- And High-Luminosity GRBs}

\titlerunning{GRB 120422A/SN 2012bz: Bridging the Gap between Low- And High-Luminosity GRBs}

\author{
S.~Schulze\inst{1,2,3},
D.~Malesani\inst{4},
A.~Cucchiara\inst{5},
N.~R.~Tanvir\inst{6},
T.~Kr\"uhler\inst{4,20},
A.~de Ugarte Postigo\inst{7,4},
G.~Leloudas\inst{8,4},
J.~Lyman\inst{9},
D.~Bersier\inst{9},
K.~Wiersema\inst{6},
D.~A.~Perley\inst{10,11},
P.~Schady\inst{12},
J.~Gorosabel\inst{7,44,45},
J.~P.~Anderson\inst{13,20},
A.~J.~Castro-Tirado\inst{7},
S.B.~Cenko\inst{14,15},
A.~De Cia\inst{16},
L.~E.~Ellerbroek\inst{17},
J.~P.~U.~Fynbo\inst{4},
J. Greiner\inst{12},
J.~Hjorth\inst{4},
D.~A.~Kann\inst{12,18},
L.~Kaper\inst{17},
S.~Klose\inst{18},
A.~J.~Levan\inst{19},
S.~Mart\'in\inst{20},
P.~T.~O'Brien\inst{6},
K.~L.~Page\inst{6},
G.~Pignata\inst{21},
S.~Rapaport\inst{22},
R.~S\'anchez-Ram\'irez\inst{7},
J.~Sollerman\inst{23},
I.~A.~Smith\inst{24},
M.~Sparre\inst{4},
C.~C.~Th\"one\inst{7},
D.~J.~Watson\inst{4},
D.~Xu\inst{16,4},
F.~E.~Bauer\inst{1,2,43},
M.~Bayliss\inst{25,26},
G.~Bj\"ornsson\inst{3},
M.~Bremer\inst{28},
Z.~Cano\inst{3},
S.~Covino\inst{27},
V.~D'Elia\inst{29,46},
D.~A.~Frail\inst{30},
S.~Geier\inst{4,31},
P.~Goldoni\inst{32},
O.~E.~Hartoog\inst{17},
P.~Jakobsson\inst{3},
H.~Korhonen\inst{33},
K.~Y.~Lee\inst{23},
B.~Milvang-Jensen\inst{4},
M.~Nardini\inst{34},
A.~Nicuesa Guelbenzu\inst{18},
M.~Oguri\inst{35,36},
S.~B.~Pandey\inst{37},
G.~Petitpas\inst{25},
A.~Rossi\inst{18},
A.~Sandberg\inst{23},
S.~Schmidl\inst{18},
G.~Tagliaferri\inst{27},
R.~P.~J.~Tilanus\inst{38,39},
J.~M.~Winters\inst{28}, 
D.~Wright\inst{40},
E.Wuyts\inst{41,42}	
}

\institute{Instituto de Astrof\'{i}sica, Facultad de F\'{i}sica, Pontificia Universidad Cat\'{o}lica de Chile, 306, Santiago 22, Chile, \email{sschulze@astro.puc.cl} 
\and Millennium Center for Supernova Science 
\and Centre for Astrophysics and Cosmology, Science and, University of Iceland, Dunhagi 5, 107 Reykjav\'ik, Iceland 
\and Dark Cosmology Centre, Niels Bohr Institute, University of Copenhagen, Juliane Maries Vej 30, 2100 Copenhagen, Denmark 
\and Department of Astronomy and Astrophysics, UCO/Lick Observatory, University of California, Santa Cruz, CA 95064, USA 
\and Department of Physics and Astronomy, University of Leicester, University Road, Leicester LE1 7RH, UK 
\and Instituto de Astrofísica de Andaluc\'ia–Consejo Superior de Investigaciones Cient\'ificas (IAA-CSIC), Glorieta de la Astronom\'ia s/n, 18008 Granada, Spain 
\and The Oskar Klein Centre, Department of Physics, Stockholm University, AlbaNova University Centre, 10691 Stockholm, Sweden 
\and Astrophysics Research Institute, Liverpool John Moores University, IC2, Liverpool Science Park Liverpool L3 5RF, United Kingdom 
\and Department of Astronomy, California and of Technology, MC 249-17, 1200 East California Blvd, Pasadena CA 91125, USA 
\and Hubble Fellow
\and Max-Planck-Institut für extraterrestrische Physik, Giessenbachstrasse 1, 85748 Garching, Germany 
\and Departamento de Astronomía, Universidad de Chile, Casilla 36-D, Santiago, Chile 
\and Astrophysics Science Division, NASA Goddard Space Flight Center, Mail Code 661, Greenbelt, MD 20771, USA 
\and Department of Astronomy, University of California, Berkeley, CA 94720-3411, USA 
\and Department of Particle Physics and Astrophysics, Faculty of Physics, Weizmann Institute of Science, Rehovot 76100, Israel 
\and Astronomical and Anton Pannekoek, University of Amsterdam, Science Park 904, 1098 XH Amsterdam, The Netherlands 
\and Th\"uringer Landessternwarte Tautenburg, Sternwarte 5, 07778 Tautenburg, Germany 
\and Department of Physics, University of Warwick, Coventry, CV4 7AL, UK 
\and European Southern Observatory, Alonso de C\'{o}rdova 3107, Vitacura Casilla 19001, Santiago 19, Chile 
\and Departamento de Ciencias Fisicas, Universidad Andres Bello, Avda. Republica 252, Santiago, Chile 
\and Research School of Astronomy and Astrophysics, Mount Stromlo Observatory, Cotter Road, Weston Creek, ACT 2611, Australia 
\and The Oskar Klein Centre, Department of Astronomy, Stockholm University, AlbaNova University Centre, 10691 Stockholm, Sweden 
\and Department of Physics and Astronomy, Rice University, 6100 South Main MS-108, Houston, TX 77005-1892, USA 
\and Harvard-Smithsonian Center for Astrophysics, 60 Garden Street, Cambridge, MA 02138, USA 
\and Department of Physics, Harvard University, 17 Oxford Street, Cambridge, MA 02138, USA 
\and INAF - Osservatorio Astronomico di Brera, via E. Bianchi 46 I-23807 Merate, Italy 
\and Institute de Radioastronomie Millim\'etrique (IRAM), 300 rue de la Piscine, 38406 Saint Martin d'H\`eres, France 
\and ASI-Science Data Center, Via del Politecnico, I-00133 Roma, Italy 
\and National Radio Astronomy Observatory, P.O. Box O, Socorro, NM 87801, USA 
\and Nordic Optical Telescope, Apartado 474, E-38700 Santa Cruz de La Palma, Spain 
\and APC, Astroparticule et Cosmologie, Universite Paris Diderot, CNRS/IN2P3, CEA/Irfu, Observatoire de Paris, Sorbonne Paris Cit\'e, 10, Rue Alice Domon et L\'eonie Duquet, 75205, Paris Cedex 13, France 
\and Finnish Centre for Astronomy with ESO (FINCA), University of Turku, V\"ais\"al\"antie 20, FI-21500 Piikki\"o, Finland 
\and Universit\`{a} degli studi di Milano-Bicocca, Piazza della Scienza 3, 20126, Milano, Italy 
\and Department of Physics, University of Tokyo, 7-3-1 Hongo, Bunkyo-ku, Tokyo 113-0033, Japan
\and Kavli Institute for the Physics and Mathematics of the Universe (Kavli IPMU, WPI), University of Tokyo, Chiba 277-8583, Japan
\and Aryabhatta Research and of observational-sciences, Manora peak, Nainital 263 129, India
\and James Clerk Maxwell Telescope, Joint Astronomy Centre, 660 North A’ohoku Place, University Park, Hilo, HI 96720, USA
\and Netherlands Organization for Scientific Research, Laan van Nieuw Oost-Indie 300, NL2509 AC The Hague, The Netherlands
\and Astrophysics Research Centre, School of Mathematics and Physics, Queen's University Belfast, Belfast BT7 1NN, UK 
\and Department of Astronomy \& Astrophysics, The University of Chicago, 5640 S. Ellis Avenue, Chicago, IL 60637, USA
\and Kavli Institute for Cosmological Physics at the University of Chicago, USA
\and Space Science Institute, 4750 Walnut Street, Suite 205, Boulder, CO 80301, USA 
\and Unidad Asociada Grupo Ciencia Planetarias UPV/EHU-IAA/CSIC, Departamento de F\'{\i}sica Aplicada I, E.T.S. Ingenier\'{\i}a, Universidad del Pa\'{\i}s Vasco UPV/EHU, Alameda de Urquijo s/n, 48013 Bilbao, Spain 
\and Ikerbasque, Basque Foundation for Science, Alameda de Urquijo 36-5, 48008 Bilbao, Spain 
\and INAF-Osservatorio Astronomico di Roma, Via Frascati 33, I-00040 Monteporzio Catone, Italy 
}

\date{Received XXXX; accepted XXXX}
\authorrunning{Schulze et al.}
\date{Received 8 January 2014; accepted XXXX}

\newcommand{\bepposax}{\textit{BeppoSAX}}
\newcommand{\hst}{\textit{HST}}
\newcommand{\swift}{\textit{Swift}}
\newcommand{\xmm}{\textit{XMM-Newton}}
\newcommand{\gcn}{GCN Circ.}

\begin{document}

\abstract
{
At low redshift, a handful of gamma-ray bursts (GRBs) have been discovered with
peak luminosities ($L_{\rm iso} < 10^{48.5}~\rm{erg\,s}^{-1}$) substantially
lower than the average of the more distant ones ($L_{\rm iso} > 10^{49.5}~\rm{erg\,s}^{-1}$).
The properties of several low-luminosity (low-$L$) GRBs indicate that they can
be due to shock break-out, as opposed to the emission from ultrarelativistic jets.
Owing to this, it is highly debated how both populations are connected, and
whether there is a continuum between them.
The burst at redshift $z=0.283$ from 2012 April 22 is one of the
very few examples of intermediate-$L$ GRBs with a $\gamma$-ray luminosity
of $L\sim10^{48.9}~\rm{erg\,s}^{-1}$ that
have been detected up to now. Together with the robust detection of its accompanying supernova
SN 2012bz, it has the potential to answer important questions on the origin of low- and
high-$L$ GRBs and the GRB-SN connection.
We carried out a spectroscopy campaign using medium- and low-resolution spectrographs
at 6--10-m class telescopes, covering the time span of 37.3 days, and a
multi-wavelength imaging campaign from radio to X-ray energies over a duration of
$\sim270$ days. Furthermore, we used a tuneable filter centred at H$\alpha$ to map
star formation in the host galaxy and the surrounding galaxies. We used these data
to extract and model the properties of different radiation components and incorporate
spectral-energy-distribution fitting techniques to extract the properties of the host
galaxy.
Modelling the light curve and spectral energy distribution from the radio to the X-rays
revealed the blast-wave to expand with an initial Lorentz factor of $\Gamma_0\sim60$,
low for a high-$L$ GRB, and that the afterglow had an exceptional low peak
luminosity-density of $\lesssim2\times10^{30}~\rm{erg\,s}^{-1}\,\rm{Hz}^{-1}$ in the sub-mm.
Because of the weak afterglow component, we were for the first time able to recover the
signature of a shock break-out that was not a genuine low-$L$ GRB. At 1.4 hours after the
burst, the stellar envelope had a blackbody temperature of $k_{\rm B}T\sim16~\rm eV$ and a
radius of $\sim7\times10^{13}\,\rm cm$.
The accompanying
SN 2012bz reached a peak luminosity of $M_{\rm V}=-19.7$ mag, 0.3 mag more
luminous than SN 1998bw. The synthesised  nickel mass of $0.58~\rm{M}_\odot$, ejecta
mass of $5.87~\rm{M}_{\odot}$, and kinetic energy of $4.10\times10^{52}~\rm erg$
were among
the highest recorded values for GRB-SNe, making it the most luminous spectroscopically
confirmed SN to data. Nebular emission lines at the GRB location were
visible, extending from the galaxy nucleus to the explosion site. The host and the
explosion site had close to solar metallicities. The burst occurred in an isolated
star-forming region with a SFR that is 1/10th of that in the galaxy's nucleus. 
While the prompt $\gamma$-ray emission points to a
high-$L$ GRB, the weak afterglow and the low $\Gamma_0$ were very atypical for such a
burst. Moreover the detection of the shock-break-out signature is a new quality for
high-$L$ GRBs.  So far, shock break-outs were exclusively detected for low-$L$ GRBs,
while GRB 120422A had an intermediate $L_{\rm iso}$ of $\sim10^{48.9}~\rm{erg\,s}^{-1}$.
Therefore, we conclude that GRB 120422A was a transition object between low- and high-$L$
GRBs, supporting the failed-jet model that connects shock-break-out driven low-$L$
and high-$L$ GRBs that are powered by ultra-relativistic jets.
}



\maketitle

\section{Introduction}

The discovery of SN 1998bw in the error-box of GRB 980425 by \citet{Galama1998a} gave
the study of the GRB-SN connection a flying start. This event remains unique in several
ways among the many hundred GRBs that have been studied since. It is still the nearest
GRB with a measured redshift and it is the least energetic GRB yet observed.
Nevertheless, SN 1998bw seems to be representative of the type of SNe that accompany the
more typical and brighter long-duration GRBs \citep[for recent reviews see][]{Woosley2006a,
Hjorth2011a}, i.e. a bright ($M_{\rm{bol},\,\rm{peak}}\lesssim-19~\rm mag$), broad-lined
(due to the expansion velocities of several $10^4~\rm{km\,s}^{-1}$) type Ic SN (i.e.
lacking of hydrogen and helium). Interestingly, in only two out of 16 cases of nearby
long-duration GRBs ($z<0.5$) no SN was found to deep limits \citep{Fynbo2006a,
DellaValle2006a, Gal-Yam2006a, Ofek2007a, Kann2011a}, though their classification
is not free of ambiguity \citep[e.g.][]{Zhang2009a, Kann2011a}.

So far, most GRBs with spectroscopically-confirmed SN associations have had a much lower
apparent luminosity than the bulk of the long-duration GRBs. GRB 030329 was the first
example of a high-luminosity GRB ($\log\,L_{\rm iso}/\left(\rm{erg\,s}^{-1}\right)=50.9$)
that was accompanied by a SN \citep{Hjorth2003a, Matheson2003a, Stanek2003a}. However,
there is a growing number of high-luminosity bursts, defined by
$\log\,L_{\rm iso}/\left(\rm{erg\,s}^{-1}\right)>49.5$ \citep{Hjorth2013a}, with a spectroscopically-confirmed
SN, such as GRBs 050525A \citep{DellaValle2006b}, 081007 \citep{DellaValle2008a,
Jin2013}, 091127 \citep{Cobb2010a,Berger2011a}, 101219B \citep{Sparre2011a}, 130215A
\citep{deUgartePostigo2013a}, 130427A \citep{Xu2013a, Levan2013a}, and 130831A
\citep{Klose2013a}.

\citet{Bromberg2011a} suggested that low-luminosity GRBs
($\log\,L_{\rm iso}/\left(\rm{erg\,s}^{-1}\right)<48.5$; \citealt{Hjorth2013a}) are driven by high-energy
emission associated with the shock break-out of their progenitor stars rather than an
emerging jet as in typical high-luminosity GRBs \citep[]{Colgate1969a, Kulkarni1998a,
Campana2006a, Soderberg2006a, Nakar2012a}. A consequence of these different energy sources
is that low-$L$ GRBs seem to be about 10-1000 times more common than high-$L$ GRBs
\citep{Pian2006a, Guetta2007a, Virgili2009a, Wanderman2010a}, but because of their
low luminosities they are primarily found at low redshifts as rare events (one every
$\sim3$ years). In contrast to high-$L$ GRBs, low-$L$ GRBs typically have
single-peak high-energy prompt light curves and can have
soft high-energy spectra with peak energies below $\sim50$ keV \citep[][but see \citealt{Kaneko2007a}]{Campana2006a,Starling2011a}. Their
optical emission is dominated by the SN emission. Until now, radio and X-ray afterglows,
but no optical afterglows have been detected for them.
The recent GRB 120422A is a particularly interesting case. It has a $\gamma$-ray luminosity
that is intermediate between
low- and high-luminosity GRBs and has a robust detection of the associated SN
\citep{Malesani2012c, Sanchez2012a, Wiersema2012a, Melandri2012a}. A study of this event
may thus answer important questions about the origin of both high- and low-$L$
GRBs.

The paper is structured as follows.
We describe the data gathering and outline the data analysis in Sect. \ref{sec:2},
and present the results on the transient following the GRB, from radio to X-ray wavelengths,
and the accompanying GRB-SN, SN 2012bz, in Sect. \ref{sec:transient}, and the properties
of  the GRB environment and the host galaxy in \ref{sec:environments}.
In Sect. \ref{sec:discussion} we compare our findings to other
events and argue that GRB 120422A represents the \textit{missing link between low- and
high-$L$ GRBs}. Finally, we summarise our findings and present our conclusions in Sect. \ref{sec:6}.

Throughout the paper we use the convention for the flux density $F_\nu \left(t\right)
\propto t^{-\alpha}\nu^{-\beta}$, where $\alpha$ is the temporal slope and $\beta$ is
the spectral slope. We refer to the solar abundance compiled in \citet{Asplund2009a}
and adopt $\rm{cm}^{-2}$ as the linear unit of column densities, $N$. Magnitudes reported
in the paper are given in the AB system and uncertainties are given at $1\sigma$ confidence
level (c.l.). We assume a $\Lambda$CDM cosmology with $H_0 = 71~\rm{km\,s}^{-1}\,\rm{Mpc}^{-1}$,
$\Omega_{\rm m} = 0.27$, and $\Omega_{\Lambda} = 0.73$ \citep{Larson2011a}.

\section{Observations and data reduction}\label{sec:2}

\begin{table*}[]
\caption{Summary of spectroscopic observations}
\centering
\scriptsize
\begin{tabular}{l@{\hspace{0.2cm}}r@{\hspace{0.2cm}}c@{\hspace{0.2cm}}c@{\hspace{0.2cm}}c@{\hspace{0.2cm}}c@{\hspace{0.2cm}}c@{\hspace{0.2cm}}c@{\hspace{0.2cm}}c@{\hspace{0.2cm}}c@{\hspace{0.2cm}}c@{\hspace{0.2cm}}c@{\hspace{0.2cm}}c@{\hspace{0.2cm}}c}
\toprule
\multicolumn{1}{c}{MJD}		& \multicolumn{1}{l}{Epoch}	& \multirow{2}{*}{Telescope/Instrument}	& \multirow{2}{*}{Arm/Grating}	& Spectral		& Resolving 	& Exposure 		& Slit 		& Position\\
\multicolumn{1}{c}{(days)}	& \multicolumn{1}{l}{(days)}&										&								& range (\AA)	& power 		& time (s)		& width		& angle\\
\midrule
56039.345					& 0.0443					& Gemini/GMOS-N						& R400+OG515	& 5942--10000	& 960		& $2\times900$	& 1\farcs0	& 180\fdg0					\\[1mm]

56039.431				 	& 0.1301					& Gemini/GMOS-N						& B600			& 3868--6632	& 844		& $2\times400$	& 1\farcs0	& 180\fdg0					\\[1mm]

\multirow{3}{*}{56040.017}	& \multirow{3}{*}{0.7160}	& \multirow{3}{*}{VLT/X-shooter}	& UVB			& 3000--5500	& 4350		& $4\times1200$	& 1\farcs0	& \multirow{3}{*}{41\fdg0}	\\
							&				 			& 									& VIS			& 5500--10000	& 8800		& $4\times1200$	& 0\farcs9	& 							\\
							&							&									& NIR			& 10000--24800	& 5100		& $16\times300$	& 0\farcs9	& 							\\[1mm]

56042.911					& 3.6112					& GTC/Osiris						& R500R			& 4800--10000	& 500		& $4\times1500$	& 1\farcs2	& 100\fdg0					\\[1mm]

\multirow{3}{*}{56044.014}	& \multirow{3}{*}{4.7139}	& \multirow{3}{*}{VLT/X-shooter}	& UVB			& 3000--5500	& 4350		& $4\times1200$	& 1\farcs0	& \multirow{3}{*}{41\fdg0}	\\
							&				 			&									& VIS			& 5500--10000	& 8800		& $4\times1200$	& 0\farcs9	& 							\\
							&							&									& NIR			& 10000--24800	& 5100		& $16\times300$	& 0\farcs9	& 							\\[1mm]

\multirow{2}{*}{56044.257}	& \multirow{2}{*}{4.9565}	& \multirow{2}{*}{Keck/LRIS}		& 400/3400		& 3000--5500	& 750 		& \multirow{2}{*}{$2\times900$}& \multirow{2}{*}{0\farcs7}	& \multirow{2}{*}{50\fdg0}		\\
							&							&									& 400/8500		& 5500--10000	& 1700		& 								& 							& 							\\[1mm]

\multirow{3}{*}{56048.061}	& \multirow{3}{*}{8.7604}	& \multirow{3}{*}{VLT/X-shooter}	& UVB			& 3000--5500	& 4350		& $4\times1200$	& 1\farcs0	& \multirow{3}{*}{41\fdg0}	\\
							&							&									& VIS			& 5500--10000	& 8800		& $4\times1200$	& 0\farcs9	& 							\\
							&							&									& NIR			& 10000--24800	& 5100		& $16\times300$	& 0\farcs9	& 							\\[1mm]

56048.304					& 9.0036 					& Gemini/GMOS-N						& R400			& 4442--8608	& 960		& $4\times1200$	& 1\farcs0	& 170\fdg0					\\[1mm]

56052.978					& 13.6772 					& Gemini/GMOS-S						& R400+GG455	& 4892--9008	& 960		& $1\times2400$	& 1\farcs0	& 180\fdg0					\\[1mm]

56053.930					& 14.6301 					& GTC/Osiris						& R500R			& 4800--10000	& 500		& $3\times1200$	& 1\farcs2	& 75\fdg0					\\[1mm]

\multirow{3}{*}{56057.996}	& \multirow{3}{*}{18.6962}	& \multirow{3}{*}{VLT/X-shooter$^a$}& UVB			& 3000--5500	& 4350		& $4\times1200$	& 1\farcs0	& \multirow{3}{*}{52\fdg0}	\\
							&							& 									& VIS			& 5500--10000	& 8800		& $4\times1200$	& 0\farcs9	& 							\\
							&							&									& NIR			& 10000--20700	& 5100		& $16\times300$	& 0\farcs9	& 							\\[1mm]

56061.996					& 22.6953 					& Gemini/GMOS-S						& R400+GG455	& 4892--9108	& 960		& $2\times2400$	& 1\farcs0	& -30\fdg0					\\[1mm]

\multirow{3}{*}{56063.999}	& \multirow{3}{*}{24.6992}	& \multirow{3}{*}{VLT/X-shooter$^a$}& UVB			& 3000--5500	& 4350		& $4\times1200$	& 1\farcs0	& \multirow{3}{*}{52\fdg0}	\\
							&				 			&									& VIS			& 5500--10000	& 8800		& $4\times1200$	& 0\farcs9	& 							\\
							&							&									& NIR			& 10000--20700	& 5100		& $16\times300$	& 0\farcs9	& 							\\[1mm]

56066.068					& 26.7680					& Magellan/LDSS3					& VPH\_ALL		& 3700--9400	& 800 		& $1\times1400$& 1\farcs2	& 141\fdg0					\\[1mm]

\multirow{3}{*}{56076.025}	& \multirow{3}{*}{36.7250}	& \multirow{3}{*}{VLT/X-shooter$^{a}$}	& UVB	& 3000--5500	& 4350		& $4\times1200$	& 1\farcs0	& \multirow{3}{*}{-143\fdg9}\\
							&				 			&									& VIS			& 5500--10000	& 8800		& $4\times1200$	& 0\farcs9	& 							\\
							&							&									& NIR			& 10000--20700	& 5100		& $16\times300$	& 0\farcs9	& 							\\[1mm]

\multirow{3}{*}{56077.000}	& \multirow{3}{*}{37.7001}	& \multirow{3}{*}{VLT/X-shooter$^a$}& UVB			& 3000--5500	& 4350		& $4\times1200$	& 1\farcs0	& \multirow{3}{*}{151\fdg1}	\\
							&				 			&									& VIS			& 5500--10000	& 8800		& $4\times1200$	& 0\farcs9	& 							\\
							&							&									& NIR			& 10000--20700	& 5100		& $16\times300$	& 0\farcs9	& 							\\[1mm]

\bottomrule
\end{tabular}
\tablefoot{Column "Epoch" shows the logarithmic mean-time after the burst in the observer
	frame. Resolving powers and spectral ranges are the nominal values from instrument
	manuals.
\tablefoottext{a}{The $K$-band blocking filter was used to increase the S/N in $JH$ band.}
}
\label{tab:obs_log_2}
\end{table*}

On 2012 April 22 at 7:12:49 UTC (hereafter called $T_0$; MJD\,=\,56039.30057), the Burst
Alert Telescope (BAT, \citealt{Barthelmy2005a}) aboard \swift~detected and localised a
faint burst \citep{Troja2012a}. Its $\gamma$-ray light curve comprised a single peak with
a duration of $T_{90}=5.4\pm1.4~\rm s$, followed by a fainter and lower-energetic emission
beginning 45 s after the trigger and lasting for 20 s.
Within 86~s, the \swift~X-Ray Telescope XRT \citep{Burrows2005a} and the UV/Optical
Telescope UVOT \citep{Roming2005a} started to observe the field and detected an
uncatalogued and rapidly decaying source at $\rm{R.A.},\,\rm{Dec.}\,(\rm{J2000})=
09^{\rm h}07^{\rm m}38^{\rm s}42\,(\pm0.01), +14^\circ01'07\farcs1\,(\pm 0.2)$
\citep{Beardmore2012a, Kuin2012a, Zauderer2012a}. Only $2''$ NE of the explosion site
there is a SDSS galaxy \citep{Cucchiara2012a, Tanvir2012a}. Spectra of the explosion site revealed several absorption
and emission lines at a common redshift of $z=0.283$, and a large number of
emission lines at the location of the SDSS galaxy at a redshift identical to that
of the GRB \citep{Schulze2012b, Tanvir2012a}.

Thanks to its low redshift and its $\gamma$-ray luminosity ($E_{\rm iso}\sim4.5\times10^{49}~\rm erg$
and $L_{\rm iso}\sim10^{49}~\rm{erg\,s}^{-1}$; \citealt{Zhang2012a}) being in between
that of high- and low-$L$ GRBs, it is an ideal target to search for the accompanying
GRB-SN, or place stringent constraints on its absence, if it is another example
of a SN-less long GRB. We therefore triggered an extensive imaging campaign with several
telescopes from mm to optical
wavelengths, as well as a large low- and medium-resolution spectroscopy campaign carried
out at 6-m to 10-m class telescopes. These campaigns began $\sim31$ min after the
trigger and ended $\sim44.6$ days later. Furthermore, we obtained an X-ray
spectrum with \xmm~12 days after the explosion. In addition to our own efforts, the
GRB-dedicated satellite \swift~observed the GRB at UV/optical and X-ray wavelengths
for 54.3 days. We incorporated these data as well as radio data obtained with the Arcminute Microkelvin Imager Large Array \citep[AMI-LA;][]{Staley2013a} to present a comprehensive study
of this event. In the following, we briefly summarise the observations and
describe how the data were analysed. A log of our observations is presented in Tables~\ref{tab:obs_log_2}, \ref{tab:obs_log_3}, \ref{tab:obs_log}, and \ref{tab:obs_log_late_time}.

\subsection{Optical and NIR spectroscopy}\label{sec:spectroscopy}

Our spectroscopic campaign began 51 min after the trigger and covered
a time span of 37.7 days. The spectral sequence comprised seven medium-resolution
spectra obtained with VLT/X-shooter \citep{Vernet2011a}; the first three spectra were obtained covering the
full spectral bandwidth from 3000 to 24800~\AA\@, while for the remaining ones  a $K$-blocking
filter (cutting the wavelength coverage at 20700~\AA\@; \citealt{Vernet2011a}) was adopted to increase the S/N in the $H$ band. These observations were complemented with ten low-resolution
spectra acquired with the Gemini Multi-Object Spectrograph (GMOS, \citealt{Hook2004a}),
mounted on Gemini-North and -South, the Gran Telescopio Canarias (GTC) OSIRIS camera, the Keck Low Resolution Imaging Spectrometer
\citep[LRIS;][]{Oke1995a} and the Magellan Low Dispersion Survey Spectrograph 3 (LDSS3).
Table~\ref{tab:obs_log_2} summarises these observations.

\begin{figure}
\centering
\includegraphics[angle=0, width=1\columnwidth]{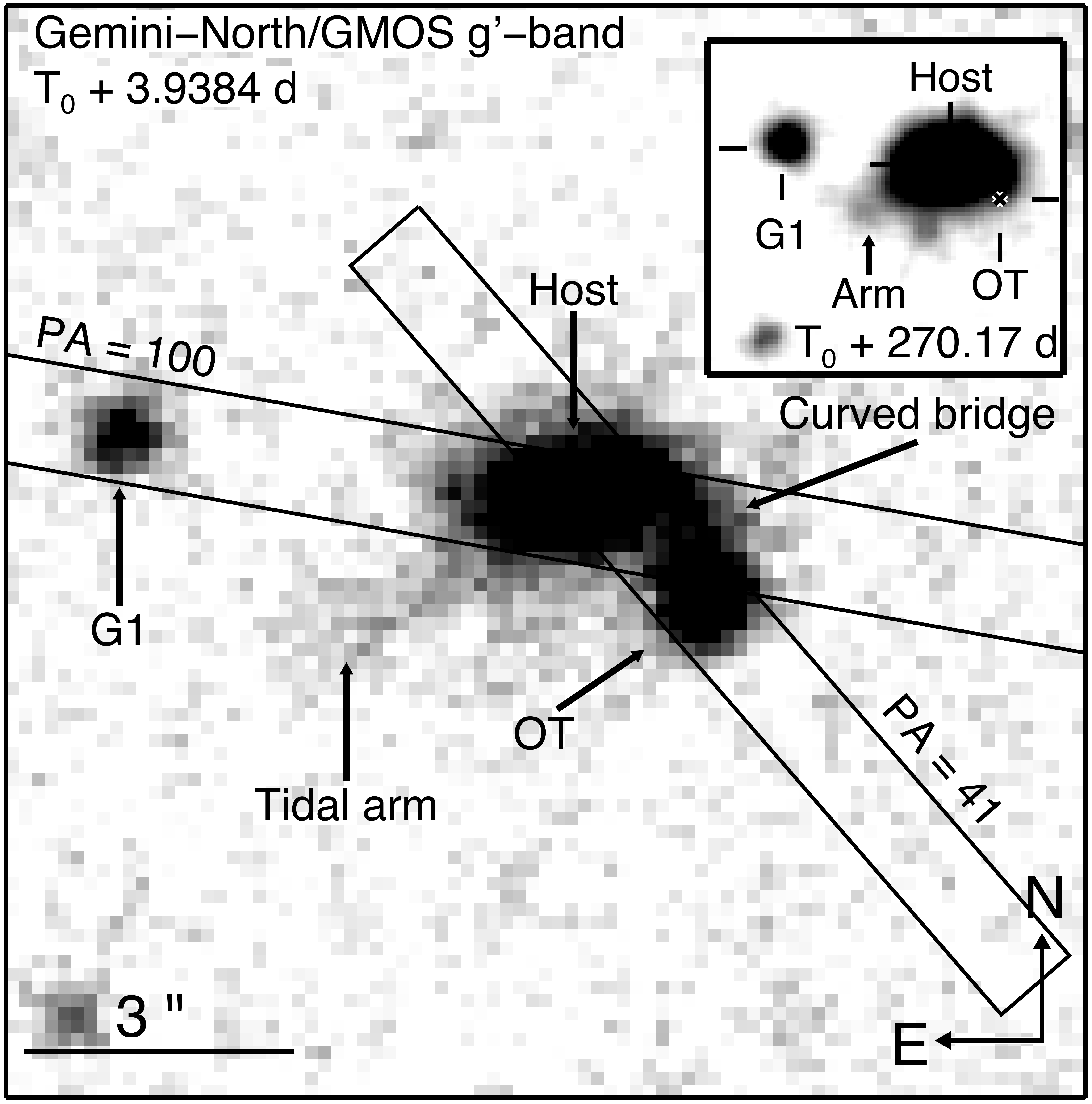}
\caption{Field of GRB 120422A ($12''\times12''$). The position
		of the optical transient (OT) accompanying GRB 120422A is marked, as well
		as of the host galaxy and the curved bridge of emission connecting the explosion site
		with the host's nucleus. Galaxy G1 has the same redshift as the GRB.
		The projected distance between the explosion site and the galaxy G1 is 28.7
		kpc. The inset shows the field observed in the $g'$-band with GMOS-N at 270.2 days after the burst.
		The image cuts were optimised to increase
		the visibility of the tidal arm that partly connects the host galaxy and G1.
		The most important slit orientations of our spectroscopic campaign (Table~\ref{tab:obs_log_2})
		are overlaid.}
\label{fig:GRB120422A_fov}
\end{figure}

Observing conditions were not always photometric, and observations were
performed irrespective of moon distance and phase. For each epoch, we centred the slit on the explosion site and  in some cases varied
the position angle to probe different parts of the host galaxy, as illustrated in
Fig.~\ref{fig:GRB120422A_fov}.

VLT/X-shooter data were reduced with the X-shooter pipeline v2.0 \citep{Goldoni2006a}.\footnote{http://www.eso.org/sci/software/pipelines/}
To extract the one-dimensional spectra of the transient and the host galaxy, we used a
customised tool that adopts the optimal extraction algorithm by \citet{Horne1986a}.
The Gemini, GTC, and Magellan spectra were reduced and calibrated using standard
procedures in \texttt{IRAF} \citep{Tody1993a}.
Keck data were reduced with a custom
pipeline that makes use of standard techniques of long-slit spectroscopy. In all cases
we chose a small aperture for studying the optical transient. For studying the emission lines, we extracted the spectral point spread
function and extracted the spectrum of the nucleus and the afterglow within
an aperture of $1\times \mathrm{FWHM}$ of each trace, e.g. the FWHMs were 1\farcs34 and 0\farcs86
for the galaxy nucleus and the explosion site, respectively, for the UVB and VIS of the
first X-shooter spectrum.

All spectra were flux-calibrated with corresponding spectrophotometric standard star
observations and the absolute flux scale was adjusted by comparing to photometry.
The data were
corrected for the Galactic reddening of $E(B-V)=0.04~\rm mag$ \citep{Schlegel1998a}.
 All wavelengths were transformed to vacuum wavelengths. In addition,
X-shooter data were corrected for heliocentric motion.
No telluric correction was applied, as it has no implications for our analysis.

\subsection{Imaging}\label{sec:imaging}

\begin{table}
\caption{Summary of mm and sub-mm observations}
\centering
\scriptsize
\begin{tabular}{l@{\hspace{1mm}}c@{\hspace{1mm}}c@{\hspace{1mm}}c@{\hspace{1mm}}c@{\hspace{1mm}}c}
\toprule
\multicolumn{1}{c}{MJD}		& \multicolumn{1}{c}{Epoch}	& \multicolumn{1}{c}{\multirow{2}{*}{Instrument}}	& \multicolumn{1}{c}{\multirow{2}{*}{Frequency}}	& Exposure	&	$F_{\nu}$\\
\multicolumn{1}{c}{(days)}	& \multicolumn{1}{c}{(days)}&													& 													& time (s)	& (mJy; $3\sigma$)\\
\midrule
56039.3291	& 0.0537		& SCUBA-2											& 350 GHz	& 5639		& $< 7.20$\\
56039.3291	& 0.0537		& SCUBA-2											& 665 GHz	& 5639		& $< 225$\\
56039.5676	& 0.2670		& AMI-LA$^a$										& 15 GHz	& 			& $< 0.62$\\
56040.1923	& 0.8917		& SMA												& 272 GHz	& 3420		& $< 3.60$\\
56041.6806	& 2.3800		& AMI-LA$^a$										& 15 GHz	& 			& $< 0.47$\\
56041.9422	& 2.6416		& PdBI												& 86.7 GHz	& 5040		& $< 0.39$\\
56041.9943	& 2.6937		& CARMA												& 92.5 GHz	& 3480		& $< 1.15$\\
56043.6806	& 4.3800		& AMI-LA$^a$										& 15 GHz	& 			& $< 0.37$\\
56046.7206	& 7.4200		& AMI-LA$^a$										& 15 GHz	&			& $< 0.24$\\
56048.8054	& 9.5048		& PdBI												& 86.7 GHz	& 5040		& $< 0.24$\\
56052.7506	& 13.450		& AMI-LA$^a$										& 15 GHz	&			& $< 0.23$\\
56067.8906	& 28.590 		& AMI-LA$^a$										& 15 GHz	&			& $< 0.46$\\
\bottomrule
\end{tabular}
\tablefoot{Column "Epoch" shows the logarithmic mean-time after the burst in the observer
	frame.
	\tablefoottext{a}{Data taken from \citet{Staley2013a}.}
	}
\label{tab:obs_log_3}
\end{table}

Following the BAT trigger, \swift~slewed immediately to the burst and UVOT took
a $v$-band settling exposure 86 s after the BAT trigger. Science observations
began at $T_0+104~\rm s$ and cycled through all filters. Follow-up
observations in the $v$ and $b$ bands continued until $T_0+2.3$~days, in the $uvw1$, $uvm2$ and
$uvw2$ UV filters until $T_0+9.7$~days, and in the
$u$ band until $T_0+54.3$ days, at which time a final set of observations of the host galaxy
was taken in all filters.\footnote{Additional UVOT data were acquired in October
2012. These data are not discussed in this paper. This has no implications on our work.}

Our ground-based imaging campaign began 31 min after the explosion and spanned a
time interval of $\sim45$ days. Due to the proximity of a $R=8.24~\rm mag$ star
($79''$ NW of the explosion site), we either moved the position of the optical
transient to the NW corner of the chip, or (most of the time) obtained short dithered
exposures to avoid excessive saturation.

Observations were carried out with the 2.56-m Nordic Optical Telescope (NOT) equipped
with ALFOSC, MOSCA, and StanCAM in the $u'g'Rr'Ii'$ bands \citep{Malesani2012b, Schulze2012a}.
These observations began at 14.29 hr post-burst and were stopped at 44.5 days
because of the small Sun distance.  Further imaging data were acquired with GMOS-N and GMOS-S
in the $u'g'r'i'z'$ bands between 31 min and 40.7 days
after the explosion \citep{Cucchiara2012a, Perley2012b}. The Gamma-ray Optical/Near-infrared
Detector (GROND, \citealt{Greiner2007a, Greiner2008a}) mounted at the MPG/ESO 2.2~m telescope
on La Silla imaged the field simultaneously in four optical ($g'r'i'z'$) and three NIR ($JHK_{\rm{s}}$) bands
starting at $T_0+16.5~\rm hr$ \citep{Nardini2012a}. Additional epochs were obtained at
nights 2, 9, 11, 20, 29, before the visibility of the field was compromised by its small
Sun distance on day 39. We monitored the optical transient in the $g'r'i'$
bands with the 60-inch Palomar telescope for 37 days beginning at $T_0+0.87~\rm day$
and in the $JHK$ bands with the Wide Field Camera (WFCAM) mounted at the United Kingdom Infrared
Telescope (UKIRT) on Mauna Kea at seven epochs between $T_0+0.06$ and 25.98 day.

We complemented these optical observations with the 10.4-m
GTC telescope equipped with OSIRIS in the $g'r'i'z'$ bands, the multi-filter imager BUSCA
mounted at the 2.2-m telescope of  Calar Alto (CAHA) in $g'$ and the
$r'$ bands,\footnote{http://www.caha.es/newsletter/news01a/busca/}
the 3.5-m CAHA telescope equipped with the Omega$_{2000}$ camera in the $z'$ band,\footnote{http://www.mpia-hd.mpg.de/IRCAM/O2000/} the LDSS3 camera
mounted at the 6-m Clay telescope telescope in the $r'$ and $i'$ bands, the Direct CCD Camera mounted on
the Irenee du Pont 2.5-m telescope at Las Campanas in the $r'$ and $i'$ bands, the 2.4-m Gao-Mei-Gu (GMG)
telescope in $i'$, and the 1.04-m and the 2-m optical-infrared Himalayan Chandra Telescope in $R_c$ and $I_c$.
 Additional NIR data were acquired with the Omega$_{2000}$
in the $YJHK_{\rm{s}}$ bands, the Near-Infrared Imager (NIRI)
mounted on Gemini-North in the $J$ and $K$ bands, and the Wide-field Infrared Camera
(WIRC) on the 200-inch Hale telescope at Palomar Observatory in the $J$ band \citep{Perley2012c}.

Very late-time observations were secured with the 2.0-m Liverpool telescope, with BUSCA mounted at the 2.2-m CAHA, and GMOS mounted at Gemini-North (Table~\ref{tab:obs_log_late_time}).
The observation with the Liverpool telescope comprises 185 images.
To minimise the data heterogeneity an observational seeing constraint of $<1\farcs1$
was imposed for all epochs. The CAHA  observation did unfortunately not go very deep. We will not discuss these data in the following.

In addition to these broad-band observations, we made use of the tuneable filters at the 10.4-m GTC to trace
the H$\alpha$ emission in the host galaxy on 2012 May 16, 25.5 days after the burst.
Observations consisted of 5$\times$600 s exposures using a 15-\AA{} wide filter tuned to the
wavelength of H$\alpha$ at the redshift of the burst ($\lambda_{\rm obs}=8420$~\AA\@),
and a $3\times100~\rm s$ exposure with a 513-{\AA\@}-wide
order-sorter filter centred at 8020 {\AA\@} to probe the continuum emission (filter f802/51). The
seeing was $\sim1^{\prime\prime}$, although the transparency was affected by extinction due
to Saharan dust suspended in the atmosphere (Calima).

In general, observing conditions were not always photometric; in particular, part of
the NOT observations suffered from poor transparency due to the Calima. Table~\ref{tab:obs_log}
summarises all observations with good data quality.

We obtained the UVOT data from the \swift~Data Archive.\footnote{http://www.swift.ac.uk/swift\_portal/}
These data had bad pixels identified, mod-8 noise corrected,
and endowed with FK5 coordinates. We used the standard
UVOT data analysis software distributed with \texttt{HEASOFT 6.12} along
with the standard calibration data.\footnote{http://heasarc.nasa.gov/lheasoft/}
Optical and NIR data were processed through standard procedures (bias
subtraction and flat field normalisation) using \texttt{IRAF}
or instrument specific software packages, i.e. the {\tt GEMINI IRAF} software
package for GMOS and NIRI, for GROND data a customised pipeline
(for details we refer to \citealt{Yoldacs2008a} and \citealt{Kruehler2008a}),
a modified version of the \texttt{WIRCSoft} package
for P200/WIRC data,\footnote{http://humu.ipac.caltech.edu/\~{}jason/sci/wircsoft/index.html} and for WFCAM data the UKIRT
pipeline.\footnote{http://casu.ast.cam.ac.uk/surveys-projects/wfcam}
Some observations suffered from variable conditions, and in
those cases individual images were weighted according to their S/N. The $i'$-
and the $z'$-band images suffer from fringing, which was corrected using a
fringe pattern computed from the science data themselves, although in some
cases the presence of the halo from the nearby bright star hampered the
process. These data resulted in a lower S/N.
Astrometric calibration was
computed against the USNO-B1 catalog \citep{Monet2003a}, yielding an $rms$ of $0\farcs4$. All
images were then registered together, yielding a relative RMS of less than
$0\farcs08$. We measure the afterglow location to be
$\rm{R.A.},\,\rm{Dec.}\,(\rm{J2000})=09^{\rm h}07^{\rm m}38^{\rm s}42, +14^\circ01'07\farcs5$.

\subsubsection{Sub-mm/mm observations}

Our sub-mm/mm observations comprised five epochs and cover a time interval of 9.48 days.
First, \citet{Smith2012a} simultaneously obtained an early epoch at $450~\mu\rm m$ and $850~\mu\rm m$ with the sub-millimetre
continuum camera SCUBA-2 \citep{Holland2013a} on the James Clerk Maxwell Telescope (JCMT). The 1.6-hr
observation began at $T_0+41.5~\rm min$ and was performed under moderate weather
conditions. The CSO 225 GHz tau, which measures the zenith atmospheric attenuation, was 0.089 initially, but generally degraded through
the run.  The elevation of GRB 120422A fell from 54\fdg6 to 30\fdg4.
In the consecutive night, \citet{Martin2012a} triggered a short 45-min snapshot
observation at the Submillimeter Array (SMA) at $T_0+21.4~\rm hr$. Receivers were tuned
to the local oscillator (LO) centre frequency of
271.8~GHz ($\lambda=1.1~\rm mm$), with the correlator configured to cover two 4~GHz
bands centred at $\pm6$~GHz from the LO frequency. All 8 SMA antennas were used in its
very extended configuration under excellent weather conditions, with an average zenith
opacity of 0.03 (precipitable water vapour of $\textrm{PWV}\sim0.5$~mm) at 225~GHz. A further observation was carried out by
\citet{Perley2012a} with the Combined Array for Research in Millimeter-Wave Astronomy
(CARMA) in D-configuration at 92.5 GHz ($\lambda=3~\rm mm$).
This observation was carried out between 23:13 UT on 24 April and 00:29 UT on April 25.
The total on-source integration time was 58 min. We finally obtained two epochs
with the Plateau de Bure  Interferometer (PdBI) at a frequency of 86.7 GHz
($\lambda=3.4~\rm mm$) in its 6 antenna compact D configuration. These
observations began at $T_0+2.6416$ and 9.5048 days and lasted for 84 min each.
AMI-LA obtained six epochs between
0.27 and 28.59 days after the burst \citep{Staley2013a}.

The SCUBA-2 data were reduced in the standard manner \citep{Chapin2013a} using \texttt{SMURF}
(Version 1.5.0) and \texttt{KAPPA} (Version 2.1-4) from the Starlink Project.\footnote{http://starlink.jach.hawaii.edu/starlink}
Observations of the SCUBA-2 calibrator Mars bracketed the GRB 120422A
observation, and observations of the calibrator CRL2688 were taken
several hours later.  The calibration observations spanned a larger
range of weather conditions than during the GRB 120422A run, and were
in general agreement with the standard values of the flux conversion
factors \citep{Dempsey2013a}, which were then used for the flux normalisation.
We reduced CARMA and SMA data with the \texttt{MIRIAD} and \texttt{MIR-IDL} software packages
\citep{Sault1995a}.\footnote{http://www.atnf.csiro.au/computing/software/miriad/\\ https://www.cfa.harvard.edu/\~{}cqi/mircook.html}
CARMA data were absolute flux calibrated with observations of 3C84 and Mars.
The calibration of the SMA data is twofold: first we used the nearby quasars J0854+201 and J0909+013
as atmospheric gain calibrators, and then J0854+201 for bandpass calibration.
Absolute flux calibration was bootstrapped from previous measurements of these quasars
resulting in an absolute flux uncertainty of $\sim30\%$.
PdBI data were reduced with the standard \texttt{CLIC} and \texttt{MAPPING}
software distributed by the Grenoble GILDAS group.\footnote{http://www.iram.fr/IRAMFR/GILDAS}
The flux calibration was secured with the Be binary star system MWC349 ($F_\nu = 1.1~\rm Jy$ at 86.7 GHz).

\subsubsection{X-ray observations}

\textit{Swift}/XRT started to observe the BAT GRB error circle
roughly 90 s after the trigger, while it was still slewing.
Observations were first carried out in windowed timing
mode for 80~s. When the count rate was $\lesssim1~\rm{ct\,s}^{-1}$, XRT
switched to photon counting mode. Observations continued
until $T_0+53.8~\rm days$, when the visibility
of the field was compromised by its small Sun distance.
We obtained the temporal and spectroscopic data from the \textit{Swift}/XRT
Light Curve and Spectrum Repository \citep{Evans2007a, Evans2009a}.
GRB 120422A was also observed by \xmm\ with a DDT, starting at 2012 May 3,
15:13 UT. At this epoch, exposures of 56841, 58421 and 58426 s
were obtained with the PN, MOS1 and MOS2 detectors, respectively.

To analyse the spectroscopic data we used \texttt{Xspec}, version 12.7.1, as part of \texttt{HeaSoft}
6.12, \xmm\ specific calibration files and for the \swift/XRT pc mode data the respective \swift\ calibration files
version 13. The X-ray emission up to $T_0+200~\rm s$
was discussed in detail in \citet{Starling2012a} and \citet{Zhang2012a}. Therefore,
we focus on the analysis of the data after that epoch. In total, XRT registered
270 background-subtracted photons between 0.3 and 10 keV; data that were flagged as
bad were excluded from analysis. We re-binned the spectrum to have at least 20 count
per bin and applied $\chi^2$ statistics.

\subsection{Photometry}\label{sec:photometry}

Measuring the brightness of the transient is complicated due to blending with its
extended, offset host galaxy. To
limit the host contribution to the transient photometry, we used point-spread function
(PSF) fitting techniques. Using bright field stars, a model of the PSF was constructed
for each individual image and fitted to the optical transient. To provide reliable fit results, all images
were registered astrometrically to a precision of better than 0\farcs08, and the
centroid of the fitted PSF was held fixed with a small margin of re-centering
corresponding to the uncertainty of the astrometric alignment of the individual
images. In addition, the PSF-fitting radius was adjusted to the specific conditions of the
observations and instrument, in particular seeing and pixel scale. The fit radius is
different for each observation, but typically in the range between 0\farcs5 and
0\farcs8. Generally, the radius was smaller under unfavourable sky conditions in an
attempt to minimise the host's effect on the fit. Naturally, this leads to a lower
S/N for these measurements than one would expect for isolated point
sources.

For images taken under adverse sky conditions (seeing $\gtrsim 1\farcs6$), with imagers
with large pixel scales (e.g. the NIR channels of GROND with 0\farcs6 per pixel), or in
filters/epochs with low S/N (e.g. most of the late NIR data), the individual contributions
of point-source and galaxy cannot be disentangled robustly. These measurements are
ignored in the following analysis. For all observations the source was close to the
centre of the field of view, and differences in the PSF between observations were,
therefore, negligible.

To measure the brightness of the transient in the UVOT images, we measured the
host galaxy flux at the position of the SN from the later UVOT observations,
where there was no longer a contribution from the GRB or SN. This additional
flux was then subtracted from our photometric measurements at the position
of the GRB.
In contrast, host-galaxy photometry was performed via aperture techniques. Here, we used
our PSF-model to subtract the transient from the deepest images in each filter with the
clearest separation between galaxy and point source, i.e. those images with the smallest
FWHM of the stellar PSF. A circular aperture radius was chosen sufficiently large
(2\farcs5, e.g. 10.7~kpc at $z=0.2825$), so that the missed emission from low surface
brightness regions does not affect our photometry significantly. In addition, we also
corroborated the galaxy photometry using elliptical Kron apertures \citep{Kron1980a}
via their implementation in \texttt{Source Extractor} \citep{Bertin1996a}.

Once a magnitude was established, it was calibrated photometrically against the
brightness of a number of field stars measured in a similar manner. Photometry was
tied to the SDSS DR8 \citep{Aihara2011a} in the optical filters
($u'g'r'i'z'$) and 2MASS \citep{Skrutskie2006a}
in the NIR ($JHK_{\rm{s}}$). For those filter bands not covered by our primary calibration
systems (e.g. $I_{\rm{C}}$ or $Y$), we used the instrument-specific band passes to
transform magnitudes into the respective filter system via synthetic photometry similar
to the procedure outlined in \citet{Kruehler2011a}.  UVOT images were calibrated
using the method described in \citet{Poole2008a}.

The photometric error was then estimated based on the contributions from photon
statistics and goodness of the PSF fit (typically between 0.5 to 15 \%), the absolute
accuracy of the primary calibration system ($\approx 2$--3\% ), the systematic scatter
of different instrument/bandpasses with respect to the primary calibrators
($\approx 3$--6\%) or the uncertainty in the colour transformation (if applicable,
$\approx 6$--9\%).

\begin{figure*}
\centering
\includegraphics[viewport=3 7 1146 1473, clip, angle=0, width=1.875\columnwidth]{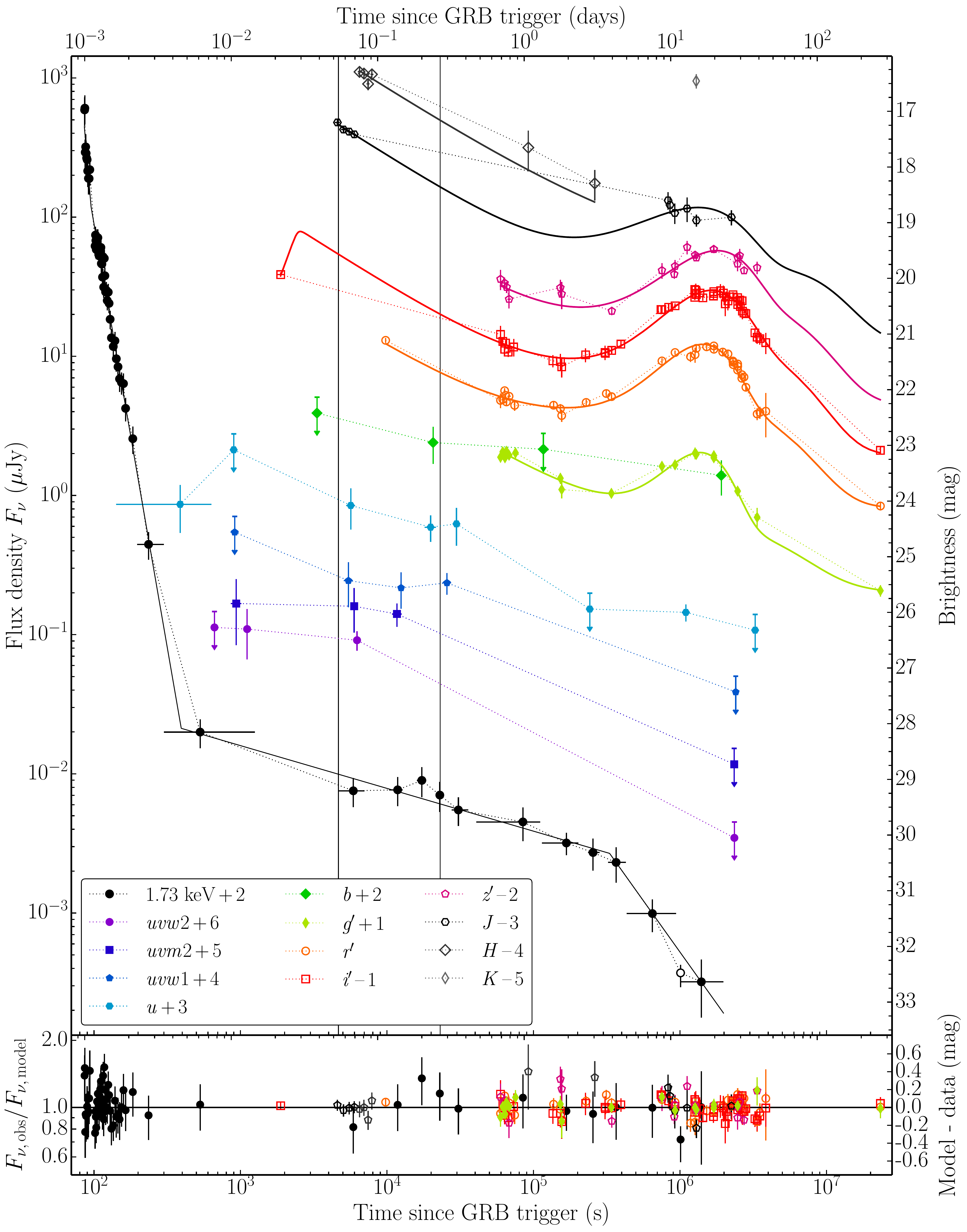}
\caption{X-ray, optical and NIR light curves of the transient following GRB 120422A.
		Arrows indicate $3\sigma$ upper limits.
		The UVOT $v$-band upper limits are very shallow and not displayed.
		Data in the $g'r'i'z'J$-bands were modelled with a SN 1998bw
		template at $z=0.283$ superposed on a power law (where the slope was identical
		in all bands) using the formalism in
		\citet{Zeh2004a}. The best-fit model parameters are shown in Table~\ref{tab:sn_prop}.
		Model light curves in bluer or redder filters are not shown since they would require
		extrapolation of the spectral range of the SN1998bw template.
		Fit residuals are displayed in the bottom panel. The \xmm\ observation was
		carried out at 980 ks (open dot). The shifts (in magnitude) of the different bands
		are given in the legend. To convert the X-ray light curve to flux density,
		we assumed a spectral slope of $\beta=0.9$ and no spectral evolution (for details
		on the SED modelling see Sect. \ref{sec:radio_X-ray_sed}). Both assumptions have no
		implications on our analysis. The \xmm\ data point was discarded
		from the light curve fit because of uncertainties in the cross-calibration between \swift/XRT and \xmm.
		The vertical lines indicate the epochs of the X-ray-to-NIR SEDs presented in
		Sect. \ref{sec:radio_X-ray_sed}.
	}
\label{fig:OT_LC}
\end{figure*}

The photometry described in the earlier paragraph inevitably contains a seeing-dependent
fraction of the host light directly at the position of the transient. This contribution
is best removed via differential imaging with deep reference frames from the same
instrument/filter combination taken after the transient has faded completely. Given the
vast number of different observers taking part in our photometry campaign, however, this
procedure was not feasible in our case for all images. We instead used reference frames from
a single telescope (Gemini-N, obtained $\sim270$ days after the explosion) in three filters.
We measure: $g'=24.62\pm0.10$, $r'=24.09\pm0.09$, and $i'=24.09\pm0.09~\rm mag$, i.e. a host light
contribution of 10\%, 7\% and 7\% in $g'r'i'$ at maximum SN light. To estimate the fraction
in different filters, we scaled the above numbers to the respective filters using the SED
of the host. We assume that this factor is similar for all data from various telescopes.
We note that the values in Table~\ref{tab:obs_log} are \textit{not} corrected for this host contribution.

\section{The transient accompanying GRB 120422A}\label{sec:transient}

Figure \ref{fig:OT_LC} displays the brightness evolution of the transient accompanying
GRB 120422A from X-rays to the NIR. During the first three days, its
brightness in the UVOT filters gradually decreases with a decay slope of $\alpha=0.2$
that is followed by a rebrightening peaking at $\sim20$ days post-burst. The time scale
and the colour evolution of the rebrightening are comparable to those of GRB-SNe
\citep[e.g.][]{Zeh2004a}. The initially decaying transient could therefore be a
superposition of the afterglow and the thermal emission of the cooling photosphere
after the SN emerged. Key to understanding the evolution of the transient
accompanying GRB 120422A is disentangling the different radiation components.
In the following sections we will present our results on each component.

\subsection{The stellar envelope cooling-phase}\label{sec:SN2012bz_SBO}

\begin{figure*}
\centering
\includegraphics[viewport=2 8 706 505, clip, angle=0, width=1\columnwidth]{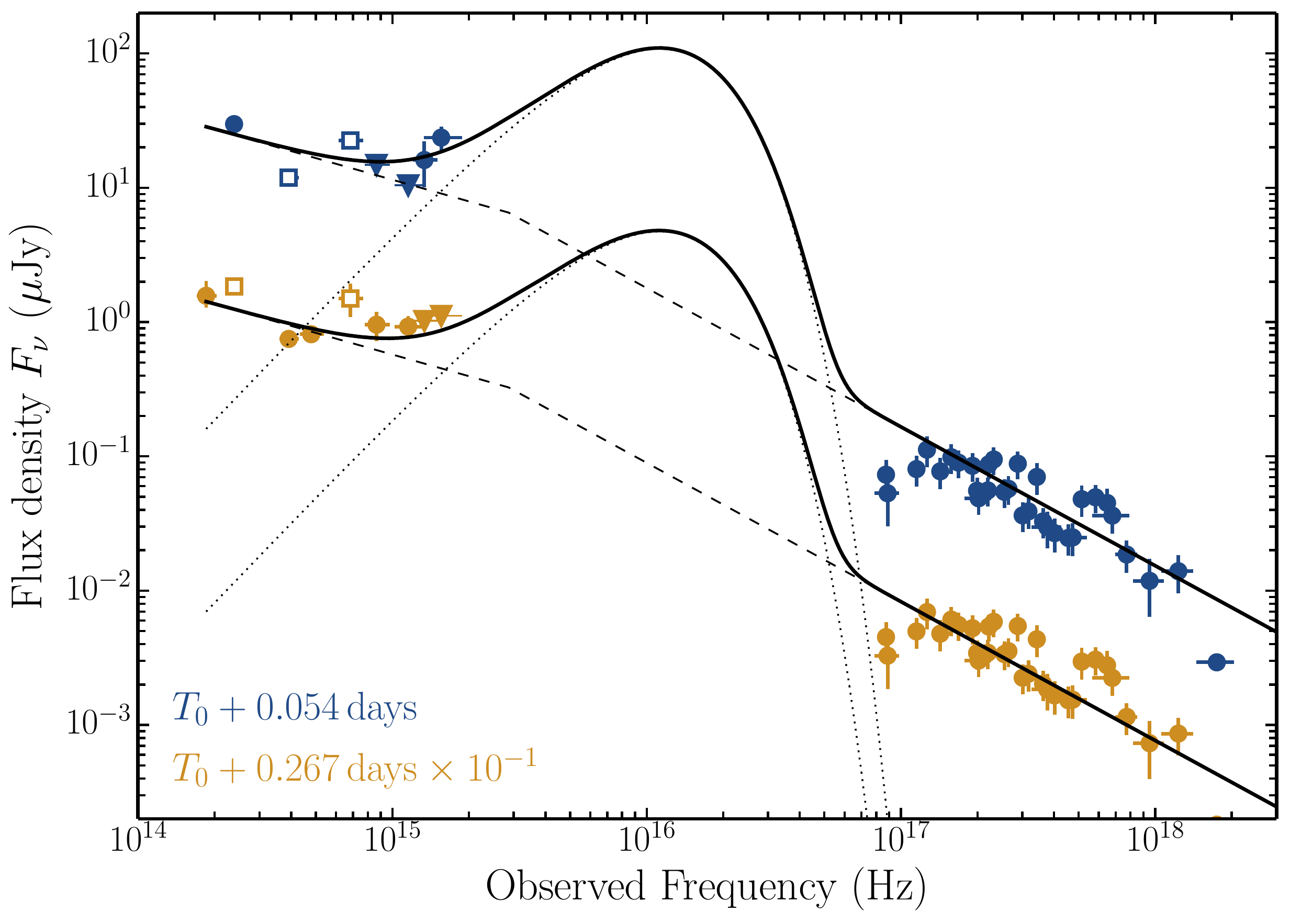}
\includegraphics[viewport=2 8 695 505, clip, angle=0, width=1\columnwidth]{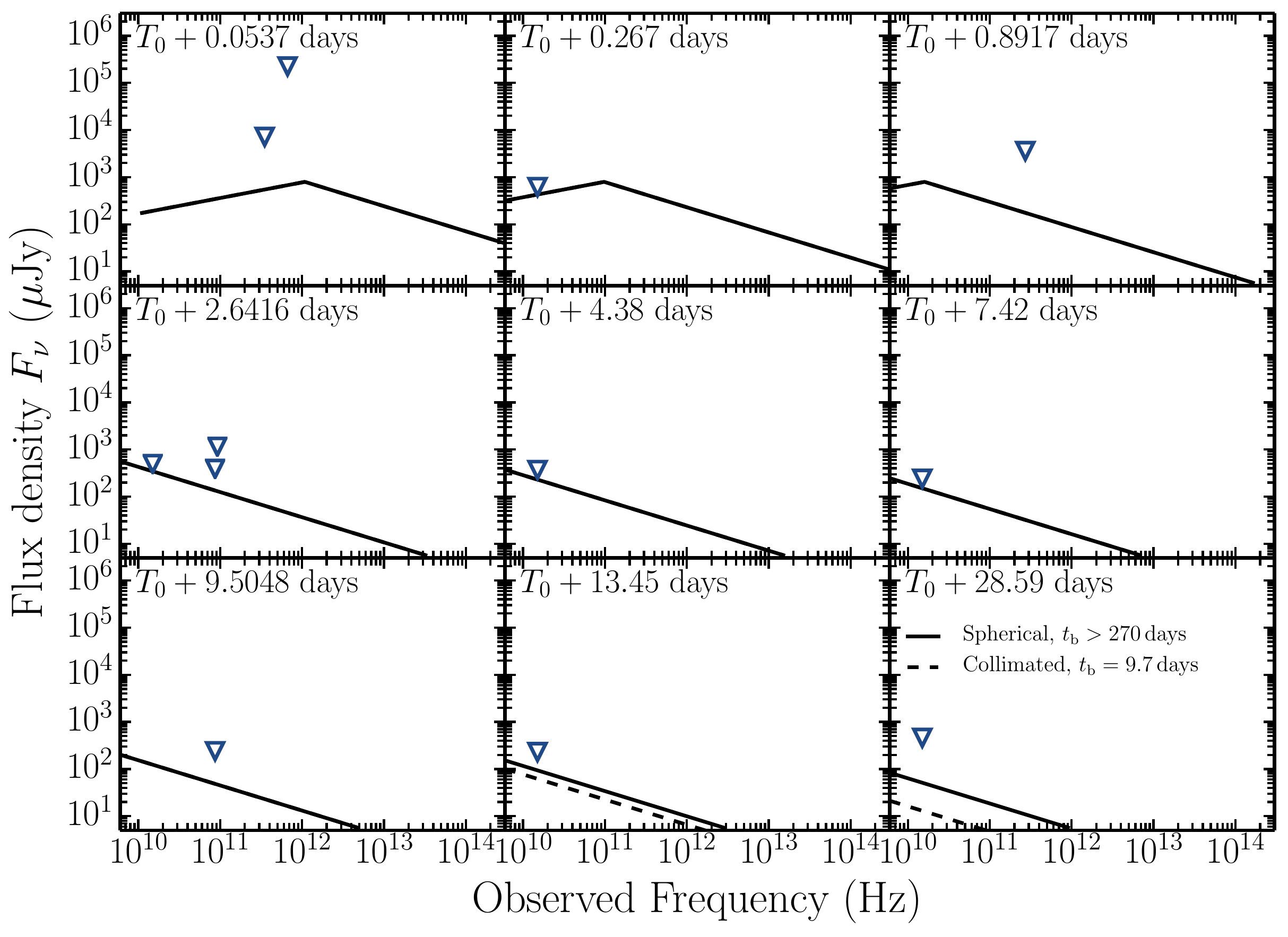}
\caption{\textbf{\textit{Left}}: Spectral energy distribution from the NIR to the X-ray
		at early epochs. The optical-to-X-ray SEDs are best described by absorbed
		broken power law (dashed lines) models modified by a blackbody (dotted lines).
		Data excluded from fits are shown as empty symbols and upper limits by triangles.
		\textbf{\textit{Right}}: Radio-to-sub-mm SEDs at the epochs listed in
		Table~\ref{tab:obs_log_3}. The NIR-to-X-ray SED from $T_0+0.267~\rm days$ was
		extrapolated to radio frequencies and evolved in time for a collimated and
		spherical expansion of the blast-wave (for details see text). The AMI-LT
		measurement from $T_0+2.38$ was shifted to 2.6416 days, assuming the injection
		frequency to be blueward of the observed bandpass and using the scaling
		relations in \citet{Sari1998a}. This has no implications on our analysis.
		}
\label{fig:SEDs}
\end{figure*}

Figure \ref{fig:SEDs} displays spectral energy distributions at 0.054 and 0.267 days after
the GRB. While afterglows have spectra formed by piecewise-connected power laws from
radio to X-rays \citep{Sari1998a}, the cooling phase of the stellar envelope
that was heated by the SN shock break-out is characterised by thermal emission
peaking in the UV.

The early UV emission is indeed well fitted with a blackbody (for
details see Sect. \ref{sec:radio_X-ray_sed}). We measure a blackbody temperature
of $kT_{\rm obs}=14~\rm{eV}$ and a blackbody radius of $9\times10^{13}~\rm{cm}$ at
$T_0+0.054~\rm{days}$. These values are consistent with expectation from the
shock-break-out model \citep[e.g.][and references therein]{Ensman1992a, Campana2006a}
and lie in the ballpark of observed values of Ib/c SNe, such as 1993J
\citep{Richmond1994a, Richmond1996a, Blinnikov1998a}, 1999ex \citep{Stritzinger2002a},
2008D \citep{Soderberg2008a, Malesani2009a, Modjaz2009a} and 2011dh
\citep{Arcavi2011a,Soderberg2012a,Ergon2013a}, and of the GRB-SNe 2006aj
\citep{Campana2006a} and 2010bh \citep{Cano2011a, Olivares2012a}.

The observed decline in the $u$ band  between its first detection and $T_0 +2.8~\rm days$
and the local minimum in the light curve before the SN rise is $\sim2$~mag. It
is comparable to that observed in GRB 060218 \citep{Campana2006a}. However, for this event,
these authors also reported a rise in brightness up to 0.57 days after the burst (shifted to
the observer frame of GRB 120422A). This initial rise is not present in our
data, although the first observation was at 86.4 s after the onset of the $\gamma$-ray
emission.

\subsection{The afterglow emission}\label{sec:afterglow}
\subsubsection{X-rays}\label{sec:X-ray}

\citet{Zhang2012a} reported that the early X-ray emission ($t<200~\rm s$) is consistent
with  high-latitude emission from the prompt emission phase
\citep[e.g.][]{Fenimore1997a,Kumar2000a,Dermer2004a}, with evidence for small-scale
deviation from power-law models \citep{Starling2012a}, possibly due to
a thermal component as seen in other GRBs \citep[e.g.][]{Campana2006a, Page2011a,
Starling2011a,Starling2012a, Sparre2012a, Friis2013a}. \citet{Friis2013a} suggested that
such a thermal component is not produced by the stellar photosphere but by the photosphere of
the GRB jet. In the following, we will focus on the emission at $>200~\rm{s}$ after the burst.

At the time of our \xmm\ observation the X-ray spectrum is adequately fit as an absorbed
power-law with a spectral slope of $\beta=94^{+0.12}_{-0.11}$, with absorption entirely
consistent with the Galactic column ($3.71 \times 10^{20}~\rm{cm}^{-2}$). The spectral
slope is consistent with that derived from the late time XRT spectrum
($\beta = 0.98\pm0.13$), and suggests no late time spectral changes
($t>4600~\rm s$). The spectral slope is typical for GRB afterglows at that phase.

The joint XRT and \xmm\ light curve is shown in Fig.~\ref{fig:OT_LC}, where we converted
the XRT observations to flux based on the mean spectral index of the system (following
\citealt{Evans2009a}), and then added the \xmm\ observations assuming their measured
spectral parameters.
The X-ray light curve is adequately fit by a multiply broken power-law with indices of
$\alpha_1 = 12.7\pm4.1$, $\alpha_2= 6.09\pm0.16$, $\alpha_3= 0.31\pm0.04$,
$\alpha_4= 1.48\pm0.40$, and break times of $t_{b,\,1}=95.3\pm3.2~\rm s$,
$t_{b,\,2}=394\pm19~\rm s$ and
$t_{b,\,3} = 330.5\pm89.0~\rm ks$, the resulting $\chi^2$/d.o.f. = 43.5/54. We note that
an early break is needed to fit the WT settling mode exposures, which has a chance
improvement  probability of $\sim 6.6 \times 10^{-5}$.

The steep-to-shallow-to-normal decay-phase
evolution is typical for X-ray afterglows of high-$L$ GRBs \citep{Nousek2006a, Evans2010a}.
In particular, the very rapid decay phase ($\propto t^{-13}$) unambiguously points
to high-latitude emission, and has not been observed for low-$L$ GRBs so far.

\subsubsection{Optical/NIR}\label{sec:opt_ag}

As mentioned before, the thermal emission of the cooling photosphere has an intrinsically
blue spectrum and does not significantly contribute to the integrated emission
in the optical and NIR. Therefore, the optical/NIR emission can be decomposed into three
distinct emission components: \textit{i}) the afterglow, which can be modelled with simple
and broken power-law models; \textit{ii}) the supernova; and \textit{iii}) the host galaxy,
which can be accounted for by a constant flux. To characterise the SN component,
we follow the approach in \citet{Zeh2004a}.  They used the multi-color light curves of the
prototypical GRB-SN 1998bw \citep{Galama1998a,Patat2001a} as templates. They derived the
SN 1998bw light curves at the given GRB redshift, and in the given observed band (including
the cosmological $k$-correction), and additionally modified the template with two
parameters. The luminosity factor $k$ determines the SN peak luminosity in a given band in
units of the SN 1998bw peak luminosity in that band. The stretch factor $s$ determines
if the light curve evolution is faster ($s<1$) or slower ($s>1$) than that of SN 1998bw,
whereby the actual evolutionary shape remains the same, and the explosion time is always
identical to the GRB trigger time. However, we limit the SN modelling to the
$g'r'i'z'J$ bands. Model light curves in bluer or redder filters require extrapolating the
spectral range of the SN1998bw template.

The results of our fits are given in Table~\ref{tab:sn_prop}. In this section we report
on the afterglow properties and on those of the SN in Sect. \ref{sec:ks_paradigm}.
The light curve fits reveal that there is indeed a power-law component, and hence provide
strong evidence for an optical/NIR afterglow  accompanying GRB 120422A. The fit with a simple power law makes the
assumption that the afterglow light curve does not steepen until $T_0+270.2~\rm{days}$.
For a collimated outflow the observer sees the edge of the jet at a certain time, resulting
in a significant steepening \citep{Sari1999a}. A jet break after 270 days has been observed
in GRB 060729 \citep[][see also \citealt{Perley2013c} for a further example of a very late
jet break]{Grupe2010a}, but a typical value is $\sim0.6$ day \citep[rest-frame; e.g.][]{Zeh2006a,Racusin2009a}. We refitted
the light curve with a smoothly broken power law \citep{Beuermann1999a}, where the
post-break decay slope was fixed to 2. The
pre-break slope is identical to the value from the simple power law fit. The jet-break time
of $9.7\pm4.4$ days (observer frame) is still large and very uncertain, but its value is
more consistent with the observed distribution in \citet{Racusin2009a}. A reason for this
large uncertainty in the break time is the brightness of the SN.

Both afterglow models over-predict the $i'$-band brightness at $T_0+1880~\rm s$
by 0.9 mag. The required rise could be either due to the crossing of the injection frequency $\nu_m$
or due to the coasting phase before the afterglow blast-wave begun decelerating. In the former case
the rise slope $\alpha_{\rm r}$ is -0.5 \citep[with $F_\nu\propto t^{-\alpha_{\rm r}}$;][]{Sari1998a},
and in the latter between $-3$ and $-2$ for constant-density medium and $>0.5$ for a
free-stellar-wind density profile \citep{Shen2012a}.

The crossing of the injection frequency $\nu_{\rm m}$ is by definition a chromatic
feature. It evolves $\propto t^{-3/2}$ \citep{Sari1998a}. This means the ratio between
break times in two different bands has to obey $t_2/t_1 = \left(\nu_2/\nu_1\right)^{-2/3}$.
The $J$ band has the earliest detection after the first $i'$ observation and is not
affected by the thermal emission from the cooling stellar photosphere. Since the $J$-band
light curve is only decaying, $\nu_{\rm m}$ crossed this band at $t<4550~\rm s$ after the
burst and hence the $i'$ band at $\lesssim3260~\rm s$. Already in the limiting
case, the expected $i'$ band magnitude is 0.24 mag brighter than the observed value.
Given the small photometric error of 0.04 mag makes the deviation statistically significant
and hence this scenario unlikely. The blast-wave's coasting into a free-stellar-wind
ambient density profile is also in conflict with our data, since we detect a clear rise and
not a shallow decay.

A steep rise of $\alpha_{\rm r}=-2$ to $-3$ is fully consistent with our data. In both cases, the
break time is $\sim2550~\rm s$ (observer frame). We hence identify the coasting phase
into a constant-density circumburst medium as the most likely scenario. Since the break
time determines the transition from the coasting to the
deceleration phase, it can be used to measure the initial Lorentz factor $\Gamma_0$ of the
decelerating blast-wave \citep{Sari1999b,Panaitescu2000a,Meszaros2006a}.
Following \citet{Molinari2007a}, we measure $\Gamma_0\sim60$ using the observed break
time and the measurement of the energy $E_{\rm iso}=4.5\times10^{49}~\rm erg$ released
during the prompt $\gamma$-ray emission.

\subsubsection{The SED from the radio to the X-rays}\label{sec:radio_X-ray_sed}

To characterise the afterglow properties in more detail, we model the joint NIR-to-X-ray
spectral energy distribution (SED). We limit this analysis to $<T_0+1~\rm day$, since
SN 2012bz started contributing a non-negligible amount of flux
to the integrated light at later times. We choose the epochs $T_0+0.054~\rm days$ and
$T_0+0.267~\rm days$ to match the dates of the sub-mm/mm observations. The optical and NIR
fluxes were obtained through interpolation between adjacent data points.\footnote{ In the
UV, there are cases where one of the adjacent data points is an upper limit but the epoch
of the SED is very close to the time of the detection ($\Delta t<0.1$ dex). In these cases
we treated the interpolated data point as detection but not as upper limit.}
Errors were estimated by interpolation. The flux scales of the XRT and XMM (MOS1, MOS2, PN)
data were adjusted to the brightness of the X-ray afterglow at the respective epochs.

The NIR-to-X-ray SEDs, shown in Fig.~\ref{fig:SEDs}, have in common that the UV emission
is dominated by radiation from the cooling stellar envelope after the shock break-out (Sect.
\ref{sec:SN2012bz_SBO}). To account for this thermal emission, we fit the NIR-to-X-ray SED
with absorbed simple and broken power-law models modified by a blackbody model using
\texttt{Xspec} v12.8.0. The blackbody model is defined by:
\begin{equation}
BB\left(E;\,C,\,T\right)=1.0344\times10^{-3}\,C\,\frac{E^2\,\Delta E}{\exp\left(E/k_{\rm B}\,T\right)-1}\nonumber
\end{equation}
where the numerical constant $C$ is defined as $R_{\rm km}^2/D_{10\,\rm kpc}^2$, where $R$ is the
blackbody radius in km, $D$ the distance in units of 10 kpc, $k_{\rm B}$ the Boltzmann constant,
$T$ the temperature in units of keV, $E$ the energy and $\Delta E$ is the width of the energy bin.

Both SEDs are best fitted by a broken power
law with $\beta_{\rm o}\sim0.5$ and
$\beta_{\rm x}\sim1$ and a break energy of $\sim4$~eV. The difference in the slopes is
consistent with the expected value for synchrotron radiation, if the cooling break is
between both bands \citep{Sari1998a}. This is a further circumstantial evidence
that the optical and X-ray emission are produced by the afterglow. Given the sparse
sampling of the optical/NIR bands, we fit both epochs simultaneously and fix the
difference in the spectral slopes to 0.5 and set the break energies to identical values.
The joint fit gives a spectral slope of $\beta_{\rm o}\sim0.46$ in the optical (i.e. $\beta_{\rm x}=0.96$),
a break energy of $\sim4$~eV, no evidence for a significant
host absorption at X-ray energies, and a blackbody temperature of $\sim16$~eV and radius
of $\sim7\times10^{13}~\rm{cm}$ at 1.4~hours after the burst. The blackbody component in
the second epoch is barely constrained because of the limited amount of UV data.
The combined fit statistics is 114.7/74 d.o.f.

The peak of an afterglow spectrum is typically at cm/sub-mm wavelengths, and usually
crosses this band within the first week. We therefore extrapolate the afterglow SED from
$T_0+0.267~\rm days$ to radio wavelengths (Fig.~\ref{fig:SEDs}) and evolve the SED to all epochs
of the radio and sub-mm observation listed in Table~\ref{tab:obs_log_3}, using the scaling
relations for the injection frequency and the peak flux density for a spherical expansion
and a post-jet beak evolution from \citet{Sari1998a,Sari1999a}. In both dynamical scenarios,
the peak flux density is $\lesssim800~\mu\rm Jy$, corresponding to a specific luminosity of
$\lesssim2\times10^{30}~\rm{erg\,s}^{-1}\,\rm{Hz}^{-1}$ before the jet break occurred.

\subsection{Supernova properties}\label{sec:sn_prop}\label{sec:SN2012bz_radioactive}
\subsubsection{Supernova spectrum}

\begin{figure*}
\centering
\includegraphics[viewport=2 7 575 955, clip, angle=0, width=1.525\columnwidth]{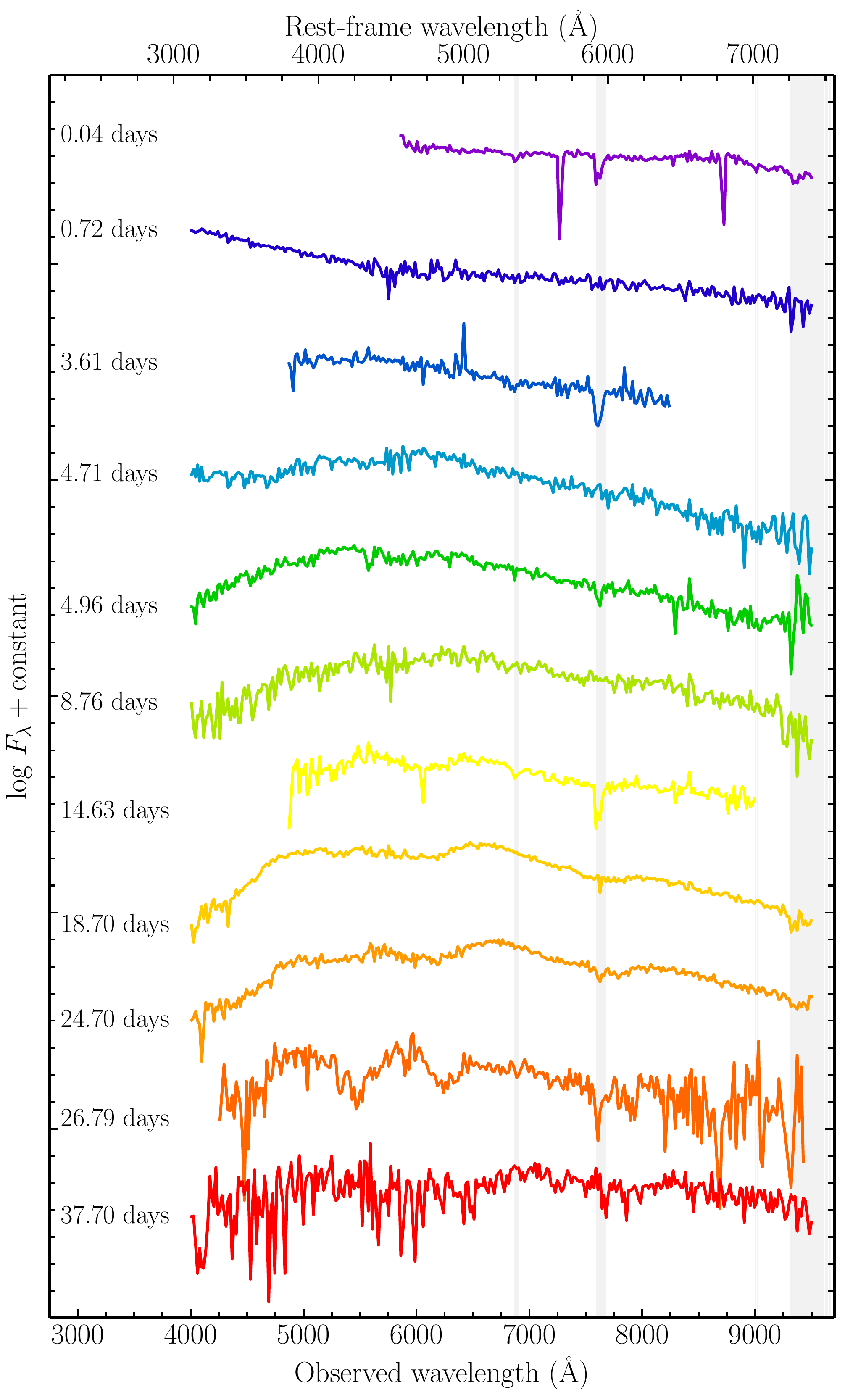}
\caption{Spectral evolution of the optical transient accompanying GRB 120422A. The
first two epochs show a smooth power-law-shaped continuum, characteristic of GRB
afterglows. After the transient started re-brightening, the shape of the spectrum
becomes redder. At 8.8 days after the GRB, the spectrum has clearly
started to resemble that of a broad-lined SN. At 18.7 days, the transformation is
complete and the spectra look similar to other GRB-SNe. All spectra were shifted vertically
by an arbitrary constant. They were rebinned (18 \AA) to increase S/N for presentation
purposes. We only display spectra with a large spectral range. Strong telluric lines
(transparency $<20\%$) are highlighted by the grey-shaded areas.
}
\label{fig:AG_SN_spec_series}
\end{figure*}

\begin{table}
\caption{Properties of the SN modelling}
\centering
\scriptsize
\begin{tabular}{ccccc}
\toprule
\multicolumn{5}{l}{\textbf{Simple power law + free host magnitude}}\\
\multicolumn{5}{l}{$\alpha_1=0.69\pm0.02$}\\
\midrule
\multirow{2}{*}{Band}	& \multirow{1}{*}{Host magnitude}	& Luminosity	& Stretch 	& \multirow{2}{*}{$\chi^2/\rm{d.o.f.}$}\\
						& (mag)								& factor $k$	& factor $s$& \\
\midrule
$g'$	&$ 24.65\pm0.12$&$	0.86	\pm	0.03	$&$	0.94	\pm	0.02	$& \multirow{5}{*}{$194.9/146$}\\
$r'$	&$ 24.06\pm0.04$&$	1.25	\pm	0.02	$&$	0.89	\pm	0.02	$\\
$i'$	&$ 24.17\pm0.08$&$	1.10	\pm	0.01	$&$	0.92	\pm	0.01	$\\
$z'$	&$ 24.31\pm0.12$&$	0.99	\pm	0.02	$&$	0.92	\pm	0.03	$\\
$J$		&$ 24.22\pm0.22$&$	1.12	\pm	0.09	$&$	0.74	\pm	0.12	$\\
$H$		& \dots		&\dots					 &\dots \\
\toprule
\multicolumn{5}{l}{\textbf{Smoothly broken power law + fixed host magnitude}}\\
\multicolumn{5}{l}{$\alpha_1=0.67\pm0.02$, $\alpha_2=2.00$ (fixed), $t_b\,\left(\rm{days}\right)=9.7\pm4.4$,}\\
\multicolumn{5}{l}{$n=10$ (fixed)}\\
\midrule
\multirow{2}{*}{Band}	& \multirow{1}{*}{Host magnitude} 		& Luminosity	& Stretch 	& \multirow{2}{*}{$\chi^2/\rm{d.o.f.}$}\\
						& (mag)							 		& factor $k$	& factor $s$& \\
\midrule
$g'$	&	24.62			&$	0.88	\pm	0.05	$&$	0.97	\pm 0.02	$& \multirow{5}{*}{$186.6/150$}\\
$r'$	&	24.09			&$	1.25	\pm 0.02	$&$	0.90	\pm 0.01	$\\
$i'$	&	24.09			&$	1.11	\pm 0.02	$&$	0.92	\pm 0.01	$\\
$z'$	&	24.15			&$	0.99	\pm 0.03	$&$	0.92	\pm 0.03	$\\
$J$		&	23.96			&$	1.06	\pm 0.09	$&$	0.68 	\pm 0.09	$\\
$H$		&	23.84			& \dots				& \dots\\
\bottomrule
\end{tabular}
\tablefoot{Best-fit parameters of the $g'r'i'z'JH$ band light curve
	fits. We modelled $g'r'i'z'J$ light curves with a SN1998bw template redshifted
	to $z=0.2825$, as described in \citet{Zeh2004a}, superposed on a simple
	power law or smoothly broken power law \citep{Beuermann1999a}, where $\alpha$
	denote the decay slopes, $t_{\rm b}$ the break time and $n$ the smoothness,
	to account for the early emission and the flux from the host
	galaxy at the explosion site. Note, for the $H$ band we used the afterglow models.
	We assumed that the afterglow component
	evolves achromatically from the $g'$ to the $H$ band.
	The supernova and afterglow light curve is equally well fitted with
	the two models. See Sect. \ref{sec:ks_paradigm} for details.
	}
\label{tab:sn_prop}
\end{table}

Our spectra  of SN~2012bz are displayed in Fig.~\ref{fig:AG_SN_spec_series}. The very
early spectra are dominated by a smooth power-law continuum, characteristic of GRB
afterglows.
At around 4.7 days, after the transient started re-brightening
(Fig.~\ref{fig:OT_LC}), the shape of the spectrum changes and it becomes redder.  By
May 1 (8.8 days after the GRB), the spectrum has clearly started to resemble that of a
supernova with broad lines \citep[Sect. \ref{sec:Spec_comp};][]{Malesani2012c,
Sanchez2012a, Wiersema2012a}. By May 10 (18.7 days after the GRB), the transformation is complete and our X-shooter
spectra from $+$18.7 and $+$24.7 days are very similar to those of other broad-lined Type Ic SNe
accompanying GRBs (Fig.~\ref{fig:Spec_comp}). The Magellan spectrum obtained 26.8 days
after the GRB has a low S/N despite showing absorption troughs at locations
consistent with the previous data and  should be interpreted with great caution.
The modelling of the spectral evolution will be presented in a forthcoming paper.

Usually, GRB-SN expansion velocities are reported using $\textrm{\ion{Si}{ii}}\lambda6355$,
with the \ion{Ca}{ii} NIR triplet at 8600~\AA{} reported sometimes as the only alternative
\citep{Patat2001a,Hjorth2003a,Chornock2010a,Bufano2012a}. 
In the case of SN~2012bz, the
\ion{Si}{ii} line is contaminated by the telluric A-band, while the Ca IR triplet is redshifted
outside the optical spectrum. For this reason, we chose to measure the expansion velocities based
on the $\textrm{\ion{Fe}{ii}}\lambda5169$ feature.
In addition, this feature appears earlier than the \ion{Si}{ii}
feature and its minimum is easier to locate as it lies between two clearly visible maxima
(Fig.~\ref{fig:AG_SN_spec_series}, \ref{fig:Spec_comp}). This makes it a potentially better
expansion velocity tracer  for GRB-SNe than \ion{Si}{ii}, which
is super-imposed on a blue continuum and it is not always easy to locate and measure, especially at early times.

\begin{figure}
\centering
\includegraphics[viewport=0 4 637 632, clip, angle=0, width=1\columnwidth]{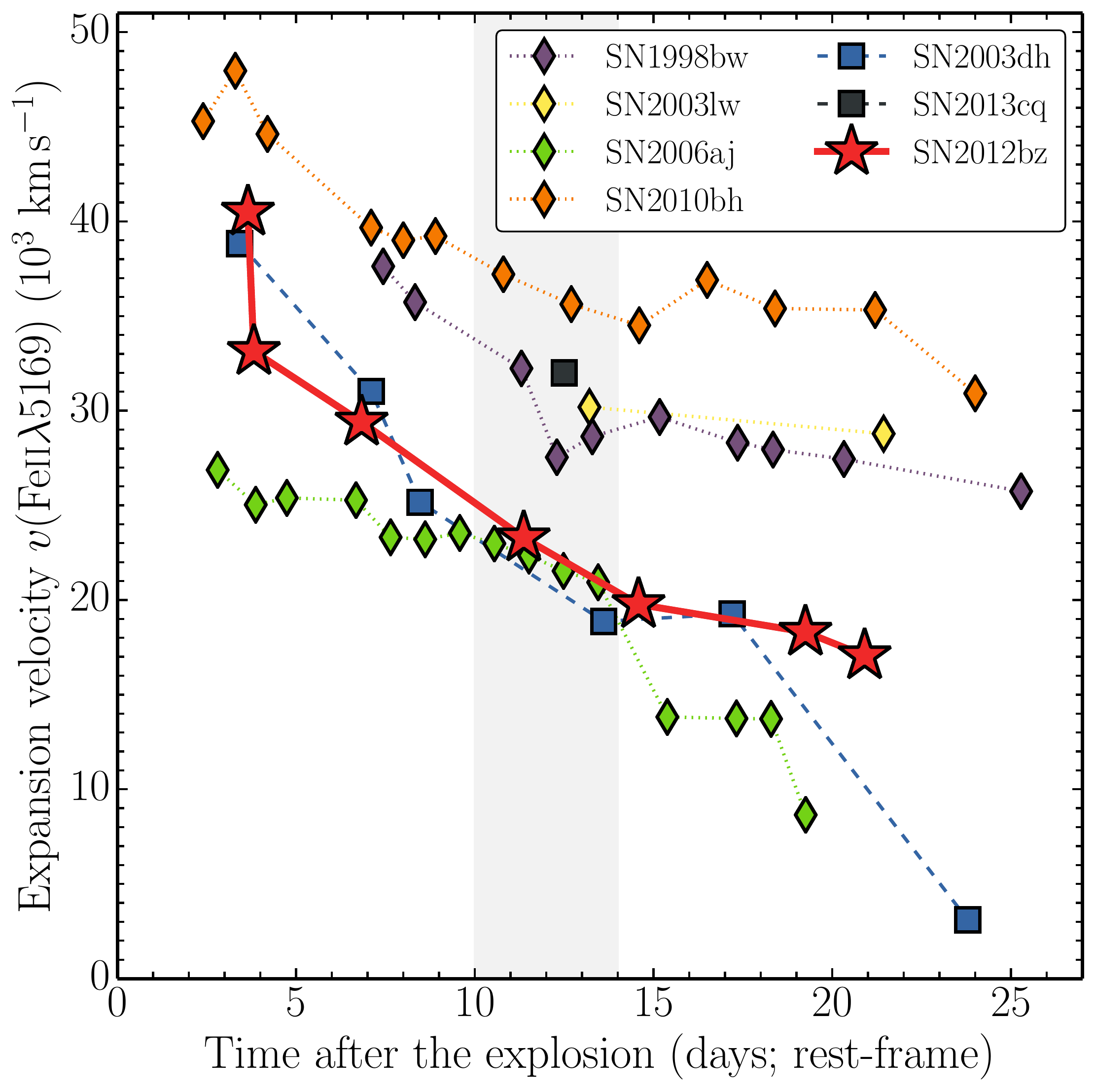}
\caption{Evolution of the expansion velocities measured from $\textrm{\ion{Fe}{ii}}\lambda5169$
	for SN~2012bz
	and six GRB-SNe of low (diamonds) and high-luminosity GRBs (boxes) with good
	spectroscopic data. Measurements were performed on our data as well on the spectra of
	\citet{Patat2001a}, \citet{Hjorth2003a}, \citet{Malesani2004a}, \citet{Pian2006a}, and
	\cite{Bufano2012a}. The value of SN 2013cq was taken from \cite{Xu2013a}. The grey shaded area displays the interval
	of observed GRB-SN peak times.
	}
\label{fig:Fe_exp_vel}
\end{figure}

We have used the fiducial rest-wavelength of 5169 \AA\@\ for \ion{Fe}{ii}, as done e.g. in
\cite{Hamuy2002a} for the expansion velocities of Type IIP SNe. We stress that even if this
identification is not correct for GRB-SNe, due to blending, these measurements are still
valuable in order to monitor the expansion velocity evolution and for comparison between
different objects as long as the measurements are done consistently. Based on these assumptions,
we present the first, to our knowledge, diagram of GRB-SNe expansion velocities, based on
$\textrm{\ion{Fe}{ii}}\lambda5169$ (Fig.~\ref{fig:Fe_exp_vel}). The velocities
(of the order of 5000--50000~km\,s$^{-1}$) 
are in the range measured  for other SNe associated with GRBs.
SN~2010bh shows the fastest explosion velocities as seen from \ion{Si}{ii}, while
SN~2006aj the slowest \citep{Chornock2010a,Bufano2012a}. SN~2012bz shows large
velocities at 3 days past explosion (the earliest spectrum where a measurement is possible)
and slowing down to 17000~km\,s$^{-1}$ $\sim20$ days later. This behaviour is very similar to
SN~2003dh associated with the high-$L$ GRB 030329 \citep{Hjorth2003a}.

\subsubsection{Absolute magnitude}\label{sec:ks_paradigm}

The luminosities of SNe are usually reported in the rest-frame $V$ band. The $r'$
bandpass (observer frame)
partly overlaps with the rest-frame $V$ band, though it is not identical. We compute
the $k$-corrected $V$-band magnitude from the $r'$-band maximum, following
\citet{Hogg2002a} and using the
X-shooter spectrum from $T_0+18.7~\rm days$ (i.e. $<2~\rm days$ after the
maximum in $r'$-band) as a weighing function. The peak luminosity of
$M_{\rm V}=-19.7~\rm mag$ is 0.3 mag brighter than SN 1998bw, using the face value
of $M_{\rm V}=-19.4~\rm mag$ from \citet[][]{Cano2011b}.

Measuring the SN luminosity by using a $k$-correction from the observed spectrum is the
most direct and accurate approach. However, the number of spectroscopically confirmed
GRB-SNe is still small. Moreover, optical spectroscopy is limited to  mostly
low redshifts ($z<0.3$), because of the prohibitively long exposures required for a
$M_{\rm V}\sim-19~\rm mag$ SN at higher redshifts. In addition, the useful wavelength
range is reduced to the red part of the observed spectrum due to line
blanketing by iron, as rest-frame UV moves into the optical $V$ band, \citep{Filippenko1997a}.
An alternative approach is to look for ``late red bumps''
in afterglows light curves, which are due to the GRB-SNe.
The best fit parameters of the SN bump with SN 1998bw templates in the $g'r'i'z'J$ bands,
as detailed in Sect. \ref{sec:opt_ag}, are displayed in Table~\ref{tab:sn_prop}. The fit
reveals that SN 2012bz is 0.3 mag more luminous
than SN 1998bw in the observed $r'$ band and the evolution is slightly faster than
that of SN 1998bw, and it is somewhat redder.

\subsubsection{The explosion-physics parameters}\label{sec:bolo}

\begin{figure}
\centering
\includegraphics[width=1\columnwidth]{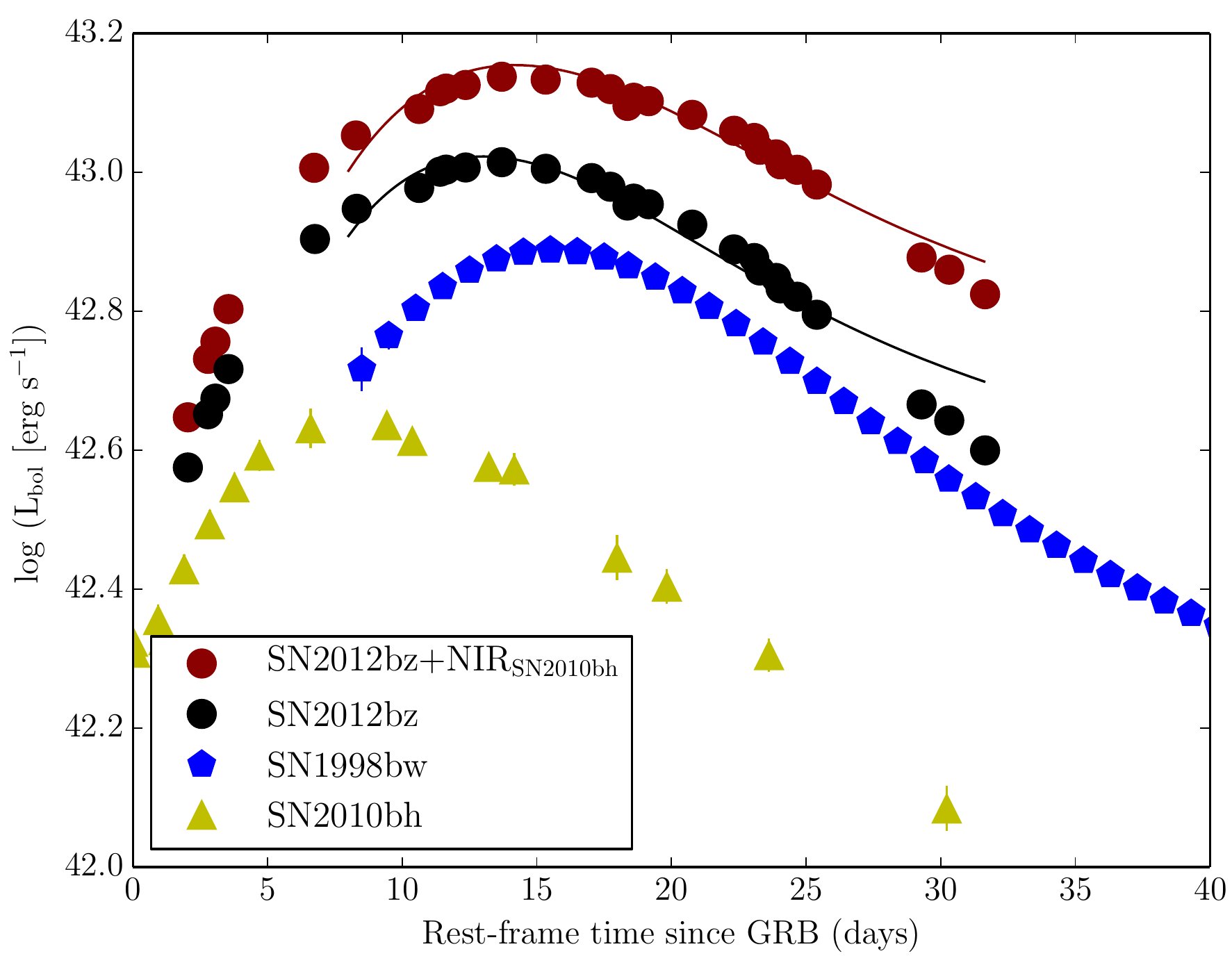}
\caption{Pseudo-bolometric light curves of SN 2012bz from direct integration of the SED
	over $g'r'i'z'$ filters, and including a NIR contribution as found for SN 2010bh.
	For comparison the $UBVRI$ light curve of SN 1998bw \citep{Clocchiatti2011a} and the
	$g'r'i'z'JH$ light curve of SN 2010bh are shown \citep{Olivares2012a}. The models for
	SN 2012bz are shown as solid lines. Early light-curve time data are not fitted as the
	analytical model does not account for other non-negligible sources of luminosity at
	these times (Sect. \ref{sec:bolo}).  Only photometric and calibration uncertainties
	are included in the error bars which are usually smaller than the size of the plot
	symbol.}
\label{fig:bolometric_lc}
\end{figure}

The peak and width of a SN light curve are determined by the explosion-physics parameters,
such as ejecta mass $M_{\rm ej}$, $^{56}$Ni mass $M_{\rm Ni}$, and kinetic energy $E_{\rm k}$
of the SN ejecta. These values are estimated from the bolometric light curve.
An estimate of the bolometric light curve was constructed using $g'r'i'z'$ photometric
points, as coverage outside these bands is limited around the SN peak.
The light curves in each filter were fitted with spline interpolations starting at 2
days past the GRB trigger, such that an estimated magnitude for all four bands was
available at each epoch of observation. Magnitudes were converted into monochromatic
fluxes at the effective (rest-frame) wavelengths of the filters for every
epoch to produce an SED.\footnote{Since we are evaluating the SED for every observation, nearby epochs
(within $<0.2~\rm day$ of each other) were first calculated individually and then averaged
when producing the final light curve for clarity.} Each SED was then integrated over the
limits of the filter wavelength range, i.e. the blue edge of $g'$ and the red edge of $z'$
($\sim3000$--8000 \AA). The SED was tied to zero flux at these limits, which were defined
as the wavelength at which the respective filter's normalised transmission curve falls
below 10\%. The integrated
fluxes were converted to luminosities using the redshift and cosmology adopted previously.
The resulting light curve (Fig.~\ref{fig:bolometric_lc}) gives a luminosity of the SN over
approximately the optical wavelength range.

Contributions to the flux outside this regime, however, are not insignificant, with the
optical accounting for $\sim50-60\%$ of the bolometric flux for stripped-envelope SNe
\citep{Lyman2013a}. Of particular importance is the contribution from the NIR, wherein
the fraction of the total luminosity emitted increases with time, reaching a comparable
contribution to the optical within 30 days \citep[e.g.][]{Valenti2008a, Cano2011a}. We
estimate this missing NIR flux by using
the fractional NIR flux of a similar event, as done in \citet{Cano2011a}. A photometric study by
\citet{Olivares2012a} of the low redshift ($z=0.059$) XRF 100316D/SN 2010bh contains
well sampled light curves in the $z'JH$ bands, extending upon our rest-frame wavelength
limits. The contribution of wavelengths $>8000$~\AA\ to the flux was determined by
first integrating SN 2010bh's de-reddened SED over the same wavelength range used for
SN 2012bz above, and then over the wavelength range redward of 8000 \AA. Thus, for
each epoch of observation, we obtain the NIR contribution as a fraction of the optical
flux. The phase of the contributions were normalised so $t=0$ was the peak of the
respective SNe, and stretched by a factor $\Delta m_{15,~(3000-8000)\,{\textup{\AA}}}$ to match the
light curve shape of the two SNe ($\Delta m_{15,~(3000-8000)\,{\textup{\AA}}}=0.78$ for SN 2012bz, 1.00 for SN 2010bh).\footnote{\citet{Phillips1993a}
introduced $\Delta m_{15}$ as the decline in the brightness between the
maximum and 15 days post maximum in $B$ band.} The fractional
values were interpolated using a smooth spline, in order to sample it at the epochs of
observations of SN 2012bz, and the appropriate amount was added to the optical flux. This
gives a NIR corrected light curve covering 3000--17000~\AA\@. No attempt was made to
account for flux missed below 3000~\AA\ due to the paucity of data constraining the UV
in such objects. However, contributions from the UV account for only $\sim 5-15\%$ of the
bolometric flux around peak \citep{Lyman2013a}.

The bolometric light curve was modelled using the simplified analytical prescription
of \citet{Arnett1982a}, updated by \citet{Valenti2008a}, to obtain estimates of the $M_{\rm Ni}$
$M_{\rm ej}$ and $E_{\rm K}$.
Since obtaining a truly bolometric light curve is unfeasible, we use our optical and
optical+NIR correction light curves as approximations. Our data cover the
photospheric phase of SN evolution, when the ejecta are considered optically thick.
The opacity is chosen to be $\kappa = 0.07~\rm{cm}^{2}\,\rm{g}^{-1}$ \citep[see][]{Cano2011a}.
To constrain the $E_{\textrm{K}}$/$M_{\textrm{ej}}$ ratio, a \emph{scale velocity} is required (see
equation 54 in \citealt{Arnett1982a}), this is taken to be the photospheric velocity
($v_{\textrm{ph}}$) at peak. \ion{Fe}{ii} lines are a good tracer of $v_{\textrm{ph}}$ \citep{Valenti2011a}, and
the peak of the pseudo-bolometric light curve occurs at $\sim$13.9 days (from fitting
low-order polynomials around peak). Using data in Fig.~\ref{fig:Fe_exp_vel} we take
20500~km\,s$^{-1}$ as an estimate of $v_{\textrm{ph}}$ at peak by linearly interpolating between the
measurements taken from spectra at epochs 11.380 days and 14.575 days.

Fitting to the optical bolometric light curve reveals the following parameters:
$M_{\textrm{Ni}}=0.40\pm0.01~\rm{M}_{\odot}$, $M_{\textrm{ej}}=4.72\pm0.04~\rm{M}_{\odot}$ and
$E_{\textrm{K}}=3.29\pm0.03\times10^{52}$~erg, and when including the NIR contribution from
SN 2010bh, we obtain $M_{\textrm{Ni}}=0.58\pm0.01~\rm{M}_{\odot}$, $M_{\textrm{ej}}=5.87\pm0.03~\rm{M}_{\odot}$
and $E_{\textrm{K}}=4.10\pm0.03\times10^{52}$~erg. The first 8 days were ignored in the fit
as contributions from other sources (GRB afterglow and cooling phase following the shock break-out) would compromise
the assumptions of the SN model. 

It is crucial to note that the errors quoted here include only the statistical uncertainties
relating to the construction of the pseudo-bolometric light curves. Systematic errors arise
from both the simplifying assumptions in the model (spherical symmetry, centrally
concentrated $^{56}$Ni mass etc.) and our choice of parameters for the fit, which
typically dominate the statistical errors. For example taking an uncertainty in $v_{\textrm{ph}}$
of 2000~km\,s$^{-1}$ translates into an error in $M_{\textrm{ej}}$ and $E_{\textrm{K}}$ of $\sim10\%$ and
$\sim25\%$, respectively. The two-component model for very energetic supernovae ($E_{\rm k}\gtrsim5\times10^{51}~\rm{erg}$)
 by \citet{Maeda2003a}
would also suggest we are only observing the outer, lower density region of the ejecta
during the photospheric phase ($\lesssim30$ days), and a fraction is hidden in a denser,
inner component during this time. Although the afterglow component is not expected to
contribute significantly around the SN peak, given that different afterglow models
do not significantly affect the $k,s$ parameters (Sect \ref{sec:ks_paradigm}), 
potential contamination by underlying host galaxy light is included in this bolometric
light curve (Sect. \ref{sec:photometry}).

\citet{Melandri2012a} modelled SN 2012bz using a scaled spectral
model for SN 2003dh to obtain estimates of the physical parameters. They obtained values
of $M_{\textrm{Ni}}\approx0.35~\rm{M}_{\odot}$, $M_{\textrm{ej}}\approx7~\rm{M}_{\odot}$ and
$E_{\textrm{K}}\approx3.5\times10^{52}$~erg using a bolometric light curve covering 3300--7400
\AA. Comparing these to our values for the optical (3000--8000 \AA) bolometric light curve,
the $M_{\textrm{Ni}}$ values are in good agreement, given our slightly extended wavelength
range, $E_{\textrm{K}}$ values are consistent, however their derived ejected mass is larger
than our measurement. Differences could be caused by the choice of photospheric velocity
$v_{\textrm{ph}}$, asymmetries, or varying opacity $\kappa$, which spectral modelling can
account for.

\section{Environments}\label{sec:environments}
Absorption and emission lines are powerful diagnostics to characterise
the gas and dust phase of interstellar media, such as the extinction,
metallicity and star-formation rate (SFR). Since long GRBs are associated with massive stars,
these diagnostics give the unique opportunity to study star-forming
regions in distant galaxies. In the following, we present our findings on the
explosion site, on the host galaxy and its large scale environment (for an independent analysis see \citealt{Levesque2012a}).

\subsection{The explosion site}\label{sec:afterglow_spectrum}

\begin{table}
\caption{Absorption and emission lines at the explosion and the host site}
\centering
\scriptsize
\begin{tabular}{l@{\hspace{0.3cm}}l@{\hspace{.3cm}}l@{\hspace{.3cm}}c@{\hspace{.3cm}}c@{\hspace{.3cm}}c}
\toprule
\multicolumn{1}{c}{\multirow{2}{*}{$\lambda_{\rm obs}$({\AA})}}	& \multicolumn{1}{l}{\multirow{2}{*}{Transition}} & \multicolumn{1}{l}{\multirow{2}{*}{redshift}} & \multicolumn{1}{c}{\multirow{2}{*}{$EW_{\rm{obs}}$(\AA)}} & $F\times10^{16}$\\
								&			&			&								& $(\rm{erg\,cm}^{-2}\rm{s}^{-1})$\\
\midrule
\multicolumn{5}{l}{Explosion site ($\left\langle z\right\rangle _{\rm abs}= 0.28253$, $\left\langle z\right\rangle _{\rm em}=0.28259$)}\\
\midrule
3586.22	&	$\textrm{\ion{Mg}{ii}}\,\lambda2796$ 	&	0.2824	&$3.25 \pm 0.42$	&\dots 	\\
3595.95	&	$\textrm{\ion{Mg}{ii}}\,\lambda2803$	&	0.2827	&$1.86 \pm 0.46$	&\dots 	\\
\dots	&	$\textrm{\ion{Mg}{i}}\,\lambda2852$		&  	\dots	& $<1.57$			& \dots		\\
4779.79	&	$[\textrm{\ion{O}{ii}}]\,\lambda3727$	&	0.28245	&\dots				&$	0.09	\pm	0.01	$\\
4783.37	&	$[\textrm{\ion{O}{ii}}]\,\lambda3729$	&	0.28245	&\dots				&$	0.16	\pm	0.01	$\\
5046.00	&	$\textrm{\ion{Ca}{ii}}\,\lambda3934$	&  	0.2825	&\dots				&$	0.75 	\pm 0.25$	\\
\dots	&	$\textrm{\ion{Ca}{ii}}\,\lambda3968$	&  	\dots	& $<1.08$			& \dots		\\
6236.87	&	H$\beta$								&	\dots	&\dots				&$	0.05	\pm	0.04	$\\
6362.18	&	$[\textrm{\ion{O}{iii}}]\,\lambda4959$	&	0.28262	&\dots				&$	0.05	\pm	0.02	$\\
6423.34	&	$[\textrm{\ion{O}{iii}}]\,\lambda5007$	&	0.28255	&\dots				&$	0.19	\pm	0.02	$\\
8419.55	&	H$\alpha$								&	0.28257	&\dots				&$	0.24	\pm	0.01	$\\
8447.92	&	$[\textrm{\ion{N}{ii}}]\,\lambda6583$	&	0.28286	&\dots				&$	0.06	\pm	0.02	$\\
8616.96	&	$[\textrm{\ion{S}{ii}}]\,\lambda6717$	&	0.28261	&\dots				&$	0.03	\pm	0.01	$\\

\midrule
\multicolumn{5}{l}{Host site ($\rm{PA}=41^\circ$; $\left\langle z\right\rangle _{\rm em}=0.28256$)}\\
\midrule
4780.16	&	$[\textrm{\ion{O}{ii}}]\,\lambda3727$	&	0.28254	&\dots&$	2.30	\pm	0.03	$\\
4783.73	&	$[\textrm{\ion{O}{ii}}]\,\lambda3729$	&	0.28254	&\dots&$	3.50	\pm	0.67	$\\
4920.46	&	H$\eta$									&	0.28255	&\dots&$	0.09	\pm	0.01	$\\
4963.37	&	$[\textrm{\ion{Ne}{iii}}]\,\lambda3869$	&	0.28262	&\dots&$	0.27	\pm	0.02	$\\
4979.74	&	H$\zeta$								&	0.28248	&\dots&$	0.20	\pm	0.01	$\\
5093.04	&	H$\epsilon$								&	0.28250	&\dots&$	0.19	\pm	0.02	$\\
5262.10	&	H$\delta$								&	0.28252	&\dots&$	0.30	\pm	0.02	$\\
5567.86	&	H$\gamma^{a}$							&	0.28242	&\dots&$	0.59	\pm	0.04	$\\
6236.89	&	H$\beta$								&	0.28260	&\dots&$	1.28	\pm	0.04	$\\
6362.15	&	$[\textrm{\ion{O}{iii}}]\,\lambda4959$	&	0.28261	&\dots&$	0.83	\pm	0.03	$\\
6423.60	&	$[\textrm{\ion{O}{iii}}]\,\lambda5007$	&	0.28261	&\dots&$	2.51	\pm	0.05	$\\
8419.75	&	H$\alpha^{b}	$						&	0.28260	&\dots&$	5.36	\pm	0.05	$\\
8446.38	&	$[\textrm{\ion{N}{ii}}]\,\lambda6583$	&	0.28262	&\dots&$	0.81	\pm	0.04	$\\
8616.90	&	$[\textrm{\ion{S}{ii}}]\,\lambda6717$	&	0.28260	&\dots&$	0.91	\pm	0.02	$\\
8635.40	&	$[\textrm{\ion{S}{ii}}]\,\lambda6731$	&	0.28260	&\dots&$	0.67	\pm	0.03	$\\
\bottomrule
\end{tabular}
\tablefoot{The reported wavelengths were derived from the first momentum of a
line profile (see Fig. \ref{fig:Spec_AG} and \ref{fig:Spec_Nuc}).
The fluxes were corrected for foreground extinction.
\tablefoottext{a}{Blended line.}
\tablefoottext{b}{This value is the total flux of both velocity components.}}
\label{tab:GRB_abs_em_lines}
\end{table}

The X-shooter spectrum, obtained on 23 April (17.2 hours post burst; see Fig. \ref{fig:Spec_AG}), exhibits
two absorption lines, which we identify as  \ion{Mg}{ii}$\lambda\lambda$2796,2803
(see Table~\ref{tab:GRB_abs_em_lines}).
After applying the heliocentric correction, we measure a mean absorption-line redshift of
$z_{\rm abs}=0.28253\pm0.00008$ (the error denotes the standard error of the mean), refining the redshift measurements of \citet{Schulze2012b} and
\citet{Tanvir2012a}.

Both lines lie in a rather noisy part of the spectrum. To measure their equivalent
widths, we rebinned the spectrum by a factor of two to increase the S/N (i.e. a wavelength
binning of 0.4 \AA), and fixed the aperture size for the weaker \ion{Mg}{ii} line to
$100~\rm{km\,s}^{-1}$ (the FWHM of the \ion{Mg}{ii}$\lambda$2796 absorption
line). Their rest-frame EWs are listed in Table~\ref{tab:GRB_abs_em_lines}. The observed line
ratio of $1.7\pm0.5$ is not well constrained. It is consistent with the theoretical expected
line ratio for an optically thin line but also for a saturated line. Assuming the weak line
regime, we can place a lower limit of $\log N \geq 13.8$ on the \ion{Mg}{ii} column
density.
When \ion{Mg}{ii} is detected, three further absorption lines are usually detected
at longer wavelengths, as well:
\ion{Mg}{i}$\lambda$2852 and \ion{Ca}{ii}$\lambda\lambda$3934,3969. Only
\ion{Ca}{ii}$\lambda$3934 is detected at $\lesssim3\sigma$ c.l.
To place limits on their rest-frame EWs, we measure the noise within an aperture of $2\times\rm{FWHM}\left(\mathrm{\ion{Mg}{ii}}\lambda2796\right)$ at the wavelength
of each line. Table~\ref{tab:GRB_abs_em_lines} displays their derived upper limits. We caution
that \ion{Mg}{ii} absorption lines can be broader than other absorption lines, hence our
upper limits might not be very stringent.

We also detect several emission lines summarised in
Table~\ref{tab:GRB_abs_em_lines} and shown in Fig. \ref{fig:Spec_AG} at a common redshift of $z_{\rm em}=0.28259\pm0.00005$,
consistent with the absorption line redshift within errors. Their fluxes were measured through direct integration.
From these measurements we derive key diagnostics of \ion{H}{ii} regions, such as extinction, SFR, and metallicity.
Balmer lines are a good diagnostic for determining the level of extinction in \ion{H}{ii}
regions. Their line ratio is purely determined by atomic constants.
The observed $3\sigma$ limit the H$\alpha$/H$\beta$ flux ratio of $>1.9\pm0.11$ is
consistent with the expected value of 2.76 for negligible extinction, assuming case B recombination.
Since we have no indication otherwise, we use $A_{V,\,\rm host}=0$ as a working hypothesis.
Knowing that, the SFR is $0.037\pm0.002~\rm{M}_\odot\,\rm{yr}^{-1}$ as measured from H$\alpha$
using the relation in \citet{Kennicutt1998a} and correcting for
a Chabrier initial mass function (IMF; \citealt{Chabrier2000a,FoersterSchreiber2009a}).
Since H$\beta$ is only marginally detected, we use the N2 diagnostic by \citet{Pettini2004a}
to measure the metallicity. This oxygen abundance of
$12 + \log\,\rm{O}/\rm{H} \geq8.57\pm0.05$ corresponds to a very high metallicity
of $Z=0.8\pm0.1~\rm{Z}_{\odot}$. The systematic error of this indicator is 0.18 dex
\citet{Pettini2004a}.

\subsection{Host Galaxy}\label{sec:host}

\subsubsection{Emission line diagnostics}\label{sec:host}

The X-shooter spectrum of the host galaxy's nucleus (obtained 0.7 days after the GRB; see Fig.
\ref{fig:Spec_Nuc}) shows no absorption but
a large number of emission lines at a common redshift of $z_{\rm{em}}=0.28256\pm0.00002$, listed in Table~\ref{tab:GRB_abs_em_lines}. Their fluxes
were measured in the same fashion as at the explosion site.

\begin{figure*}
\centering
\includegraphics[viewport=5 1 1210 417, clip, angle=0, width=2\columnwidth]{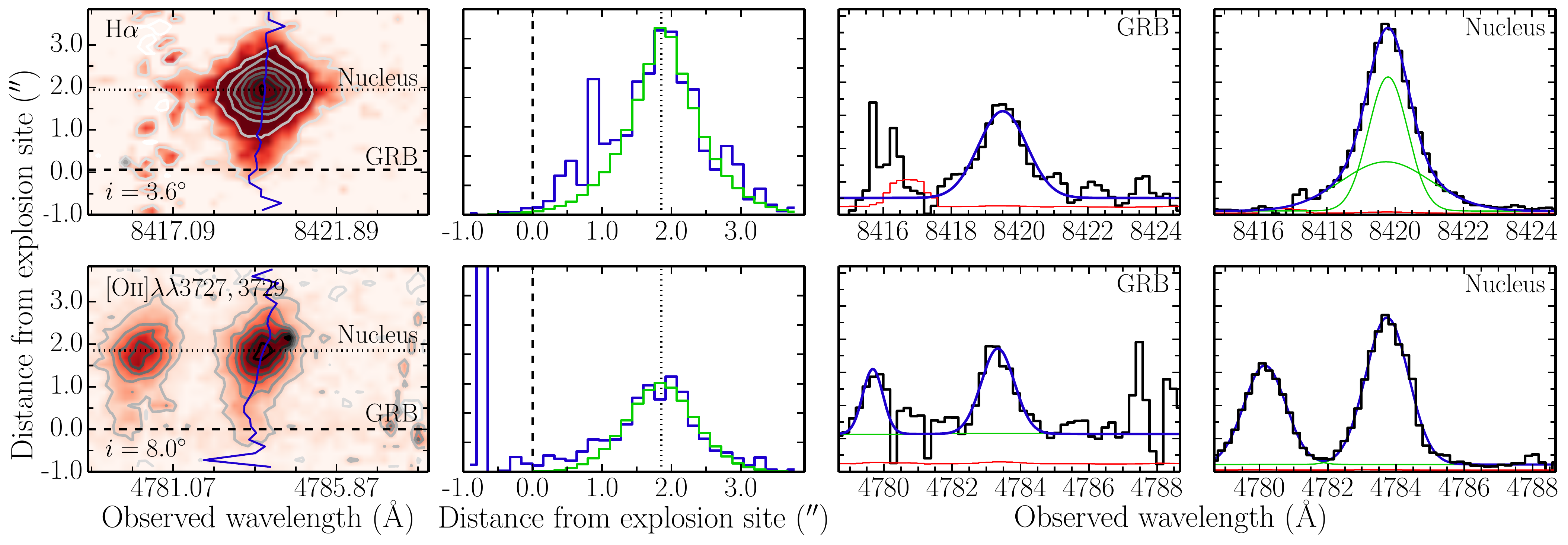}
\caption{Part of the rectified and wavelength-calibrated 2-D X-shooter spectrum ($\rm{PA}= 41^\circ$) obtained
	0.72 days after the explosion (Table~\ref{tab:obs_log_2}). The first column shows
	the 2-D profile of H$\alpha$ and the [\ion{O}{ii}] doublet. The blue lines
	trace the position of maximum flux. The inclination angles $i$ (defined as the angle between the major axis and a vertical line)
	are displayed in the lower left corner. Contour lines are overlaid to guide the eye.
	The cross dispersion profiles
	are displayed in the second column. The coding of the vertical lines  is identical to that
	in the first column. For illustration purposes we fitted the profiles with a Sersic
	function, where the wings left of the
	centres of lines were excluded from the fit.
	The line profiles in dispersion direction are shown in the last two
	columns. The green line the fit of individual components and the blue line
	of the compound.  At $z=0.283$, an angular distance of $1''$ translates into a projected
	distance of 4.3 kpc. The error spectra in the third and forth columns are overlaid in red.
}
\label{fig:host_em_lines}
\end{figure*}

Interestingly, the H$\alpha$
emission line is significantly broader than any other emission line in the spectrum,
i.e. $\mathrm{FWHM}(\rm{H}\alpha)=1.83\pm0.01$~\AA{} but $\mathrm{FWHM}(\rm{H}\beta)=1.22\pm0.05$~\AA.
To elucidate the origin of the broadening, we followed \citet[][and references therein]{Chatzopoulos2011a}
and assumed
three distinct models: \textit{a}) thermal broadening, \textit{b}) single Thompson scattering of free
electrons, and \textit{c}) multiple scattering of hot free electrons. In the first scenario the
proper motion of atoms leads to broadening that results in a Gaussian-shaped line profile. Since the
flux of an emission line stems from the total flux of all star-forming regions, we additionally assume
that there are two velocity components. The second is typical
for a broad-lined region in an AGN, producing exponential line profiles ($\propto \exp^{-\Delta v / \sigma}$,
where $\Delta v$ is the Doppler shift from the line centre and $\sigma$ is the velocity
dispersion). The third describes dense media and produces Lorentzian line profiles.
The top right panel in Fig.~\ref{fig:host_em_lines} shows the H$\alpha$ emission line
together with the best fit model. The best fit model consists of two Gaussians centred
at identical wavelengths ($\lambda_1=8419.71\pm0.06$ and $\lambda_2=8419.79\pm0.02$~\AA),
whose FWHMs are $3.42\pm0.03$ ($\Delta v=121.8~\rm{km\,s}^{-1}$) and  $1.41\pm0.03$~\AA\ ($\Delta v=50.2~\rm{km\,s}^{-1}$), and amplitudes of
$2.20\pm0.31$ and $3.21\pm0.19\times10^{-16}~\rm{erg\,cm}^{-2}\,\rm{s}^{-1}$, respectively. The width of the narrow component is
consistent with the width of the other lines. 

To check for AGN contribution, we put the integrated line measurements of H$\alpha$
H$\beta$, \ion{N}{ii}$\lambda$6584, and \ion{O}{iii}$\lambda$5007 in the BTP diagnostic plot
\citep[Fig. \ref{fig:BPT};][]{Baldwin1981a}. The emission lines ratios are fully consistent with being due to
star-formation. Knowing that, we use the Balmer decrements to to measure the extinction.
The Balmer decrements of the narrow component between H$\alpha$, H$\beta$, H$\gamma$, and H$\delta$ are all consistent with negligible dust extinction. However, we detect no
significant flux  from the broader component at the position of H$\beta$  to constrain its extinction. The inferred SFRs (computed
in the same way as for the explosion site) are: $0.48\pm0.03~\rm{M}_\odot\,\rm{yr}^{-1}$ and $>0.33\pm0.05~\rm{M}_\odot\,\rm{yr}^{-1}$ for the narrow and broad component, respectively.
The N2 metallicity indicator is calibrated on integrated measurements. Therefore, we measure
an integrated oxygen abundance of $8.43\pm0.01$. The metallicity of $Z=0.55~\rm{Z}_\odot$ is a bit lower than of the explosion site but they are consistent with each other within
$2.5\sigma$.

\citet{Levesque2012a} carried out an independent study of the emission-line properties of
the explosion site, the curved bridge connecting the galaxy's nucleus with the explosion site
(Fig. \ref{fig:GRB120422A_fov}),
and the host's nucleus using Magellan's low-resolution LDSS3 spectrograph with two
different  position angles ($\rm{PA}=50^\circ$ and $141^\circ$). Their line measurements
fundamentally deviate from our measurements. Specifically, their values of the radially
extended emission lines H$\alpha$ and the [\ion{O}{ii}] doublet are larger by 47\% and
$<42\%$, respectively,\footnote{The spectral resolution in \citet{Levesque2012a} is not sufficient
to resolve the [\ion{O}{ii}] doublet. Their reported [\ion{O}{ii}]$\lambda$3727 should
rather refer to the flux of the doublet.} of H$\beta$ and of [\ion{O}{iii}]$\lambda\lambda4959,5008$
by 118--184\%, but that of [\ion{N}{iii}]$\lambda6584$ agrees with our
measurement. At the explosion site their measurements for the [\ion{O}{ii}] doublet are
by a factor of 15.8 larger, for H$\alpha$ by a factor of 5.2, and [\ion{O}{iii}]$\lambda5007$
by a factor of 6 but that of [\ion{N}{iii}]$\lambda6584$ by a factor of $<2.5$.
The differences might be partly instrumental, because LDSS3 does not have an atmospheric dispersion corrector.
Another reason is how they extracted
the 1-D spectra. They used a fixed aperture of $1\farcs14$ (the plate scale being 0\farcs19/px). We, in contrast, based the
aperture size on the FWHM of the spectral PSF of the galaxy nucleus and the
explosion site, i.e. FWHM(nucleus)=1\farcs34 and FWHM(GRB)=0\farcs86 with
the plate scale being 0$\farcs$15/px. As described in \ref{sec:spectroscopy}, we also ensured
absolute flux-calibration and checked for differential flux losses by scaling the explosion site spectrum to the brightness of the
optical transient at that epoch and that of the galaxy nucleus to the brightness of the host
galaxy. \citet{Levesque2012a} only applied a relative flux calibration, which can be affected
by (differential) flux losses. Furthermore, comparing the emission-line profiles from
their spectra with their actual line measurements and uncertainties, e.g. for H$\alpha$,
casts doubts on their reported values and on their inferences.

We would like to point out that two of their results are in conflict with ours, which
are based on higher S/N and higher spectral-resolution data. Firstly, the fact that
their spectral resolution was not sufficient lead them conclude the presence of a
non-negligible amount of reddening at the galaxy's nucleus of $E(B-V)=0.24~\rm mag$.
Our data revealed that there are two dominant populations of star-forming regions at
the nucleus. After accounting for that, there are no indications for dust reddening.
They also argue for a dust reddening at the explosion site ($E(B-V)=0.31~\rm mag$),
which we can rule out. This in return significantly overestimates their SFR measurements:
\citet{Levesque2012a} $\gtrsim2.7~\rm{M}_\odot\,\rm{yr}^{-1}$, our total measurement
$\gtrsim0.8~\rm{M}_\odot\,\rm{yr}^{-1}$. As we will show in the following section, the
SED fit of the  host galaxy is in full agreement with our spectroscopy results, but in
conflict with \citet{Levesque2012a}.  Secondly, our data does not show any evidence for
asymmetries in the emission line profiles (see Fig.~\ref{fig:host_em_lines},
\ref{fig:Spec_AG} and \ref{fig:Spec_Nuc}), in contrast to
\citet{Levesque2012a}. The nominal values for  the skewness parameter are: 0.06--0.21
with a significance of $<2.1\sigma$ at the explosion site and $-0.03$--0.1 with a
significance of $<1.7\sigma$ at the galaxy's nucleus.

\subsubsection{Morphology and SED}

The X-shooter spectrum from 2012 April 23 ($\rm{PA}=41^\circ$; Fig.~\ref{fig:GRB120422A_fov}) reveals that the most prominent nebular lines, i.e. the
[\ion{O}{ii}] doublet and H$\alpha$, extend from  the galaxy's nucleus to the explosion
site and slightly beyond (see Fig.~\ref{fig:host_em_lines}). To obtain a better understanding
of the peculiarity of the explosion site and the host morphology, we extracted their cross-dispersion profiles by
fitting each row with a Gaussian (i.e. slicing the galaxy in chunks of $0\farcs15\times0\farcs9$,
which is equivalent to an area of $0.64\times3.9~\rm{kpc}^2$ at $z=0.283$). The largest
fluxes are recorded at the galaxy's nucleus (second column in Fig.~\ref{fig:host_em_lines}),
while the flux at GRB site is very low. Since both lines of sights are not affected by reddening
in the host galaxy, the difference in H$\alpha$ flux directly translates into a SFR difference, i.e the
explosion site does not show an enhanced SFR with respect to its surroundings and the nucleus.
A fit of the cross-dispersion profiles with a Sersic function (column 2 in Fig.~\ref{fig:host_em_lines}) reveals
an excess from the nucleus towards the GRB site.  The excess in [\ion{O}{ii}] is more diffuse and extends to larger galactocentric radii. A possible explanation could be that this nebular line is in
general less tightly correlated with star-formation and affected by differences in ionisation, metallicity, and dust content (for a detailed  discussion see e.g. \citealt{Kewley2004a}).
We also note that the 2-D profiles are slightly slanted. We measure a velocity difference
between the galaxy's nucleus and the explosion site of 7 and $22.6~\rm{km\,s}^{-1}$ at H$\alpha$ and
[\ion{O}{ii}]$\lambda3729$, respectively. Strictly speaking these are lower limits because this X-shooter
spectrum does not fully cover the nucleus.

\begin{figure}
\centering
\includegraphics[viewport=0 6 1029 646, clip, angle=0, width=1\columnwidth]{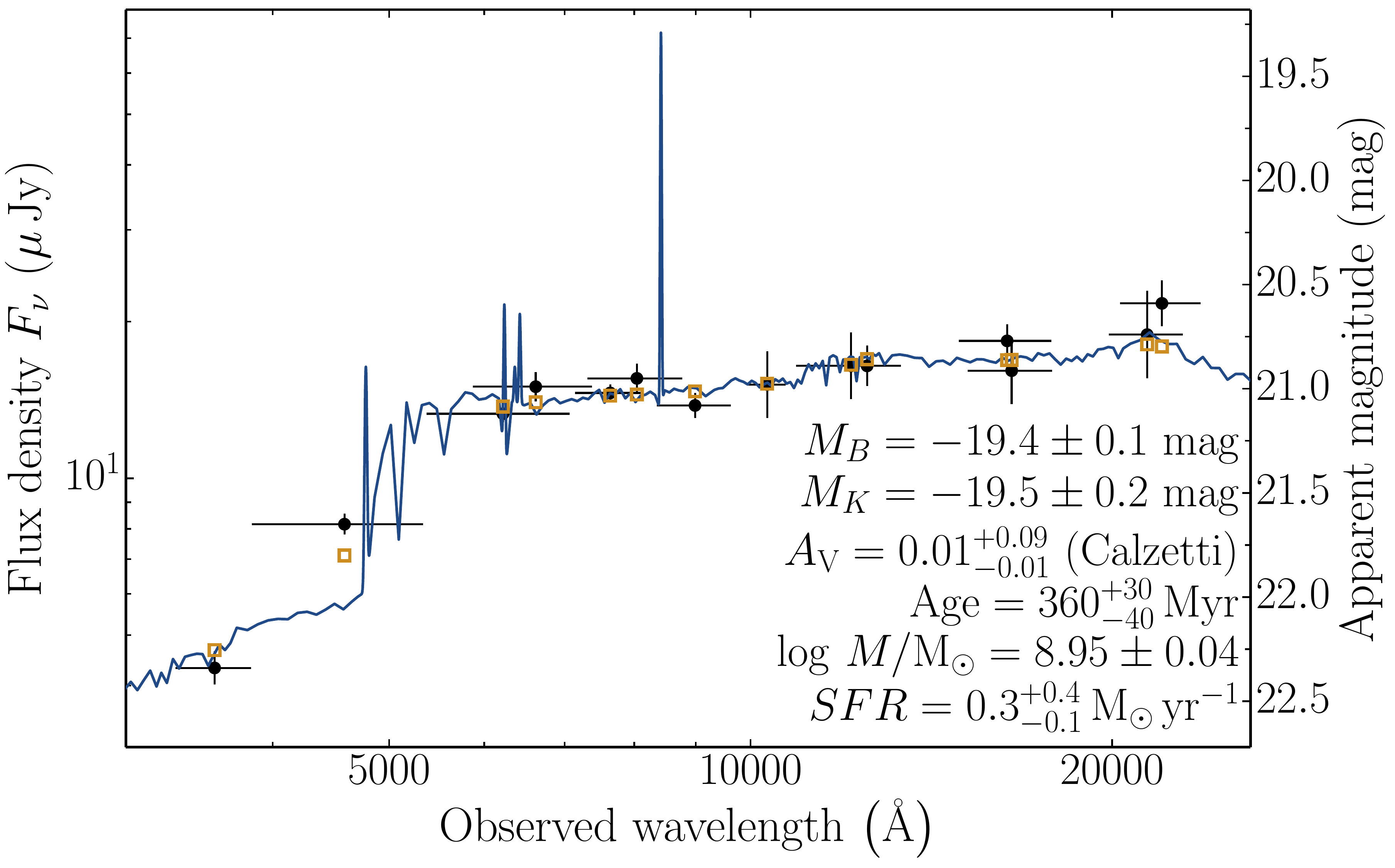}
\caption{Spectral energy distribution of the GRB host galaxy from 3500
	to 21460 \AA. The solid line displays the best fit model of the SED
	with \texttt{Le Phare} ($\chi^2=21.4$, number of filters = 14). The green points are
	the model predicted magnitudes.}
\label{fig:Host_SED}
\end{figure}

\begin{figure*}
\centering
\includegraphics[angle=0, width=2\columnwidth]{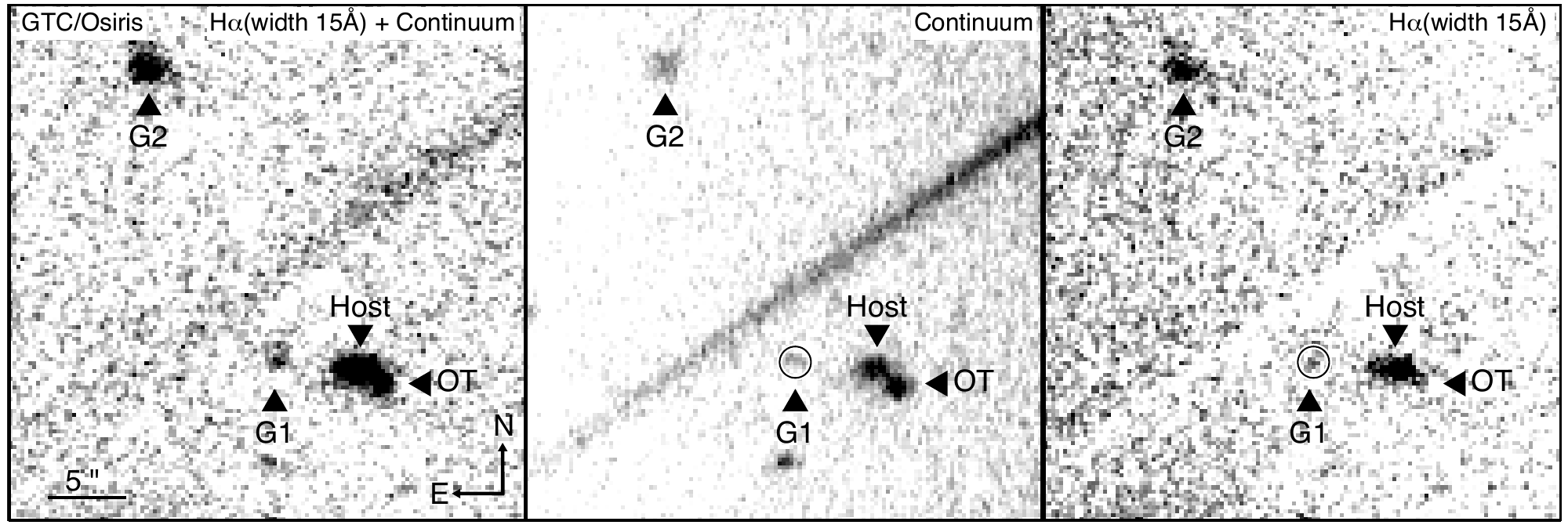}
\caption{Galaxy environment of GRB 120422A. The field of view has a size of $33''\times33''$.
	The left panel shows the H$\alpha$ image (15~\AA\ wide), which includes the emission
	line and the continuum. The middle panel shows the continuum centred at 8020 \AA\
	(6250 \AA\ rest frame) to avoid the emission from H$\alpha$ (obtained with a
	513-\AA-wide narrow-band filter). The right panel is the subtraction of the left and
	middle panel, i.e. a pure H$\alpha$ image. The host and G1 are at the same
	redshift as the GRB (OT). The galaxy G2 is possibly at the same redshift. Their projected
	distances are 7.3, 28.7 and 107.8 kpc from the explosion site, respectively
	(see Table~\ref{tab:galaxy_environment} for details). The diagonal stripes are produced by
	the $R=8.24~\rm mag$ foreground star that is $79''$ NW from the explosion site.}
\label{fig:GRB120422A_Halpha}
\end{figure*}

An image of the host galaxy obtained 3.6 days after the explosion,  shown in Fig.~\ref{fig:GRB120422A_fov},
reveals a curved bridge of emission connecting
the transient with the host. The curved bridge was also covered by the slit of the X-shooter
afterglow spectrum from 2013 April 23 (Fig.~\ref{fig:host_em_lines}). Even there stars are
formed at a rate that is in between that of the galaxy nucleus and the explosion site.
The GRB could therefore have occurred either in a morphologically disturbed/irregular
galaxy, within an interacting companion, or a spiral arm (however no counter arm is visible
on the far side of the galaxy).

Table~\ref{tab:host_mag} lists the brightness of the GRB host galaxy from 360 to 2140 nm.
We modelled the SED with
\texttt{Le Phare} \citep{Arnouts1999a,
Ilbert2006a},\footnote{http://www.cfht.hawaii.edu/\~{}arnouts/LEPHARE}
using a grid of galaxy templates based on \citet{Bruzual2003a} stellar population-synthesis
models with the Chabrier IMF and a
Calzetti dust attenuation curve \citep{Calzetti2000a}. For a description of the galaxy
templates, physical parameters of the galaxy fitting and their error estimation, we
refer to \citet{Kruehler2011b}. To account for zeropoint offsets in the cross
calibration and absolute flux scale, a systematic error contribution of 0.05 mag was
added in quadrature to the uncertainty introduced by photon noise.
Figure \ref{fig:Host_SED} displays the observed host SED and its best fit. The observed
SED is best described by a low-mass, barely-extinguished star-forming galaxy with a very
young starburst (see Table~\ref{tab:host_prop}). The extracted attenuation and
SFR are consistent with results from emission-line diagnostics.

\begin{table}
\caption{Brightness of the GRB host galaxy in the optical/NIR}
\centering
\scriptsize
\begin{tabular}{ccc}
\toprule
Filter			& $\lambda_{\rm center}\,(\rm{nm})$&		Brightness (mag)\\
\midrule
$u'$ 			&357.88	&$ 22.34 \pm 0.08$\\
$g'$ 			&458.98	&$ 21.65 \pm 0.05$\\
$r'$ 	 		&621.96	&$ 21.12 \pm 0.04$\\
$R$ 			&662.30	&$ 20.99 \pm 0.07$\\
$i'$ 			&764.01	&$ 21.02 \pm 0.05$\\
$I$ 			&804.08	&$ 20.95 \pm 0.07$\\
$z'$ 			&898.93	&$ 21.08 \pm 0.06$\\
$Y_{\rm CAHA}$	&1032.28&$ 20.98 \pm 0.16$\\
$J_{\rm CAHA}$	&1212.41&$ 20.89 \pm 0.16$\\
$J_{\rm UKIRT}$ &1250.24&$ 20.89 \pm 0.10$\\
$H_{\rm UKIRT}$ &1635.35&$ 20.77 \pm 0.08$\\
$H_{\rm CAHA}$	&1649.59&$ 20.91 \pm 0.16$\\
$K_{\rm CAHA}$	&2138.97&$ 20.74 \pm 0.21$\\
$K_{\rm UKIRT}$ &2200.45&$ 20.59 \pm 0.11$\\
\bottomrule
\end{tabular}
\tablefoot{The brightness was measured within a
circular aperture (diameter 2\farcs5). The brighness was measured in
$u'RI$ with NOT, in $g'r'i'z'$ with GROND, in $YJHK$ with CAHA, and in $JHK$
with UKIRT.}
\label{tab:host_mag}
\end{table}

\subsection{GRB host galaxy environment}\label{sec:galaxy_environment}

\begin{figure}
\centering
\includegraphics[viewport=4 4 836 645, clip, angle=0, width=1\columnwidth]{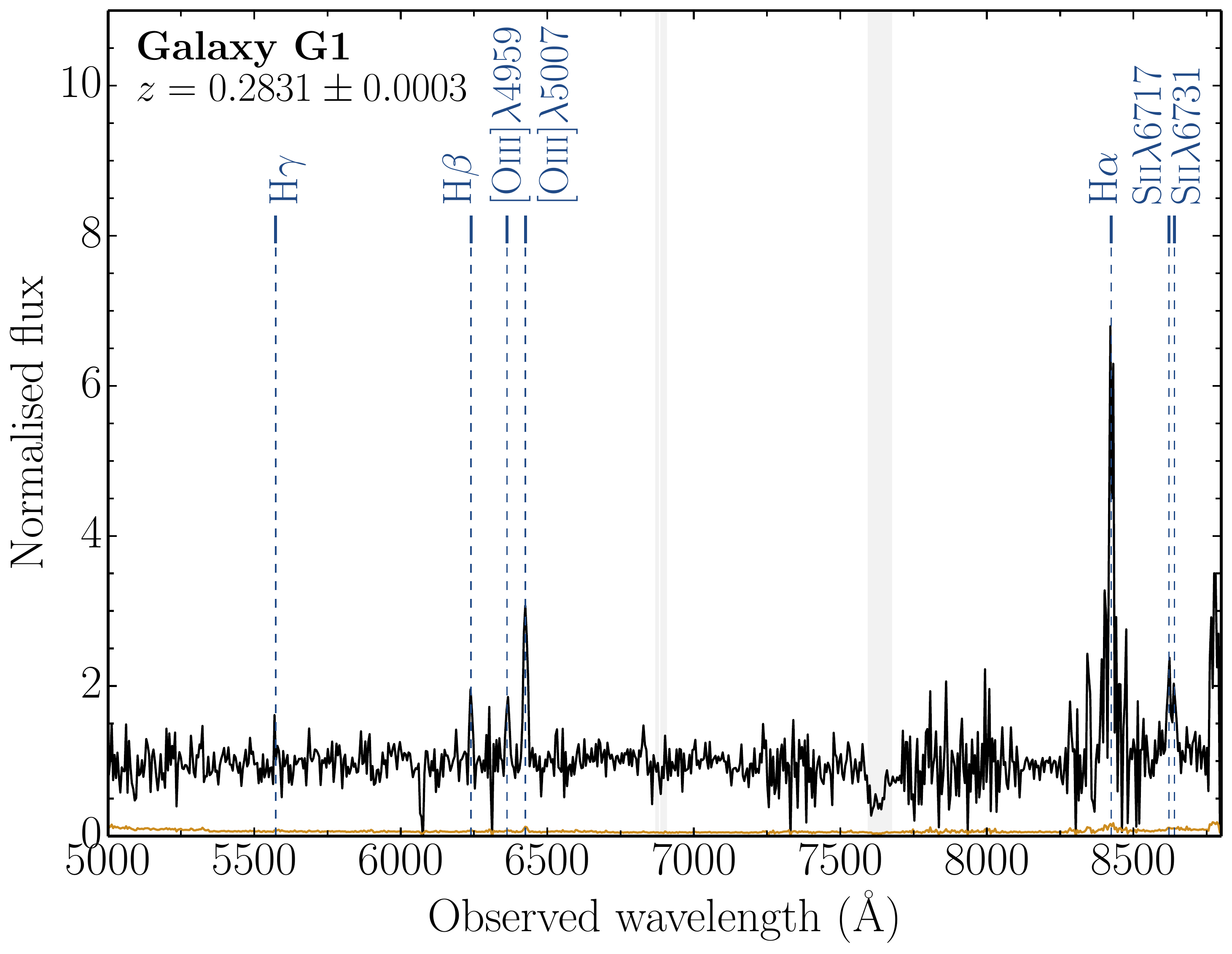}
\caption{Normalised spectrum of galaxy G1 (see Fig.~\ref{fig:GRB120422A_Halpha}), obtained with GTC/Osiris 3.6 days after the GRB.
	Several emission lines are detected at $z=0.2831$. H$\alpha$ is partly blended with
	a sky emission line. The error spectrum is shown in orange. The positions of telluric
	bands are are highlighted by grey-shaded areas.}
\label{fig:G1}
\end{figure}

In the previous section we briefly mentioned the possibility that the true host  could be a
smaller galaxy interacting/merging with the $r'=21~\rm mag$ galaxy. To explore this
scenario further, we studied the nature of other objects in the vicinity of the GRB to find
evidence for a galaxy over-density or galaxy interaction. Our GTC spectrum from 2012 April 25
elucidated that object G1 is at the same redshift as the GRB ($z=0.2831$; Fig.~\ref{fig:GRB120422A_fov},
\ref{fig:G1}, Table~\ref{tab:galaxy_environment}). The angular distance of 7\farcs1
corresponds to a projected distance of 28.7 kpc at $z=0.283$ from the host galaxy's
nucleus. Intriguingly, we detect a curved arm of emission, though not fully recovered,
that  connects G1 with the GRB host in our deep Gemini and Liverpool Telescope images
(Fig.~\ref{fig:GRB120422A_fov}). The blue colour of the tidal arm points to recent star
formation.  Together with G1's blue colour, we have compelling evidence that both
galaxies are interacting. This could be an indication that the arm connecting the host's
nucleus with the GRB site is not a spiral arm but another tidal arm due to interaction
of the $r'=21~\rm mag$ galaxy with another galaxy. Deep \hst\ observations are
needed to answer this question.

\begin{table}
\caption{Coordinates and distances from the optical transient to galaxies with emission
consistent with  H$\alpha$ at $z=0.283$ that are detected with the tuneable filters}
\centering
\scriptsize
\begin{tabular}{cccc}
\toprule
Galaxy		& R.A. (J2000)		& Dec. (J2000)		& projected Distance (kpc)	    \\
\midrule
Host	&	09:07:38.5	& +14:01:08.46		&	7.3\\
G1		&	09:07:38.9	& +14:01:09.12		&	28.7\\
G2		&	09:07:39.4	& +14:01:27.83		&	107.8\\
G3		&	09:07:42.9	& +14:00:15.40		&	355.8\\
\bottomrule
\end{tabular}
\label{tab:galaxy_environment}
\end{table}

To map the star-formation activity inside the host galaxy and identify more galaxies at the
GRB's redshift up to distances of hundreds of kpc, we acquired a deep image with the
tuneable filters ($\mathrm{FWHM}=15$~\AA) centred at H$\alpha$ at $z=0.283$ with
the GTC 25.5 days after the GRB. Figure \ref{fig:GRB120422A_Halpha} shows a
$33''\times33''$-wide post stamp. The left panel was obtained with the 15~\AA\ wide
tuneable filter, i.e. it contains the emission from the H$\alpha$ line and the continuum emission.
The continuum, displayed in the middle panel, was imaged with a broad-band filter centred at 8020 \AA\ (width 513 \AA)
that does not cover H$\alpha$. The SN continuum is not highly variable, neither in the
spectral range of the broadband filter nor at H$\alpha$
(Fig.~\ref{fig:AG_SN_spec_series});
the same is true for the host galaxy (Fig.~\ref{fig:Host_SED}). Hence, the difference image
of both observations shows the pure H$\alpha$ emission (right panel).

We detect four galaxy candidates that have emission consistent with H$\alpha$ at $z=0.283$
(Fig.~\ref{fig:GRB120422A_Halpha}, Table~\ref{tab:galaxy_environment}). We identify the
closest one, located at 7.8 kpc of the GRB, as the host. The galaxy G1 (23 kpc from the
centre of the host galaxy), already identified with the GTC spectrum from 25 April, is a satellite
galaxy. The galaxies G2 and G3 (not shown in Fig.~\ref{fig:GRB120422A_Halpha}) could be members of the same galaxy group, however a spectrum
or an additional observation tuned to the wavelength of another emission line are needed
for confirmation.

\section{Discussion} \label{sec:discussion}

\subsection{SN 2012bz}
In Sect. \ref{sec:SN2012bz_SBO} and \ref{sec:SN2012bz_radioactive}, we presented the
properties of the GRB-SN. The initial UV/optical emission until 10 ks is dominated by the
thermal emission of the cooling stellar envelope after the shock break-out. About
1.4~hours after the GRB, the stellar envelope had a temperature of 16 eV and a radius of
$7\times10^{13}~\rm{cm}$. By modelling the radioactively
powered light curve we obtained: $M_{\rm{Ni}}=0.40~\rm{M}_\odot$,
$M_{\rm{ej}}=4.72~\rm{M}_\odot$ and $E_{\rm k}=3.29\times10^{52}~\rm{erg}$,
and when the NIR emission is included, the nickel and ejecta masses to increased by 45 and
25\%, respectively, and the kinetic energy by 25\%. These values are among the highest
recorded values for GRB-SNe \citep{Cano2013a}. {We computed the intrinsic $V$-band
luminosity through direct integration over the SN spectrum. SN 2012bz has a absolute
$V$-band magnitude of $-19.7$ mag, making it 0.3 mag more luminous than SN 1998bw.
The phenomenological modelling of the SN light curve
gave a similar value. In the $r'$ band that overlaps with the rest-frame $V$ band we
inferred the SN to be 0.3 mag brighter than SN 1998bw but a slightly faster evolution.} In the following we
will discuss the SN properties in the context of other GRB-SNe.

\subsubsection{SN 2012bz in the context of other GRB-SNe}\label{sec:Spec_comp}

\begin{figure}
\centering
\includegraphics[viewport=3 7 421 589, clip, angle=0, width=1\columnwidth]{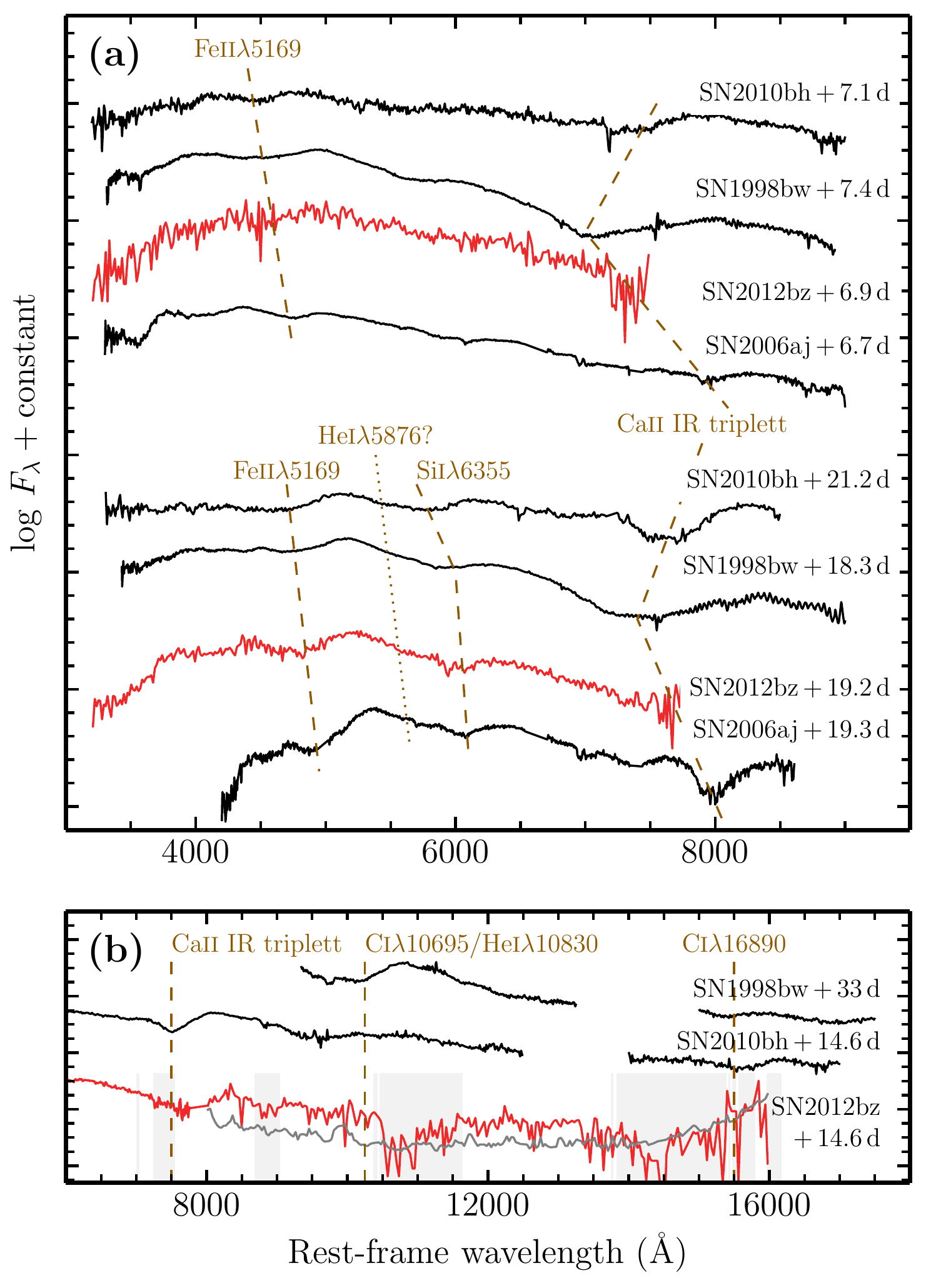}
\caption{(a) Comparison of SN 2012bz (red) to that of low-$L$ GRB SNe (black) at two
	different phases, $\sim7$ and $\sim20$ days past explosion, respectively. All
	comparisons are made in the rest frame. The dashed lines connect the approximate
	minima for the \ion{Fe}{ii} and \ion{Si}{ii} features and the spectra are shown in an
	expansion velocity sequence from the fastest (SN~2010bh) to the slowest (SN~2006aj).
	A less significant (but real) feature that has been proposed to be \ion{He}{i}
	is also identified. (b) NIR arm of the X-shooter spectrum of SN~2012bz at maximum
	light (red).  The thin grey line is the error spectrum. The \ion{Ca}{ii} IR triplet
	at the redshift of SN~2012bz is located between the VIS and NIR arms. For comparison
	NIR spectra of SN~1998bw and SN~2010bh are shown along with identification of the
	most prominent lines \citep{Patat2001a,Bufano2012a}. Unfortunately these features
	fall in unfavourable noisy regions of our spectrum.
	Positions of atmospheric features
	(shifted to the redshift of GRB 120422A) are highlighted by the
	grey-shaded areas.
	}
\label{fig:Spec_comp}
\end{figure}

Figure~\ref{fig:Spec_comp} shows the comparison of SN~2012bz at two different phases
for which simultaneous spectra of SNe~1998bw, 2006aj and 2010bh are available, all of
which accompanied low-$L$ GRBs \citep{Patat2001a, Pian2006a,Bufano2012a}. Overall,
the spectra are very similar and show the same features, although line strengths and
expansion velocities vary from object to object. We have illustrated this by annotating the
main features as they have been identified in the past \citep[e.g.][]{Patat2001a}: \ion{Fe}{ii},
usually visible between 4500--5000 \AA; \ion{Si}{ii}, around 5600--6100 \AA; the
\ion{Ca}{ii} IR triplet (that for SN~2012bz is in a noisy part of the spectrum between the
VIS and the NIR arms; see lower panel); and possibly \ion{He}{i}, at around 5500 \AA\
\citep[e.g.][]{Bufano2012a}. We stress that these SNe have very large explosion velocities
and that their broad lines are likely the result of blending, hampering the identification
of the dominating line species producing the absorption feature.

Nevertheless, spectra of different GRB-SNe displayed in
Fig.~\ref{fig:Spec_comp} are remarkably similar, reinforcing the idea that the nature of these
blends, whatever it is, is the same for all GRB-SNe, pointing towards similar explosions.
Differences do however exist in the expansion velocities  (see Fig.~\ref{fig:Fe_exp_vel}). The spectra are displayed
in an `expansion velocity sequence' going from the `fastest' \citep[SN 2010bh; see also the
discussion in ][]{Chornock2010a,Bufano2012a} to the `slowest' (SN 2006aj). This is at least
true for the \ion{Fe}{ii} and \ion{Si}{ii} lines and, in that respect, SN~2012bz seems intermediate
and more similar to SN 1998bw. The Ca IR triplet shows a different velocity behaviour, not
correlated with the one determined by the other elements, and SN~1998bw is clearly faster at
all phases.

It is interesting to point out that the notch that has been possibly identified as
\ion{He}{i} by \citet{Bufano2012a} (Fig.~\ref{fig:Spec_comp} panel a) is also visible in SN~2012bz,
and as a matter of fact in most optical GRB-SNe spectra with
sufficient S/N to the left of the main \ion{Si}{ii} trough. A powerful diagnostic to test
the presence of \ion{He}{i} is NIR spectroscopy \citep{Patat2001a,Bufano2012a}. Our
X-shooter NIR spectra are unfortunately of low S/N and for this reason we focus our
analysis only on the one obtained at maximum light (Fig.~\ref{fig:Spec_comp} lower panel).
Still, however, this spectrum is dominated by a weak continuum, while most prominent features
are located in unfavourable regions (the error spectrum is displayed). For comparison, we
have also plotted an X-shooter spectrum of SN~2010bh obtained at a similar phase
\citep{Bufano2012a}. SN~1998bw does not have a contemporaneous spectrum but we show the one
obtained at $T_0+33~\rm day$, where the identified features are more clearly visible
\citep{Patat2001a}. Both the locations where one would expect to see $\textrm{\ion{He}{i}}\lambda10830$
or $\textrm{\ion{C}{i}}\lambda\lambda10695,16890$ are located in very
noisy atmospheric regions of our spectra, at the redshift of SN 2012bz, preventing us from
drawing any meaningful conclusion.


\begin{figure}
\centering
\includegraphics[viewport=0 6 884 647, clip, width=1\columnwidth, angle=0]{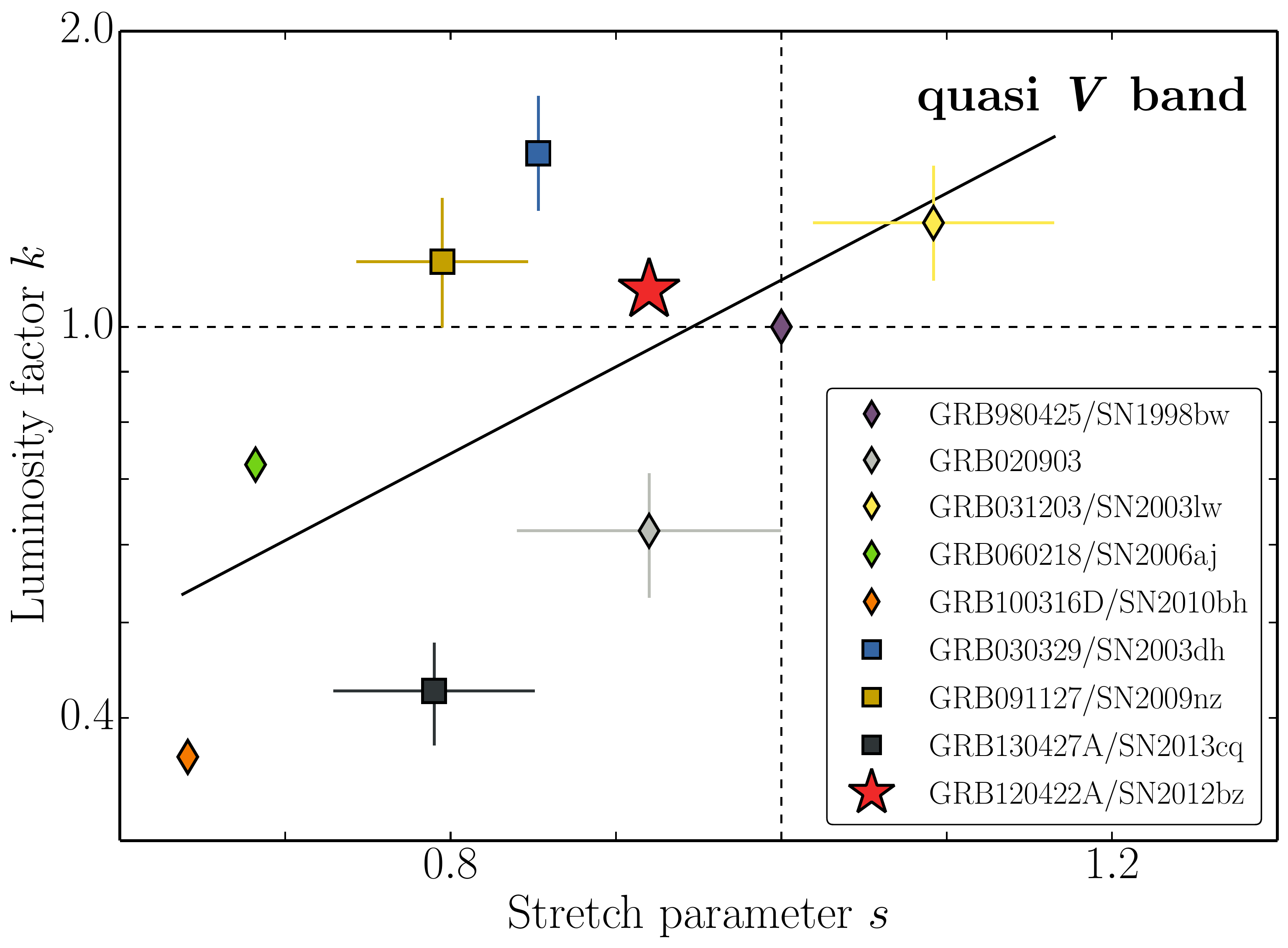}
\caption{The SN luminosity factor $k$ (peak luminosity in units of  SN 1998bw's peak
luminosity) vs. the stretch factor $s$ (time dilation vs.  SN 1998bw's peak time) in
the quasi rest-frame $V$ band. Quasi rest-frame $V$ band means that the observed
bandpass partially  overlaps with the rest-frame $V$-band. Low-$L$ GRBs are displayed
by diamonds, high-$L$ GRBs by squares, and the intermediate-$L$ GRB 120422A
by a star. \textbf{References}: GRBs 020903, 030329, 031203, 060218
\citep{Ferrero2009a}, GRBs 091127 and 100316D (Kann in priv. comm.), and
GRB 130427A (here).
}
\label{fig:SN_ks_plot}
\end{figure}

\subsubsection{A Philips-type relation for GRB-SNe?}

GRB-SN spectroscopy is in most cases limited to $z\lesssim0.3$, the redshift
domain that is observationally dominated by low-$L$ GRBs. At higher redshifts, the
detection of a GRB-SNe mostly depends on the detection of late red bumps,
that are modelled with a SN1998bw template. In the past years, the sample of GRBs
with detected late red bumps has been substantially increased
\citep{Ferrero2006a, Thoene2011a, Cano2013a}. We will use these samples to
compare the luminosity factor $k$ and the stretch factor $s$ of low- and high-$L$
GRBs. Among these, we only select those with meaningful values, i.e. GRB-SN that
were ranked better than "C" in \citet{Hjorth2011a}.\footnote{For an updated list see:\newline
http://www.dark-cosmology.dk/GRBSN/GRB-SN\_Table.html .}
Furthermore, most values were only obtained in one band, mostly in the observed
$R$ band. Since GRB-SNe have up to now been identified between $z=0.0085$ and $z\simeq1$, the observed $R$ band probes different regions in the rest-frame. GRB-SNe
emit most of its energy in the rest-frame $V$ band. Therefore, we only use those values
where the observed bandpass partly overlaps with the rest-frame $V$ band (in the
following called quasi $V$ band).

Figure \ref{fig:SN_ks_plot} displays the parameter space of the nine GRB-SNe that
fulfilled both criteria. Supernovae of low-$L$ and high-$L$ GRBs occupy the same
parameter space. Intriguingly, there is a trend between the luminosity and the stretch
factor.\footnote{Such a correlation was searched for GRB-SNe in \citet{Ferrero2006a}
and \citet{Cano2011a}, but not found because all data were used disregarding which
rest-frame waveband they probed, or on much less data  that also did not probe the
same rest-frame waveband \citep{Stanek2005a}.} This is in line with
recent findings by \citet{Hjorth2013a}, who independently reported on a correlation
between the peak magnitude and the width of the peak.
To estimate the
correlation degree and significance we applied a Monte Carlo technique. In this
method, every data point is represented by a 2-D Gaussian, where the centre of peaks
in each dimension are the parameter estimates, and the corresponding $1\sigma$ errors
are the width of the distributions. From these, we construct 30000 resamples of the
observed data sets, each of which is obtained by a random sampling with replacement
from the original data set. Note, SN 1998bw was excluded since it is
the reference value. For each of these data sets we do a linear regression fit,
using the model $\log_{10}\,k=A\times s + B$, and determine the correlation coefficients. The best
fit values and their uncertainties are given by the centre and the width of the
distribution functions. The best fit values are: $A=0.89\pm0.24$, $B=-0.84\pm0.19$.
The Pearson's correlation
coefficient,
Spearman's rank, and Kendal's $\tau$ give significances of $\sim1.3\sigma$.
Despite the correlation being statistically not significant, the trend is similar to
the Phillips relation \citep{Phillips1993a} that builds the foundation for using
Type Ia SNe as standard candles in cosmology.

The degree of correlation is affected by several systematics. First of all, none of
the displayed $k,s$ values represent the true rest-frame $V$ band. To obtain the
rest-frame $V$ band values a more sophisticated approach is needed, which is beyond
the scope of this paper. Systematic differences can arise from the fact that all
GRB-SNe are broad-lined Type Ic SNe, but the evolution and the strength of absorption
features depend on the specific GRB-SN (see Fig.~\ref{fig:Fe_exp_vel}, \ref{fig:Spec_comp}), which we think could be responsible for $\sim20\%$
of the observed scatter. Uncertainties in the line-of-sight extinction are the
second largest source of error affecting $k$ but not $s$. For instance, the afterglow
data of GRB 020903 are not good enough to build a SED for estimating the line-of-sight
extinction. The extinction towards GRB 060218/SN 2006aj and GRB 100316D/SN 2010dh are
very high and uncertain \citep{Cano2011a,Bufano2012a,Olivares2012a}. Furthermore, there
are different approaches how a 1998bw-template light curve for specific band is
constructed. Specifically, for GRB120422A we measure a difference of 0.10 mag in the
observed $r'$-band peak magnitude between the methods by \citet{Zeh2004a} and
\citet{Cano2013a}. The host contribution was taken into account either
by image subtraction, subtraction of the nominal host flux, or by adding the host
magnitude as a free parameter to the light curve fit for all GRB-SNe, except for GRB 130427A.
Last but not least, the SN fit depends on how well the afterglow component
is modelled. This affects $k$ as well as $s$.


\subsection{The afterglow of GRB 120422A}

Our analysis in Sect. \ref{sec:afterglow} reveals: \textit{i}) the optical (redward of
$B$ band) and the NIR emission of the transient accompanying GRB 120422A to be
afterglow-dominated between $\sim2$ and $86~\rm ks$, \textit{ii}) the X-ray emission
to be consistent with synchrotron radiation at all times (except for some small
deviations within the first 200 s after the burst; \citealt{Starling2012a}),
\textit{iii}) an initial Lorentz factor of $\Gamma_0\sim60$, \textit{iv}) an afterglow
peak luminosity-density of $\lesssim 2\times10^{30}~\rm{erg\,s}^{-1}\,\rm{Hz}^{-1}$,
and \textit{v}) a constant-density circumburst medium. Like in the SN discussion, we
will put the afterglow in context with low and high-$L$ GRBs.

Finding a constant-density medium around a massive star is not so surprising.
\citet{Schulze2011a} showed that most GRBs are found in constant-density-medium environments.
Simulations by \citet{vanMarle2006a} showed that a complex mass-loss history or differences
in the ram pressure can stall a free-stellar-wind density profile closer to the
progenitor star and make the disturbed density profile look like a constant density medium.

Measurements of the initial Lorentz factor are limited to a small number of bursts with
rapid follow-up. Typical values are about few hundred for high-$L$ GRBs
\citep{Molinari2007a, Ferrero2009a, Greiner2009a, Perley2011a}. For low-$L$ GRBs,
measurements exists for 060218 \citep{Soderberg2006a} and 100316D
\citep{Margutti2013a}.
For both bursts, the inferred Lorentz factor were $\Gamma=1.5$--2.3 at 1
(GRB 100316D) and 5 days (GRB 060218) after the burst. These measurements were obtained
when the blast-wave had already decelerated. According to \citet[][and references therein]{Zhang2004a},
a blast-wave's Lorentz factor evolves as $\Gamma\propto t^{-3/8}$ and $\Gamma\propto t^{-1/4}$ for
a constant-density and a free-stellar-wind ambient density profile during the deceleration phase, respectively. Given
the time when the Lorentz factors were obtained, the initial
Lorentz factors were at most one order of magnitude larger, still smaller than that of
GRB 120422A. This re-assures us in the identification of this phase transition
and also illustrates the decrease in the blast-wave's velocity from high- to
low-$L$ GRBs.

\begin{figure}
\centering
\includegraphics[viewport=5 2 413 271, clip, angle=0, width=1\columnwidth]{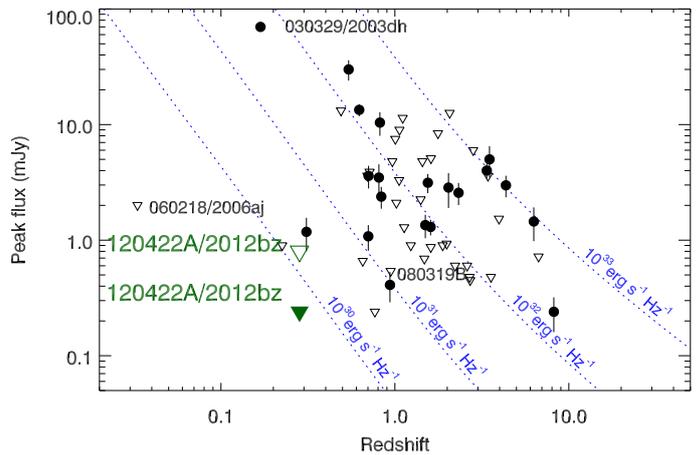}
\caption{Peak flux-density measured at sub-mm/mm wavelengths vs. redshift. Triangles
indicate
	$3\sigma$ detection limits. The deepest observed limit on the peak flux-density of GRB
	120422A is displayed by the filled green triangle, while the limit from the SED modelling
	is highlighted by the empty green triangle. Dotted lines display flux-density levels for equal luminosity at
	varying redshifts. Several interesting bursts are highlighted in the figure:
	the high-$L$ GRBs 030329 and 080319B, the low-$L$ GRB 060218, and the short GRB
	050509B, i.e. a burst that originated from the coalescence of a binary system of
	compact objects. Figure adapted from \citet{deUgartePostigo2012a}.}
\label{fig:submm}
\end{figure}

As mentioned in Sect. \ref{sec:radio_X-ray_sed}, the peak of an afterglow synchrotron
spectrum crosses the cm to sub-mm range within the first week. During the first week
an afterglow can also exhibit variability that can affect the peak flux and value of
the injection frequency. Several sub-mm observations during the first week after the
burst can be used as proxy for the peak luminosity without the need for modelling the
broad-band afterglow \citep{deUgartePostigo2012a}. Figure \ref{fig:submm} displays the inferred
sub-mm peak fluxes as a function of redshift. The observed limit on GRB 120422A's
peak flux-density from the sub-mm/mm observations and the limit from the SED modelling
point to faint afterglow. The limit of $\lesssim10^{30}~\rm{erg\,s}^{-1}\,\rm{Hz}^{-1}$
is exceptionally deep for high-$L$ GRBs. For example, the afterglow of GRB 030329, a burst
with spectroscopically confirmed SN and $E_{\rm iso}>10^{51}~\rm erg$, had a $\sim200$-times
larger peak luminosity density and the afterglow of GRB 080319B,
a burst with photometric evidence for a SN \citep{Tanvir2010a}, was about $\sim20$-times more luminous than 120422A.
Intriguingly, the peak luminosity density is in the expected range of low-$L$ GRBs,
such as GRB 060218.

\begin{figure}
\centering
\includegraphics[viewport=2 8 625 549, clip, angle=0, width=1\columnwidth]{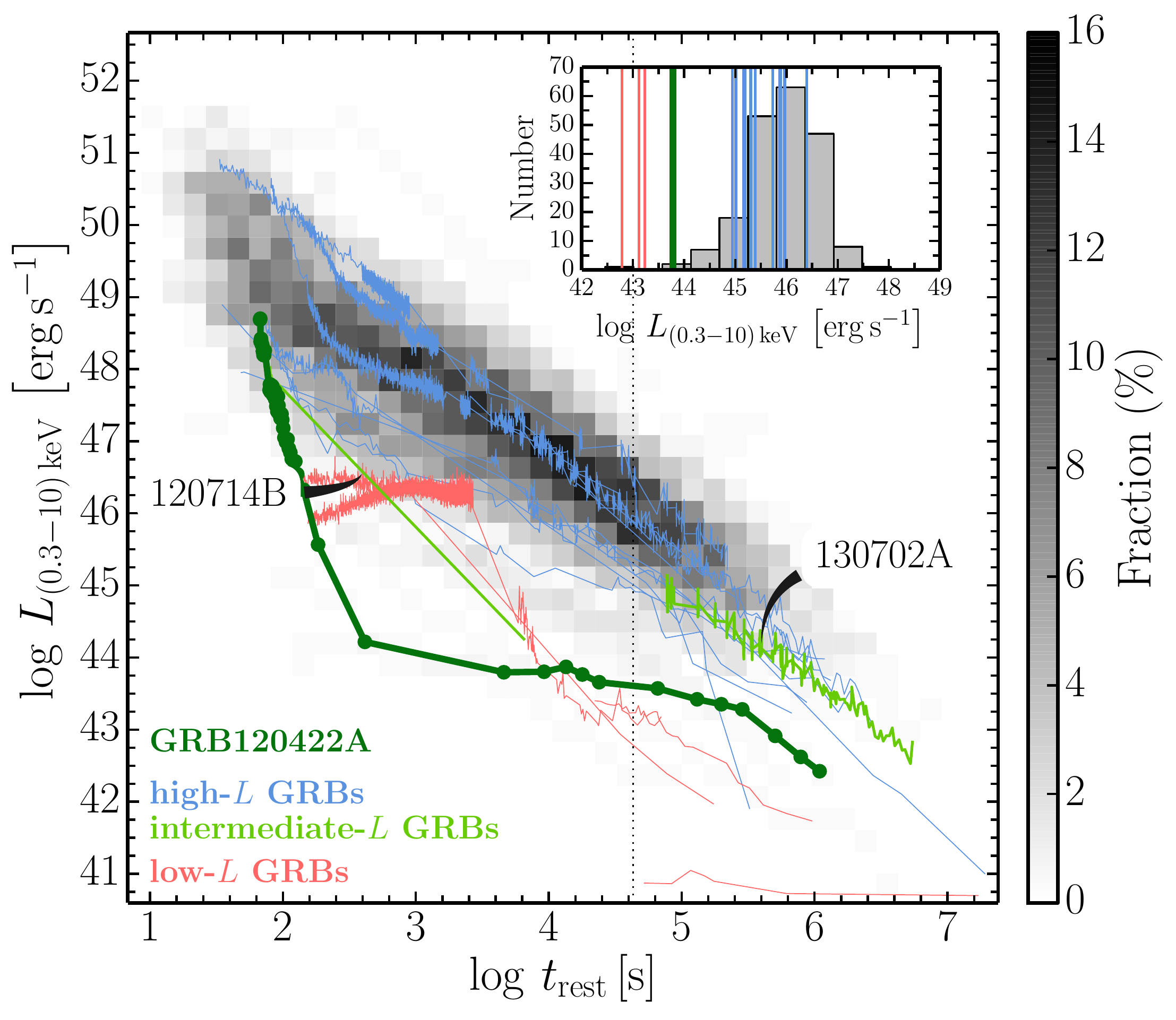}
\caption{X-ray light curves of low, intermediate and high-$L$ SN-GRBs. Overlaid is the
evolution of the observed luminosity distribution of 258 long \swift\ GRBs for which a
SN search was not feasible or unsuccessful (i.e. GRBs\,060505 and 060614) that were
discovered between December 2004 and June 2013. The colour table on the right side
translates a grey shade at a given luminosity and time into a fraction of bursts. The
inset displays the observed luminosity distribution at 0.5 days (dotted vertical line).
The vertical lines in the inset show the luminosity of intermediate-$L$ GRB
120422A and the low-$L$ GRBs\, 031203, 060218 and 100316D.
}
\label{fig:120422A_XLC}
\end{figure}

Current sub-mm observations are limited to a small number of GRBs ($\sim5\%$ of all GRBs)
and have only been successful in detecting bright afterglows. The number of \swift\ GRBs
with measured redshift is $\sim27\%$, i.e. $\sim5$-times larger than the sub-mm/radio
recovery rate and almost all GRBs with redshift information have a detected X-ray afterglow.
\citet{deUgartePostigo2012a} reported on the presence of a correlation between the
sub-mm flux density and the X-ray flux density at 0.5 days, as expected if they are
co-spatial. Hence, we can re-address the question on the faintness of GRB 120422A's
afterglow by exploring the X-ray luminosity distribution.

We download the 0.3--10-keV light curves of long \swift\ GRBs (i.e. $T_{90}\geq2~\rm s$)
with detected X-ray afterglow (requiring at least two X-ray detections) and known redshift
that were discovered between December 2004 and June 2013 from the \swift\ Burst Analyser
\citep{Evans2010a}. Because of the small number of low-$L$ GRBs in the \swift\ sample,
we include all pre-\swift-era GRBs with detected supernova, which are mostly low-$L$ GRBs. We retrieve their light curves
from the \bepposax\ GRB Afterglow Database \citep{dePasquale2006a, Gendre2006a}.\footnote{http://www.asdc.asi.it/GRBase/} The differences in the observed bandpasses were taken
into account to compute the X-ray luminosity in the 0.3--10-keV bandpass.
In total, $\gtrsim270$ GRBs fulfil both criteria. Following \citet{Hogg2002a} and assuming
a top-hat response function, the luminosity between 0.3 and 10 keV is given by
\begin{equation}
L_{\left(0.3-10\right)\,\rm{keV}}=4\pi\,d^{2}_L\left(z\right)\left(1+z\right)^{\beta-1}F_{\left(0.3-10\right)\,\rm{keV}},\nonumber
\end{equation}
where $d_L$ is the luminosity distance and $\beta$ is the spectral slope. The Burst
Analyser provides information on the spectral slope for each data point, inferred from
hardness ratio \citep{Evans2010a}. Sometimes the given slope is highly variable or has
unphysical values (for limitations of the Burst Analyser see \citealt{Evans2010a}),
especially at late times ($t>1000~\rm s$) when statistics are poor. To minimise the
impact of such deviations, we set the spectral slope to the median late-time spectral
slope (i.e. $t>1000~\rm s$) if deviations are $<3\sigma$ and if the slope is larger than
$4$. For pre-\swift\ GRBs only the time-average spectral slope is available.
Next we resampled the rest-frame X-ray light-curves
to a grid defined by the observed luminosities of and the time interval spanned
by all X-ray afterglows, and interpolated between adjacent data points in case of
orbit gaps. The resulting luminosity distribution as a function of rest-frame time
is shown in Fig.~\ref{fig:120422A_XLC} (the grey shaded area). Highlighted in this
plot are GRBs with detected SNe, colour-coded according to their time-averaged
$\gamma$-ray luminosity.

The inset in Fig.~\ref{fig:120422A_XLC} shows the luminosity distribution of 196 X-ray
afterglows at 0.5 days after the burst. High-$L$ GRBs with detected SNe occupy the
same parameter space like all high-$L$ for which no SN search was feasible. This
supports the discovery made by \citet{Xu2013a} that even bursts with the largest
energy releases during the prompt emission are accompanied by SNe \citep[see also][]{Tanvir2010a}.
Low- and intermediate-$L$ GRBs lie at the very faint end of the luminosity
distribution. Specifically, GRB 120422A is 2.1 dex fainter than
the mean value, while the afterglows of the brightest low-$L$ GRBs are by a factor of only a few
less luminous than that of GRB 120422A.

\subsection{The host galaxy and galaxy environment}

Our analysis in Sect. \ref{sec:environments} revealed that \textit{i}) the extinction
at the explosion site as well as the galaxy nucleus is negligible, \textit{ii}) there
are two populations of \ion{H}{ii} regions in the nucleus, \textit{iii}) the GRB appears to have not occurred in a region
of enhanced SFR,  \textit{iv}) the metallicities of the nucleus and the explosion site are high and
identical to  within errors, \textit{v}) the host galaxy is interacting with a galaxy at a projected distance
of 23 kpc, and \textit{vi}) the host is possibly embedded in a galaxy group.
First we will discuss the GRB environment in the context of all GRBs, then the host galaxy
and finally the host environment.

The GRB environment appears to be rather average. The lower limit on the \ion{Mg}{ii} column
density of $\log\,N>13.8$ is $\sim1$ dex lower than that of an average GRB environment
\citep{Christensen2011a}, but such a low column density was reported for other GRBs before:
050922C $\log N = 14.6 \pm0.3$ \citep{Piranomonte2008a}, 121019B
$\log N = 13.43^{+0.08} _{-0.10}$ \citep{Sparre2011a}, and even lower GRBs 070125: $\log N = 12.96^{+0.12} _{-0.18}$
\citep{DeCia2011a} and 071003: $\log N > 12.6$ \citep{Perley2008a}. To quantify the integrated absorption-line strength of the
interstellar medium in GRB host galaxies, \citet{deUgartePostigo2012b} introduced
the line-strength parameter ($LSP$) that is derived from detected absorption lines.
The observed $LSP$ of $-0.15\pm0.40$ is small, but considering the large error, the value
is consistent with the mean for GRB environments.
In contrast to that, the high metallicity is exceptional (see \citealt{Christensen2008a,
Thoene2008a, Levesque2011a}), though the dark GRB 020819A occurred in an even higher-metallicity
environment \citep{Levesque2010b}. Moreover, the negligible reddening in conjunction
with the very high metallicity is remarkable.
A possible reason for this high metallicity could be
the limited spatial resolution. For example, based on \hst\ observations \citet{Fynbo2000a}
showed that the stellar cluster, of which GRB 980425A's progenitor was part, was very
compact (the radius being 2.25 pc) and faint, and at lower spatial resolution it would
merge with a much brighter Wolf-Rayet star hosting complex 800 pc away. In contrast,
the extracted X-shooter spectrum averages over an area of $0.64\times3.9~\rm{kpc}^2$.
We also note that emission-line metallicities average over all star-forming regions along the
line-of-sight, not exclusively the GRB explosion site. On the other hand, if the explosion
site has indeed such a high metallicity, this might challenge current GRB progenitor
models that predict a low-$Z$ cut-off \citep[][and references therein]{Woosley2011a}.

\begin{table}
\caption{Properties of the host galaxy and GRB hosts at $z<1.5$}
\centering
\scriptsize
\begin{tabular}{l@{\hspace{0.3cm}}c@{\hspace{0.3cm}}c@{\hspace{0.3cm}}c}
\toprule
\multicolumn{1}{l}{Parameter}			& Host					& GHostS						& TOUGH			\\
\midrule
Sample size								&						& 74					& 20			\\
Redshift								&0.28256				& $0.78^{+0.23}_{-0.33}$& $0.83^{+0.25}_{-0.44}$\\
$M_{\rm{UV,est}}\,(\rm{mag})$			&$-18.0$				& $<-18.6\pm1.2$		& $<-18.4\pm1.4$\\
$M_{K_{\rm s,est}}\,(\rm{mag})$			&$-19.8$				& $<-20.6^{+0.6}_{-0.9}$& $<-19.4^{+0.6}_{-0.5}$\\
$M_{\rm{B,SED}}\,(\rm{mag})$			&$-19.4\pm0.1$			& $-20.5^{+1.1}_{-1.0}$ & \dots			\\
$M_{K_{\rm s,SED}}\,(\rm{mag})$			&$-19.5\pm0.2$			& $-20.2^{+1.0}_{-0.9}$ & \dots		\\
$A_V\,(\rm{mag})$						&$0.01^{+0.09} _{-0.01}$& $0.6^{+0.6}_{-0.2}$ 	& \dots			\\
$\log M_\star (\rm{M}_\odot)$ 			&$8.95\pm0.04$			& $9.3\pm0.5$ 			& \dots			\\
Age (Myr)								&$360^{+30} _{-40}$		& $1119^{+896}_{-325}$	& \dots			\\
SFR $(\rm{M}_\odot\,\rm{yr}^{-1})$		&$0.3^{+0.4} _{-0.1}$	& $2.4^{+4.0}_{-1.7}$ 	& \dots			\\
$Z$ $(\rm{Z}_\odot)$					& 0.6					& $0.5^{+0.2}_{-0.1}$ 	& \dots			\\
Offset (kpc)							& 7.3					& $1.9^{+1.2}_{-1.3}$ 	& \dots\\
\bottomrule
\end{tabular}
\tablefoot{Host properties of 120422A, and the median values of GRB hosts as compiled
in the GHostS database (date: 2013 December 3) and of the homogeneous, optically unbiased TOUGH survey
\citep[][incl. results from
 Schulze et al. in prep.]{Hjorth2012a}.
The errors of the comparison samples indicate the distance from the median values
to the 25 and 75\% percentiles. The age represents the age of the starburst. The stellar mass of GRB
120422A's host was calculated assuming the Chabrier IMF. The UV and $K_{\rm s}$ luminosities,
marked by 'est' were computed using the method in Schulze et al. (in prep.) and \citet{Laskar2011a},
respectively. Measurements designated with 'SED' were obtained from SED fitting.
The GRB offsets of the sub-set of bursts in the GHostS sample are compiled in \citet{Bloom2002b} and
the age distribution and results from SED fitting for the GHost sample were taken from \citet{Savaglio2009a}.
}
\label{tab:host_prop}
\end{table}

\begin{figure}
\centering
\includegraphics[viewport=0 7 636 424, clip, width=1\columnwidth, angle=0]{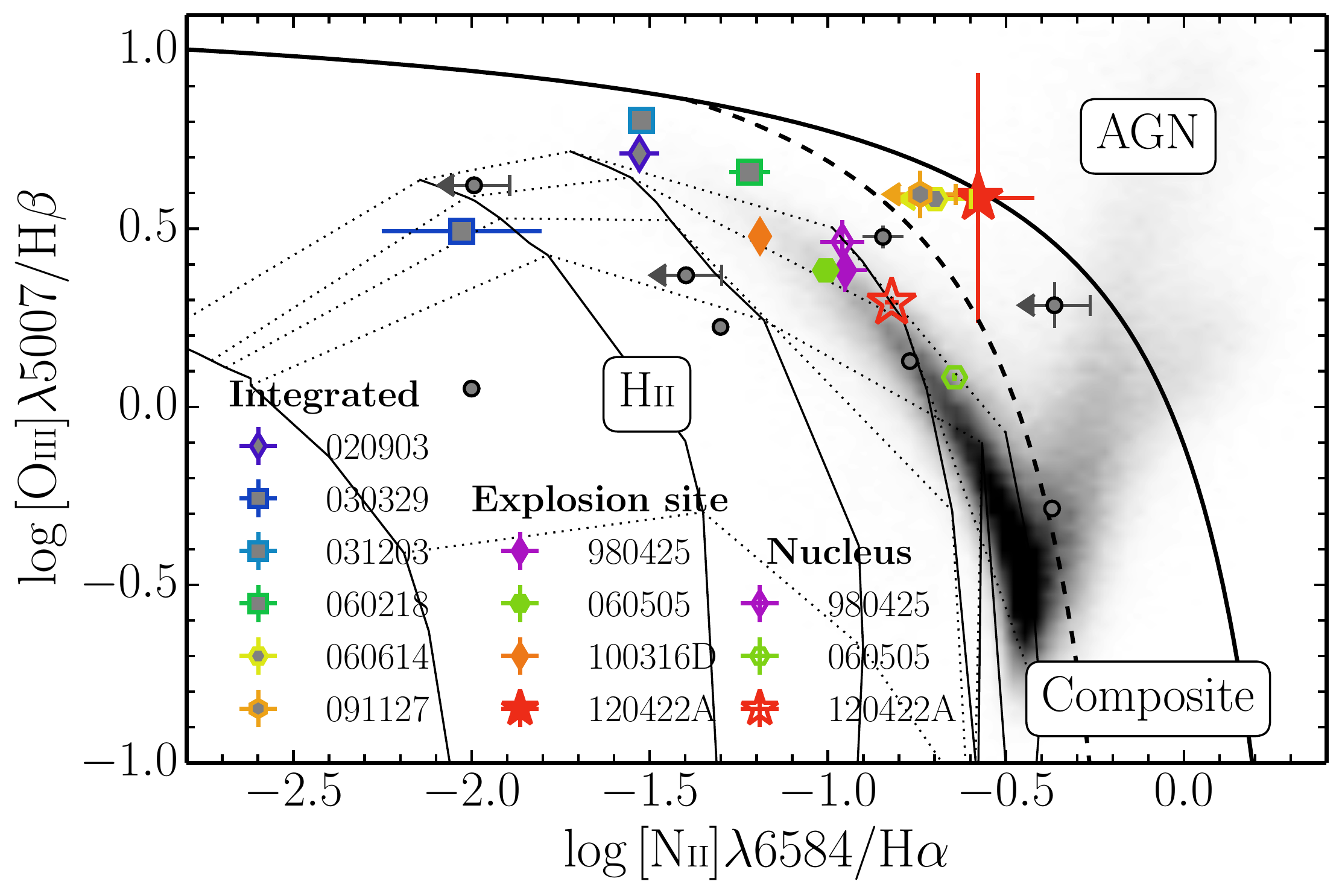}
\caption{Emission-line ratios for low-$L$ ($\diamondsuit$) and high-$L$ ($\Square$) GRB hosts
and explosion sites. Long GRBs for which a SN search was not feasible are shown as circles.
Limits are signified by arrows. For comparison, the emission-line ratios are overlaid for a wide population
of field galaxies from the SDSS DR9 \citep{Ahn2012a} sample as density plot. Among these data we selected
those whose emission lines were detected at $>5\sigma$ c.l. The discerning line between
star-formation and AGN-dominated galaxies is shown as thick solid line and was taken from
\citet{Kewley2001a} and the region of composite galaxies is encircled by the thick solid the
thick dashed lines  \citep{Kauffmann2003a}. Evolutionary models, calculated by \citet{Dopita2006a}, that link emission-line
regions at ages from 0.1 to 5 Myr (shown as dotted lines; ages: 0.1, 1, 2, 3, and 4 Myr with
the youngest stellar populations having the highest [\ion{O}{iii}]/H$\beta$ line ratios) and
different metallicities (shown as thin solid  lines; $Z=0.05$, 0.2, 0.5, 1.0, and
$2.0\,Z_\odot$; metallicity increases from left to right) are displayed. Error are shown,
if available. In some cases these are smaller than the size of the respective plot symbol.
Figure adapted from \citet{Christensen2008a}.
}\tablebib{
\citet{DellaValle2006a}: the SN-less GRB 060614;
\citet{Hammer2006a}: 980425, \citet{Han2010a}: 990712, 020903, and 030329;
\citet{Levesque2010a}: 991208, 010921, 050826, and 070612; \citet{Levesque2010b}: 020819B;
\citet{Thoene2008a}: the SN-less GRB 060505;
\citet{Wiersema2007a}: 060218; \citet{Kruehler2012a}: 080605A; \citet{Vergani2011a}: 091127; \citet{Levesque2011a}: 100316D}
\label{fig:BPT}
\end{figure}

To address this peculiarity further, we compare the [\ion{O}{iii}]/H$\beta$ vs
[\ion{N}{ii}]/H$\alpha$ line ratios with those of other GRB hosts (see Fig. \ref{fig:BPT}).
In addition, we distinguish between spatially-resolved and integrated line measurements.
The emission-line ratio of the host's nucleus is not different from other GRB hosts.
All hosts are located in the region that is dominated by \ion{H}{ii} regions.  Compared
to models by \citet{Dopita2006a}, the observed line ratios always point stellar
populations with an age of a few million years and metallicities between 0.05 and
$2\,Z_\odot$, in contrast to the bulk of emission-line galaxies in the SDSS DR9
\citep{Ahn2012a}. GRB~120422A's host is among the most metal-rich GRB hosts. The large
uncertainties in the line measurements of the explosion site do not allow to draw a
firm conclusion on its peculiarity.

Comparing the integrated host properties with GRB samples is not straightforward. Most samples are
heterogeneous and biased towards a particular GRB population, e.g. GRBs with negligible
reddening or bright afterglows. On the other hand, optically unbiased samples are limited
to observations in a few bands, from which only a few host properties can be extracted.
Keeping these limitations in mind, we compare 120422A's host with the GHostS database, built by \citet{Savaglio2009a},
that contains well-sampled multi-band SEDs and host spectra for a larger number of hosts,
and the optically-unbiased, homogeneous GRB host (TOUGH) sample
by \citet{Hjorth2012a}, which is in most cases limited to observations in $R$ and $K_{\rm s}$
bands. Based on recent findings by \citet{Perley2013a} that GRB hosts at $z<1.5$ are bluer
and significantly less massive than higher redshifts, we limit the comparison to hosts at $z<1.5$.

Table~\ref{tab:host_prop} lists the host properties of 120422A and the median
values for both comparison samples. In context of the
GHostS sample, GRB 120422A's host is in the lower half of the luminosity, mass and
SFR distribution. Only its very low extinction (0.5 mag less than the GHostS median)
in combination with high metallicity are peculiar. We caution that the completeness
of both properties is very low. Another peculiarity is the exceptional large distance
between the explosion site and the galaxy's nucleus. The offset of 7.3~kpc is among
the largest values reported in \citet{Bloom2002b}.
Based on the
optically-unbiased
TOUGH sample,  a slightly different picture can be drawn. Since only observations in two
filters are available for most hosts, we assume that the rest-frame NIR and UV
can be approximated by power laws, similar to the approaches in \citet{Laskar2011a} and in
Schulze et al. (in prep.). We measure a UV luminosity at 1700~\AA\ of $-18.0~\rm{mag}$ and a $K_{
\rm{s}}$-band luminosity of $-19.8~\rm mag$ (see Table~\ref{tab:host_prop}). These values
are consistent with the ensemble median values.
%

Observations with the tuneable filters revealed that host is interacting with another
galaxy that is at a projected distance of 23 kpc from GRB 120422A host galaxy's nucleus
and that there are possibly two further star-forming galaxies at the same redshift within
$\sim360~\rm{kpc}$. In general, little is known about the galaxy environments of GRB host
galaxies. Several GRB fields show an increased galaxy density, e.g. GRBs\,000301C, 000926
\citep{Fynbo2002a}, 011211 \citep{Fynbo2003a}, 021004 and 030226 \citep{Jakobsson2005a},
030115 \citep{Levan2006a}, and GRB 050820A \citep{Chen2011a}, but nearest burst, GRB
980425, does not \citep{Foley2006a}. The comparison is also limited due to the lack of
information for SN fields.

\subsection{The missing link between low- and high-$L$ GRBs}

The division between low- and high-$L$ GRBs is not entirely operational. Both
populations have very distinct properties. Low-$L$ GRBs are thought to be driven
by shock break-outs, producing (quasi-)spherical explosions whose $\gamma$-ray
light curves are smooth and single-peaked \citep{Campana2006a, Starling2011a},
spectra that can have peak energies of only a few keV \citep[e.g.][but see Kaneko2007a]{Campana2006a, Starling2011a},
and mildly relativistic outflows ($\Gamma<10$; e.g. \citealt{Soderberg2006a}
\citealt{Margutti2013a}). In contrast, high-$L$ GRBs are powered
by ultra-relativistic collimated outflows with Lorentz factors of a few hundred
\citep{Molinari2007a, Ferrero2009a, Greiner2009a, Perley2011a}, and $\gamma$-ray
light curves that can exhibit variability on the milli-second domain
\citep{Bhat2012a}. The rate of low-$L$ GRBs in the nearby Universe exceeds
that of high-$L$ GRBs by a factor of 10–1000 \citep{Pian2006a, Guetta2007a,
Virgili2009a, Wanderman2010a}.
However, a recent work by \citet{Lazzati2012a} based on relativistic jet
simulations proposed a non-uniform distribution of the central engine's
on-time to account for the differences.

GRB 120422A is one of the very few examples of intermediate-luminosity GRBs.
Its $\gamma$-ray light
curve exhibits an initial spike (starting at $T_0 - 3~\rm{s}$ and ending at
$\sim T_0 + 20~\rm s$; \citealt{Barthelmy2012a}) followed by a weaker and
softer extended component (starting at $T_0+45~\rm s$ and ending at to
$T_0+65~\rm s$; \citealt{Barthelmy2012a}), as observed in
other high-$L$ GRBs before \citep{Bostanci2013a}. In addition, the X-ray emission is
not dominated by thermal emission from the cooling photosphere after the
shock break-out, like the low-$L$ GRBs 060218 and 100316D \citep{Campana2006a, Nakar2012a}.
In contrast, the properties of the longer-lasting transient accompanying the GRB point
in a different direction. The blast-wave had a very low initial Lorentz factor
of $\Gamma_0 \sim 60$ and the produced afterglow had an unprecedentedly low peak
luminosity of $L_{\nu,\,\rm max}\lesssim2\times10^{30}\,\rm{erg\,s}^{-1}\,\rm{Hz}^{-1}$
for a high-$L$ GRB.
Thanks to the weak afterglow, the signature of the shock break-out was
for the first time detected for a non-low-$L$ GRB.

The failed-jet model predicts high-$L$ GRBs to transition into low-$L$ bursts
as the jet produced by the central engine gets weaker, e.g. because of
a lower kinetic energy in the outflow, a central engine that is active
of a shorter period, or a less collimated outflow \citep{Bromberg2011a, Lazzati2012a}.
According to this model, the weaker a jet, the more it gets decelerated
in the stellar envelope until it is choked. Examples for choked jets are the
Type Ib/c SNe 2002ap, 2012au \citep{Soderberg2006b,Milisavljevic2013a,Margutti2013a} and
GRB 100316D whose jet barely broke through
the stellar cocoon \citep{Margutti2013a}. The fact that we do detect the
thermal emission from the cooling photosphere after the shock break-out
raises the questions of how much energy GRB~120422A's jet already transferred
into the stellar envelope and how much more energy it could lose before
getting choked. As coined by \citet{Hjorth2013a}, GRB~120422A is indeed a transition
object between shock-break-out driven low-$L$ and high-$L$ GRBs that are powered by
ultra-relativistic jets.

To fully connect GRBs of low and high luminosities, it has to be shown
that even the most luminous bursts ($L_{\rm iso} \sim 10^{54}~\rm \rm{erg\,s}^{-1}$) are
accompanied by SNe. These very energetic bursts have however been found at
redshifts where SN searches are getting unfeasible, i.e. $z\gtrsim1$.
Serendipitously, one of the most energetic bursts, GRB 130427A, occurred
at $z=0.34$ \citep{Perley2013a, Xu2013a}. During its prompt
$\gamma$-ray emission that had a duration of $T_{90}=276\,\rm s$, this burst released
$8.1\times 10^{53}~\rm erg$; \citealt{Maselli2013a}). This translates into
a time-averaged $\gamma$-ray luminosity of $L_{\rm iso}\sim 10^{51.6}~\rm {erg\,s}^{-1}$.
Thanks to its low redshift, an accompanying broad-lined Type Ic SN was
spectroscopically detected with properties similar to those of low-$L$ GRBs
\citep[][see also \citealt{Tanvir2010a}, who reported on the
photometric SN discovery for an almost 1 dex more luminous GRB]{Xu2013a}.
In addition, \citet{Maselli2013a} showed that its afterglow properties are
similar to those of very energetic, high-redshift GRBs, making it a
genuine very high-$L$ GRB.

Combining the findings on GRBs 120422A and 130427A lets us conclude that low-
and high-$L$ GRBs are not distinct populations of stellar explosions. They
are due to the gravitational collapses of very massive stars and are accompanied
by SNe. Their central engines are the same. Only the properties of their prompt
emissions and of their afterglows (shock break-out vs. jet dominated) differ,
depending on whether the jet can successfully drill through the stellar cocoon.
This does not make them disjunct phenomena.

\section{Summary and conclusions}\label{sec:6}

We carried out extensive imaging and spectroscopy campaigns to study the intermediate
luminosity GRB 120422A that shares properties of low-$L$ and high-$L$ GRBs. Our detailed
analysis focussed on the GRB-SN 2012bz, the GRB afterglow, the host galaxy, and its
environment.

We showed that SN 2012bz is the most luminous spectroscopically-confirmed GRB-SN to date,
the peak luminosity being $M_{\rm V}=-19.7~\rm mag$. The explosion physics parameters of
$M_{\rm Ni}=0.58~\rm{M}_{\odot}$, $M_{\rm ej}=5.87~\rm{M}_{\odot}$, and $E_{\rm k}=4.10\times10^{52}~\rm{erg}$
are among the largest values known to date. However, the exact values highly depend on the NIR
contribution. \citet{Cano2011a} showed that the nickel and the ejecta masses and the
kinetic energy can be underestimated by 25--45\% if the NIR is not included in the
modelling of the bolometric light curve. For future GRB-SN studies it is imperative
to secure NIR data to place more stringent constraints on GRB progenitor models.
As an alternative to a campaign with long wavelength coverage, the method
presented in \citet{Lyman2013a} would allow to construct the bolometric light curve
from two optical light curves with well-determined $k$-corrections.

The spectral sequence of SN covers a time span of $\sim40$~days. All spectra are similar
to that of other GRB-SNe. Differences exist in the strength of the absorption features
and expansion velocities. For the first time, \ion{Fe}{ii}$\lambda$5169 was used to
trace the evolution of the GRB-SN expansion velocity. The velocities and their evolution are
not very different from \ion{Si}{ii} measurements. \ion{Fe}{ii}$\lambda$5169 has
the advantage of being easier to identify and is detectable at earlier times
We find an intriguing trend between the peak luminosity $k$ and SN stretch factor $s$,
similar to the Philips relation. Its significance is poor but several systematics affect
the result.

GRB120422A was accompanied by one of the least luminous afterglows detected to date.
Its blast-wave expanded with a low initial Lorentz factor of $\Gamma_0\sim60$
into a constant density medium and produced an very weak afterglow
$L_{\nu,\,\rm max}\lesssim2\times10^{30}~\rm{erg\,s}^{-1}\,\rm{Hz}^{-1}$. Thanks to the weak
afterglow, we recovered emission from the cooling photosphere after the shock break-out,
which was only observed for the low-$L$ GRBs 060218 and 100316D. GRB 120422A's photosphere
had a temperature of $\sim16~\rm eV$ and a radius of $7\times10^13~\rm cm$ at 1.4 hours after
the GRB, typical values of SNe with detected shock break-out signatures. This fundamentally
new quality for a non-low-$L$ GRB questions whether 120422A is a genuine high-$L$ GRB.
Considering the properties of the prompt emission and the afterglow makes us conclude
that GRB 120422A is the missing link between shock-break-out driven low-$L$ and high-$L$
GRBs that are powered by an ultra-relativistic jet, hence providing evidence for the
failed-jet model for low-$L$ GRBs.

The GRB occurred in an almost typical host galaxy. Its close to solar metallicity along
with its negligible extinction makes it peculiar. Thanks to the large offset of
7.3 kpc from the nucleus, we perform spatially resolved spectroscopy. Surprisingly,
even at the explosion site the metallicity is close to solar, while the SFR is not enhanced
with respect to its immediate environment and only 1/10th of that of the galaxy's
nucleus. Based on the N2 indicator we measure $Z=0.8\pm0.1~\rm{Z}_\odot$ at the explosion
site. This does
not necessarily mean that the GRB-hosting star-forming region had these properties. The
X-shooter spectrum was only sensitive to a region of $4.0\times3.9~\rm{kpc}^2$.
What needs to be stressed is that emission-line measurements from low-resolution
spectra should be taken with caution. Our medium resolution data revealed
that the H$\alpha$ line resolved in two components. This can lead in low-resolution
data to an overestimation of the extinction and SFR.

Our narrowband imaging (width 15 \AA) showed that the host is possibly interacting
with a galaxy that lies at a projected distance of 23 kpc away. We
identified two additional putative galaxy group members. In contrast to previous studies of
GRB galaxy environments, tuneable filters allow us to more efficiently identify
star-forming galaxies at a GRB's redshift. In particular, this approach
is complementary to SED fitting techniques, which are limited to bright galaxies but not
necessarily highly star-forming galaxies. Both
approaches are needed to address the long-standing question on the  peculiarity of GRB
host galaxies and how galaxy interaction affects the production of GRB progenitors
at low and high metallicities.

\begin{acknowledgements}

We thank Shri Kulkarni (Caltech) for obtaining the Keck spectrum. SS thanks
Tsvi Piran (The Hebrew University, Israel), Nir Sapir, Eli Waxman (Weizmann Institute
of Science, Israel), Milena Bufano (Universidad Andr\'es Bello, Chile) for many
productive and valuable discussions.

SS acknowledges support by a Grant of Excellence from the Icelandic Research Fund,
from the University of Iceland Research Fund, from the Dark Cosmology Centre, where part
of this study was performed, and from CONICYT through FONDECYT grant 3140534.
We acknowledge support from Basal-CATA PFB-06/2007 (FEB, SS), Iniciativa Cientifica
Milenio grant P10-064-F (Millennium Center for Supernova Science), by Project IC120009
"Millennium Institute of Astrophysics (MAS)" of Iniciativa Cient\'ifica Milenio del
Ministerio de Economía, Fomento y Turismo de Chile, with input from "Fondo
de Innovaci\'{o}n para la Competitividad, del Ministerio de Econom\'{\i}a, Fomento y
Turismo de Chile" (FEB, GP, JAP, SS), CONICYT-Chile FONDECYT 1101024 (FEB). JAP
acknowledges support by CONICYT through FONDECYT grant 3110142.

TK and HK acknowledge support by the European Commission under the Marie Curie
Intra-European Fellowship Programme in FP7. JPUF and BMJ acknowledges support from the
ERC-StG grant EGGS-278202.

KLP acknowledges financial support by the UK Space Agency for the \swift\ project.

GL is supported by the Swedish Research Council through grant No. 623-2011-7117.

JL acknowledges the UK Science and Technology Facilities Council for research studentship
support.

The research activity of AdUP, CT, and JG is supported by Spanish research project
AYA2012-39362-C02-02. AdUP acknowledges support by the European Commission under the
Marie Curie Career Integration Grant programme (FP7-PEOPLE-2012-CIG 322307).

AJCT acknowledges support from the Spanish research project AYA2009-14000-C03-01 and
AYA2012-39727-C03-01

DAK acknowledges support by the DFG cluster of excellence "Origin and
Structure of the Universe" and funding by the Th\"uringer Landessternwarte Tautenburg.

AR acknowledges support by the Th\"uringer Landessternwarte Tautenburg.

PS acknowledges support through the Sofja Kovalevskaja Award from the Alexander von
Humboldt Foundation of Germany.

ANG and SK acknowledge support by DFG KL 766/16-1. S. Schmidl acknowledges support
by the Th\"uringer Ministerium f\"ur Bildung, Wissenschaft und Kultur under FKZ 12010-514.

The Dark Cosmology Centre is funded by the Danish National Research Foundation.

This work made use of data supplied by the UK \swift\ Science Data Centre at the
University of Leicester. This research has made use of the GHostS database
(www.grbhosts.org), which is partly funded by
\textit{Spitzer}/NASA grant RSA Agreement No. 1287913.

Based in part on observations collected at the European Organisation for Astronomical
Research in the Southern Hemisphere, Chile, as part of the program 089.A-0067, the
Gemini Observatory, which is operated by the Association of Universities for Research
in Astronomy, Inc., under a co-operative agreement with the NSF on behalf of the Gemini
partnership, as part of the programs GN-2012A-Q-9, GN-2012A-Q-39, GN-2012B-Q-5,
GS-2012A-Q-30, GS-2012A-Q-38, GS-2012A-Q-30, and GN-2012B-Q-5, the Nordic Optical
Telescope (NOT), operated by the Nordic Optical Telescope Scientific Association at the
Observatorio del Roque de los Muchachos, La Palma, Spain, of the Instituto de
Astrofisica de Canarias, as part of the program P45-002 (PI: P. Jakobsson) and ITP10-04
(PI: R. Kotak, QUB), the Gran Telescopio Canarias (GTC), installed in the Spanish
Observatorio del Roque de los Muchachos of the Instituto de Astrof\'{i}sica de Canarias,
in the island of La Palma, with Magellan as part of CN2012A-059, with the IRAM Plateau
de Bure Interferometer, the James Clerk Maxwell Telescope, as part of the program M12AI12,
with \xmm, an ESA science mission with instruments and contributions directly funded
by ESA Member States and NASA.

Some of the data presented herein were obtained at the W.M. Keck Observatory, which is
operated as a scientific partnership among the California Institute of Technology, the
University of California and the National Aeronautics and Space Administration. The
observatory was made possible by the generous financial support of the W.M. Keck
Foundation.

The United Kingdom Infrared Telescope is operated by the Joint Astronomy Centre on
behalf of the Science and Technology Facilities Council of the U.K.

The James Clerk Maxwell Telescope is operated by the Joint Astronomy Centre on behalf
of the Science and Technology Facilities Council of the United Kingdom, the National
Research Council of Canada, and the Netherlands Organisation for Scientific Research.
Additional funds for the construction of SCUBA-2 were provided by the Canada Foundation
for Innovation.

The Submillimeter Array is a joint project between the Smithsonian Astrophysical
Observatory and the Academia Sinica Institute of Astronomy and Astrophysics and is
funded by the Smithsonian Institution and the Academia Sinica.

IRAM is supported by INSU/CNRS (France), MPG (Germany) and IGN (Spain).

Support for CARMA construction was derived from the Gordon and Betty Moore Foundation,
the Kenneth T. and Eileen L. Norris Foundation, the James S. McDonnell Foundation,
the Associates of the California Institute of Technology, the University of Chicago,
the states of California, Illinois, and Maryland, and the National Science Foundation.
Ongoing CARMA development and operations are supported by the National Science Foundation
under a cooperative agreement, and by the CARMA partner universities.

Part of the funding for GROND (both hardware as well as personnel) was generously
granted from the Leibniz-Prize to Prof. G. Hasinger (DFG grant HA 1850/28-1).

Funding for SDSS-III has been provided by the Alfred P. Sloan Foundation, the
Participating Institutions, the National Science Foundation, and the U.S. Department
of Energy Office of Science. The SDSS-III web site is http://www.sdss3.org/.
SDSS-III is managed by the Astrophysical Research Consortium for the Participating
Institutions of the SDSS-III Collaboration including the University of Arizona,
the Brazilian Participation Group, Brookhaven National Laboratory, Carnegie Mellon
University, University of Florida, the French Participation Group, the German
Participation Group, Harvard University, the Instituto de Astrofisica de Canarias,
the Michigan State/Notre Dame/JINA Participation Group, Johns Hopkins University,
Lawrence Berkeley National Laboratory, Max Planck Institute for Astrophysics, Max
Planck Institute for Extraterrestrial Physics, New Mexico State University, New York
University, Ohio State University, Pennsylvania State University, University of
Portsmouth, Princeton University, the Spanish Participation Group, University of
Tokyo, University of Utah, Vanderbilt University, University of Virginia, University
of Washington, and Yale University.
\end{acknowledgements}

\newpage
\newpage

\appendix
\newpage
\newpage
\section{Photometry of the optical transient}
\begin{table}[h!]
\caption{Log of optical and NIR observations}
\centering
\scriptsize
\begin{tabular}{r@{\hspace{1mm}}r@{\hspace{1mm}}c@{\hspace{1mm}}c@{\hspace{1mm}}c@{\hspace{1mm}}c@{\hspace{1mm}}c}

\toprule
\multicolumn{1}{c}{MJD}		& \multicolumn{1}{c}{Epoch}	& \multirow{2}{*}{Instrument}	& \multicolumn{1}{c}{\multirow{2}{*}{Filter}}	& Exposure	&Brightness\\
\multicolumn{1}{c}{(days)}	& \multicolumn{1}{c}{(s)}	& 								&												& time (s)	&(mag$_{\rm{AB}}$)\\
\midrule
56039.308	&	664.2	& \swift/UVOT		&$	uvw2$&	38.9		&$	>20.27					$\\
56039.313	&	1110.2	& \swift/UVOT		&$	uvw2$&	38.9		&$	20.30^{+0.55} _{-0.36}	$\\
56039.373	&	6244.8	& \swift/UVOT		&$	uvw2$&	332.2		&$	20.50^{+0.19} _{-0.16}	$\\
56066.505	&	2350484	& \swift/UVOT		&$	uvw2$&	16823.3		&$	>24.05					$\\[2mm]

56039.311	&	935.9	& \swift/UVOT		&$	uvm2$&	77.8		&$	20.84^{+0.75} _{-0.44}	$\\
56039.370	&	5967.3	& \swift/UVOT		&$	uvm2$&	196.6		&$	20.89^{+0.47} _{-0.33}	$\\
56039.436	&	11694.8	& \swift/UVOT		&$	uvm2$&	885.6		&$	21.03^{+0.23} _{-0.19}	$\\
56066.471	&	2347530	& \swift/UVOT		&$	uvm2$&	16365		&$	>23.73					$\\[2mm]

56039.311	&	912.7	& \swift/UVOT		&$	uvw1$&	58.3		&$	>20.56					$\\
56039.364	&	5454.6	& \swift/UVOT		&$	uvw1$&	393.2		&$	21.43^{+0.48} _{-0.33}	$\\
56039.445	&	12479.4	& \swift/UVOT		&$	uvw1$&	645.3		&$	21.56^{+0.37} _{-0.28}	$\\
56039.598	&	25671.6	& \swift/UVOT		&$	uvw1$&	1771.2		&$	21.47^{+0.21} _{-0.18}	$\\
56067.110	&	2402743	& \swift/UVOT		&$	uvw1$&	11529.3		&$	>23.43					$\\[2mm]

56039.305	&	387.2	& \swift/UVOT		&$	u	$&	245.8		&$	21.06^{+0.51} _{-0.35}	$\\
56039.311	&	898.9	& \swift/UVOT		&$	u	$&	52.9		&$	>20.08					$\\
56039.366	&	5659.9	& \swift/UVOT		&$	u	$&	393.2		&$	21.08^{+0.43} _{-0.31}	$\\
56039.530	&	19853.4	& \swift/UVOT		&$	u	$&	1770.3		&$	21.47^{+0.27} _{-0.21}	$\\
56039.645	&	29800.8	& \swift/UVOT		&$	u	$&	651.6		&$	21.41^{+0.39} _{-0.29}	$\\
56042.104	&	242229.9& \swift/UVOT		&$	u	$&	11277.9		&$	>22.94					$\\
56051.997	&	1096947	& \swift/UVOT		&$	u	$&	69897.4		&$	23.00^{+0.17} _{-0.14}	$\\
56077.011	&	3258148	& \swift/UVOT		&$	u	$&	27766.2		&$	>23.32					$\\[2mm]

56039.339	&	3328.3	& \swift/UVOT		&$	b	$&	451.6		&$	>20.42					$\\
56039.539	&	20641.6	& \swift/UVOT		&$	b	$&	1523.7		&$	20.95^{+0.39} _{-0.28}	$\\
56040.652	&	116790.9& \swift/UVOT		&$	b	$&	1444		&$	>21.07					$\\
56061.463	&	1914851	& \swift/UVOT		&$	b	$&	7250.3		&$	21.54^{+0.36} _{-0.27}	$\\
56093.296	&	4665241	& \swift/UVOT		&$	b	$&	449.7		&$	>19.82					$\\[2mm]

56039.302	&	90.4	& \swift/UVOT		&$	v	$&	9			&$	>17.52					$\\
56039.311	&	912		& \swift/UVOT		&$	v	$&	77.7		&$	>18.83					$\\
56039.541	&	20785.9	& \swift/UVOT		&$	v	$&	1607.4		&$	>20.28					$\\
56041.111	&	156431	& \swift/UVOT		&$	v	$&	1282.9		&$	>20.27					$\\
56093.300	&	4665526	& \swift/UVOT		&$	v	$&	449.8		&$	>19.17					$\\[2mm]

56039.990	&	59563	&	GROND			&$	g'	$&$	4\times	115	$&$	22.21	\pm	0.10	$\\
56040.015	&	61702	&	GROND			&$	g'	$&$	4\times	369	$&$	22.12	\pm	0.05	$\\
56040.036	&	63556	&	GROND			&$	g'	$&$	4\times	369	$&$	22.15	\pm	0.07	$\\
56040.040	&	63883	&	Gemini/GMOS		&$	g'	$&$	1\times 60	$&$	22.22	\pm	0.06	$\\
56040.057	&	65383	&	GROND			&$	g'	$&$	4\times	369	$&$	22.11	\pm	0.08	$\\
56040.089	&	68118	&	GROND			&$	g'	$&$	8\times	369	$&$	22.18	\pm	0.06	$\\
56040.173	&	75361	&	P60				&$	g'	$&$	900			$&$	22.14	\pm	0.10	$\\
56041.067	&	152661	&	GROND			&$	g'	$&$	16\times115	$&$	22.59	\pm	0.10	$\\
56041.104	&	155811	&	GROND			&$	g'	$&$	16\times115	$&$	22.79	\pm	0.16	$\\
56043.239	&	340286	&	Gemini/GMOS		&$	g'	$&$	1\times100	$&$	22.86	\pm	0.06	$\\
56048.018	&	753173	&	GROND			&$	g'	$&$	8\times	369	$&$	22.38	\pm	0.10	$\\
56050.011	&	925358	&	GROND			&$	g'	$&$	8\times	369	$&$	22.35	\pm	0.08	$\\
56053.969	&	1267371	&	GTC/Osiris		&$	g'	$&$	1\times 100	$&$	22.14	\pm	0.11	$\\
56054.249	&	1291574	&	Gemini/GMOS		&$	g'	$&$	1\times 30	$&$	22.16	\pm	0.07	$\\
56059.005	&	1702443	&	GROND			&$	g'	$&$	8\times	369	$&$	22.22	\pm	0.04	$\\
56059.020	&	1703734	&	Gemini/GMOS		&$	g'	$&$	1\times 120	$&$	22.19	\pm	0.07	$\\
56067.938	&	2474262	&	GROND			&$	g'	$&$	4\times	369	$&$	22.82	\pm	0.05	$\\
56078.217	&	3362396	&	GROND			&$	g'	$&$	24\times115	$&$	23.29	\pm	0.20	$\\
\bottomrule
\end{tabular}
\tablefoot{Magnitudes are
	corrected for Galactic extinction ($E(B-V) = 0.03\,\rm{mag}$). Column
	"Epoch" shows the logarithmic mean-time after the GRB in the observer
	frame. We only display the total observing time of the \swift/UVOT and P60 data (see Sect. \ref{sec:imaging} for details).
	As described in Sect. \ref{sec:photometry}, photometry was tied to the
	SDSS DR8 standard  ($'g'r'i'z'$)
	and to the 2MASS standard ($JHK_s$). For those filters
	not covered by our primary calibration systems ($RI_{\rm{C}}i_iY$) we used the instrument-specific band passes	to transform magnitudes
	into the respective filter system.}
\label{tab:obs_log}
\end{table}

\begin{table}
\centering
\scriptsize
\begin{tabular}{r@{\hspace{1mm}}r@{\hspace{1mm}}c@{\hspace{1mm}}c@{\hspace{1mm}}c@{\hspace{1mm}}c@{\hspace{1mm}}c}

\toprule
\multicolumn{1}{c}{MJD}		& \multicolumn{1}{c}{Epoch}	& \multirow{2}{*}{Instrument}	& \multicolumn{1}{c}{\multirow{2}{*}{Filter}}	& Exposure	&Brightness\\
\multicolumn{1}{c}{(days)}	& \multicolumn{1}{c}{(s)}	& 								&												& time (s)	&(mag$_{\rm{AB}}$)\\
\midrule
56039.414	&	9793	&	Gemini/GMOS		&$	r'	$&$	1\times60	$&$	21.11	\pm	0.04	$\\
56039.990	&	59563	&	GROND			&$	r'	$&$	4\times	115	$&$	22.19	\pm	0.14	$\\
56040.015	&	61702	&	GROND			&$	r'	$&$	4\times	369	$&$	22.17	\pm	0.07	$\\
56040.036	&	63556	&	GROND			&$	r'	$&$	4\times	369	$&$	22.02	\pm	0.05	$\\
56040.047	&	64459	&	Gemini/GMOS		&$	r'	$&$	1\times60	$&$	22.10	\pm	0.05	$\\
56040.057	&	65383	&	GROND			&$	r'	$&$	4\times	369	$&$	22.22	\pm	0.07	$\\
56040.089	&	68118	&	GROND			&$	r'	$&$	8\times	369	$&$	22.12	\pm	0.08	$\\
56040.161	&	74357	&	P60				&$	r'	$&$	900			$&$	22.28	\pm	0.10	$\\
56040.888	&	137134	&	NOT/MOSCA		&$	r'	$&$	4\times 300	$&$	22.28	\pm	0.08	$\\
56041.068	&	152661	&	GROND			&$	r'	$&$	16\times 115$&$	22.34	\pm	0.09	$\\
56041.104	&	155811	&	GROND			&$	r'	$&$	16\times 115$&$	22.47	\pm	0.12	$\\
56041.949	&	228856	&	NOT/MOSCA		&$	r'	$&$	12\times300	$&$	22.23	\pm	0.06	$\\
56042.938	&	314237	&	NOT/ALFOSC		&$	r'	$&$	24\times150	$&$	22.07	\pm	0.06	$\\
56043.247	&	340990	&	Gemini/GMOS		&$	r'	$&$	1\times100	$&$	22.12	\pm	0.06	$\\
56048.018	&	753173	&	GROND			&$	r'	$&$	8\times	369	$&$	21.48	\pm	0.05	$\\
56048.967	&	835203	&	NOT/StanCAM		&$	R	$&$ 8\times 150	$&$	21.45	\pm 0.10	$\\
56050.011	&	925358	&	GROND			&$	r'	$&$	8\times	369	$&$	21.33	\pm	0.04	$\\
56052.958	&	1179975	&	Gemini/GMOS		&$	r'	$&$	1\times 30	$&$	21.41	\pm	0.10	$\\
56053.886	&	1260196	&	NOT/StanCAM		&$	R	$&$	12\times150	$&$	21.25	\pm 0.06	$\\
56053.972	&	1267581	&	GTC/Osiris		&$	r'	$&$ 1\times100	$&$	21.37	\pm	0.14	$\\
56054.258	&	1292332	&	Gemini/GMOS		&$	r'	$&$	1\times30	$&$	21.26	\pm	0.05	$\\
56056.895	&	1520199	&	NOT/ALFOSC		&$	r'	$&$	8\times 150 $&$	21.23	\pm	0.07	$\\
56059.032	&	1704797	&	Gemini/GMOS		&$	r'	$&$	1\times 120 $&$	21.27	\pm	0.05	$\\
56059.005	&	1702443	&	GROND			&$	r'	$&$	8\times	369	$&$	21.21	\pm	0.04	$\\
56061.962	&	1957915	&	Gemini/GMOS		&$	r'	$&$	1\times 30	$&$	21.32	\pm	0.06	$\\
56063.912	&	2126451	&	NOT/ALFOSC		&$	r'	$&$	15\times 90	$&$	21.36	\pm	0.08	$\\
56065.897	&	2297894	&	NOT/MOSCA		&$	r'	$&$	16\times 90	$&$	21.48	\pm	0.05	$\\
56065.968	&	2304100	&	Gemini/GMOS		&$	r'	$&$	1\times 100	$&$	21.51	\pm	0.04	$\\
56066.028	&	2309230	&	Magellan/LDSS3	&$	r'	$&$	3\times180	$&$	21.55	\pm	0.04	$\\
56067.938	&	2474262	&	GROND			&$	r'	$&$	4\times	369	$&$	21.65	\pm	0.04	$\\
56067.957	&	2475887	&	NOT/MOSCA		&$	r'	$&$	15\times90	$&$	21.54	\pm	0.08	$\\
56067.979	&	2477779	&	DuPont/CCD		&$	r'	$&$	4\times500	$&$	21.58	\pm	0.05	$\\
56069.966	&	2649489	&	Gemini/GMOS		&$	r'	$&$	1\times100	$&$	21.73	\pm	0.04	$\\
56069.902	&	2643961	&	NOT/MOSCA		&$	r'	$&$ 15\times90	$&$	21.79	\pm	0.07	$\\
56070.957	&	2735091	&	Gemini/GMOS		&$	r'	$&$	1\times100	$&$	21.77	\pm	0.08	$\\
56071.907	&	2817199	&	NOT/ALFOSC		&$	r'	$&$	10\times90	$&$	21.96	\pm	0.07	$\\
56078.217	&	3362396	&	GROND			&$	r'	$&$	24\times115	$&$	22.43	\pm	0.11	$\\
56079.955	&	3512502	&	Gemini/GMOS		&$	r'	$&$	1\times30	$&$	22.41	\pm	0.11	$\\
56083.915	&	3854715	&	NOT/ALFOSC		&$	r'	$&$	20\times90	$&$	22.39	\pm	0.47	$\\[2mm]

56039.322	&	1880	&	Gemini/GMOS		&$	i'	$&$	1\times240	$&$	20.93	\pm	0.04	$\\
56039.896	&	51451	&	NOT/MOSCA		&$  I	$&$ 12\times300	$&$ 22.19	\pm 0.08	$\\
56039.990	&	59563	&	GROND			&$	i'	$&$	4\times	115	$&$	22.01	\pm	0.18	$\\
56040.015	&	61702	&	GROND			&$	i'	$&$	4\times	369	$&$	22.13	\pm	0.07	$\\
56040.036	&	63556	&	GROND			&$	i'	$&$	4\times	369	$&$	22.27	\pm	0.10	$\\
56040.047	&	64459	&	Gemini/GMOS		&$	i'	$&$	1\times	60	$&$	22.16	\pm	0.06	$\\
56040.078	&	67209	&	GROND			&$	i'	$&$	12\times369	$&$	22.32	\pm	0.09	$\\
56040.149	&	73289	&	P60				&$	i'	$&$	900			$&$	22.23	\pm	0.17	$\\
56040.871	&	135683	&	NOT/MOSCA		&$	i'	$&$	4\times300	$&$	22.48	\pm	0.12	$\\
56041.067	&	152661	&	GROND			&$	i'	$&$	16\times115	$&$	22.50	\pm	0.16	$\\
56041.104	&	155811	&	GROND			&$	i'	$&$	16\times115	$&$	22.59	\pm	0.19	$\\
56041.924	&	226705	&	NOT/MOSCA		&$	i'	$&$	6\times300	$&$	22.37	\pm	0.13	$\\
56042.866	&	308045	&	GTC/Osiris		&$	i'	$&$	1\times10	$&$	22.33	\pm	0.14	$\\
56042.885	&	309735	&	NOT/ALFOSC		&$	i'	$&$	13\times300	$&$	22.35	\pm	0.06	$\\
56043.255	&	341689	&	Gemini/GMOS		&$	i'	$&$	1\times100	$&$	22.29	\pm	0.06	$\\
56043.874	&	395133	&	NOT/ALFOSC		&$	i'	$&$	12\times150	$&$	22.18	\pm	0.08	$\\
56047.890	&	742159	&	NOT/ALFOSC		&$	i'	$&$	3150$\tablefootmark{a}&$	21.56	\pm	0.08	$\\
56048.018	&	753173	&	GROND			&$	i'	$&$	8\times	369	$&$	21.56	\pm	0.08	$\\
56048.987	&	836905	&	NOT/StanCam		&$  i_i	$&$ 8\times 150	$&$	21.52	\pm 0.08	$\\
56050.011	&	925358	&	GROND			&$	i'	$&$	8\times	369	$&$	21.49	\pm	0.05	$\\
56053.916	&	1262736	&	NOT/StanCAM		&$	i_i	$&$12\times 150	$&$	21.34	\pm 0.05	$\\
56053.903	&	1261613	&	GTC/Osiris		&$	i'	$&$	1\times100	$&$	21.20	\pm	0.09	$\\
56063.974	&	1267805	&	GTC/Osiris		&$	i'	$&$	1\times100	$&$	21.27	\pm	0.06	$\\
56054.163	&	1284132	&	P60				&$	i'	$&$	1800		$&$	21.21	\pm	0.14	$\\
56054.264	&	1292809	&	Gemini/GMOS		&$	i'	$&$	1\times30	$&$	21.29	\pm	0.05	$\\
56055.175	&	1371560	&	P60				&$	i'	$&$	3600		$&$	21.26	\pm	0.11	$\\
56055.884	&	1432840	&	NOT/StanCAM		&$	i_i	$&$ 8\times150	$&$	21.35	\pm 0.05	$\\
56058.889	&	1692472	&	NOT/ALFOSC		&$	i'	$&$	11\times150	$&$	21.32	\pm	0.07	$\\
56059.005	&	1702443	&	GROND			&$	i'	$&$	8\times	369	$&$	21.27	\pm	0.03	$\\
56059.005	&	1702494	&	Gemini/GMOS		&$	i'	$&$	1\times120	$&$	21.24	\pm	0.04	$\\
56061.176	&	1890072	&	P60				&$	i'	$&$ 3240		$&$	21.22	\pm	0.14	$\\
56062.181	&	1976858	&	P60				&$	i'	$&$	3600		$&$	21.27	\pm	0.15	$\\
56062.886	&	2037796	&	NOT/ALFOSC		&$	i'	$&$	15\times90	$&$	21.46	\pm	0.22	$\\
56063.180	&	2063213	&	P60				&$	i'	$&$	3600		$&$	21.33	\pm	0.10	$\\
56063.891	&	2124602	&	NOT/ALFOSC		&$	i'	$&$	10\times120	$&$	21.41	\pm	0.17	$\\
56065.921	&	2300003	&	NOT/MOSCA		&$	i'	$&$	10\times120	$&$	21.29	\pm	0.05	$\\
\bottomrule
\end{tabular}
\tablefoottext{a}{The image is a stack of images with different exposure times. The shown time is the sum of the single images.}
\hfill\center{Tab. \ref{tab:obs_log} --- continued}
\end{table}

\begin{table}
\centering
\scriptsize
\begin{tabular}{r@{\hspace{1mm}}r@{\hspace{1mm}}c@{\hspace{1mm}}c@{\hspace{1mm}}c@{\hspace{1mm}}c@{\hspace{1mm}}c}

\toprule
\multicolumn{1}{c}{MJD}		& \multicolumn{1}{c}{Epoch}	& \multirow{2}{*}{Instrument}	& \multicolumn{1}{c}{\multirow{2}{*}{Filter}}	& Exposure	&Brightness\\
\multicolumn{1}{c}{(days)}	& \multicolumn{1}{c}{(s)}	& 								&												& time (s)	&(mag$_{\rm{AB}}$)\\
\midrule
56066.040	&	2310267	&	Magellan/LDSS3	&$	i'	$&$	3\times180	$&$	21.33	\pm	0.04	$\\
56067.934	&	2473900	&	NOT/MOSCA		&$	i'	$&$	10\times120	$&$	21.35	\pm	0.05	$\\
56067.938	&	2474262	&	GROND			&$	i'	$&$	4\times	369	$&$	21.47	\pm	0.06	$\\
56068.014	&	2480838	&	DuPont/CCD		&$	i'	$&$	4\times500	$&$	21.41	\pm	0.04	$\\
56068.923	&	2559358	&	NOT/MOSCA		&$	i'	$&$	19\times60	$&$	21.39	\pm	0.06	$\\
56069.1911	&	2582542	&	P60				&$	i'	$&$	720			$&$	21.50	\pm	0.09	$\\
56070.0313	&	2655131	&	Magellan/LDSS3	&$	i'	$&$	3\times300	$&$	21.40	\pm	0.06	$\\
56070.1766	&	2667689	&	P60				&$	i'	$&$	2340		$&$	21.59	\pm	0.15	$\\
56070.9694	&	2736187	&	Gemini/GMOS		&$	i'	$&$	1\times120	$&$	21.62	\pm	0.04	$\\
56071.8946	&	2816120	&	NOT/ALFOSC		&$	i'	$&$	13\times90	$&$	21.63	\pm	0.08	$\\
56076.8994	&	3248536	&	NOT/ALFOSC		&$	i'	$&$	20\times90	$&$	21.98	\pm	0.09	$\\
56078.2172	&	3362396	&	GROND			&$	i'	$&$	24\times115	$&$	22.06	\pm	0.12	$\\
56079.8939	&	3507261	&	NOT/ALFOSC		&$	i'	$&$	30\times90	$&$	22.08	\pm	0.11	$\\
56083.8841	&	3852020	&	NOT/ALFOSC		&$	i'	$&$	20\times90	$&$	22.15	\pm	0.21	$\\[2mm]

56039.9900	&	59563	&	GROND			&$	z'	$&$	4\times	115	$&$	22.01	\pm	0.20	$\\
56040.0360	&	63541	&	GROND			&$	z'	$&$	12\times369	$&$	22.09	\pm	0.11	$\\
56040.0599	&	65605	&	Gemini/GMOS		&$	z'	$&$	1\times	60	$&$	22.16	\pm	0.11	$\\
56040.0890	&	68118	&	GROND			&$	z'	$&$	8\times	369	$&$	22.37	\pm	0.17	$\\
56041.0675	&	152661	&	GROND			&$	z'	$&$	8\times	369	$&$	22.17	\pm	0.16	$\\
56041.1039	&	155811	&	GROND			&$	z'	$&$	8\times	369	$&$	22.28	\pm	0.27	$\\
56043.2633	&	342383	&	Gemini/GMOS		&$	z'	$&$	1\times	100	$&$	22.59	\pm	0.07	$\\
56048.0179	&	753173	&	GROND			&$	z'	$&$	8\times	369	$&$	21.86	\pm	0.15	$\\
56049.8781	&	913895	&	CAHA/Omega$_{2000}$	&$ z'$&$20\times90	$&$	21.93	\pm	0.08	$\\
56050.0107	&	925358	&	GROND			&$	z'	$&$	8\times	369	$&$	21.78	\pm	0.12	$\\
56052.2596	&	1119664	&	UKIRT/WFCAM		&$	z'	$&$	4\times 360	$&$	21.44	\pm	0.13	$\\
56053.9770	&	1268047	&	GTC/Osiris		&$	z'	$&$	3\times70	$&$	21.58	\pm	0.07	$\\
56054.2708	&	1293429	&	Gemini/GMOS		&$	z'	$&$	1\times	30	$&$	21.63	\pm	0.06	$\\
56058.9733	&	1699722	&	Gemini/GMOS		&$	z'	$&$	1\times 120	$&$	21.48	\pm	0.04	$\\
56059.0048	&	1702443	&	GROND			&$	z'	$&$	8\times	369	$&$	21.48	\pm	0.07	$\\
56067.9101	&	2471861	&	NOT/MOSCA		&$	z'	$&$	12\times120	$&$	21.74	\pm 0.15	$\\
56067.9379	&	2474262	&	GROND			&$	z'	$&$	4\times	369	$&$	21.62	\pm	0.07	$\\
56068.8970	&	2557129	&	NOT/MOSCA		&$	z'	$&$	13\times120	$&$	21.59	\pm 0.12	$\\
56070.9776	&	2736894	&	Gemini/GMOS		&$	z'	$&$	1\times	120	$&$	21.86	\pm	0.06	$\\
56078.2172	&	3362396	&	GROND			&$	z'	$&$	24\times115	$&$	21.81	\pm	0.12	$\\[2mm]

56049.9204	&	917552	&	CAHA/Omega$_{2000}$	&$Y$&$20\times90	$&$	21.93	\pm	0.17	$\\[2mm]

56039.3537	&	4589	&	UKIRT/WFCAM		&$	J	$&$	360			$&$	20.20	\pm	0.06	$\\
56039.3590	&	5048	&	UKIRT/WFCAM		&$	J	$&$	360			$&$	20.33	\pm	0.07	$\\
56039.3644	&	5514	& 	UKIRT/WFCAM		&$	J	$&$	360			$&$	20.36	\pm	0.07	$\\
56039.3698	&	5979	&	UKIRT/WFCAM		&$	J	$&$	360			$&$	20.41	\pm	0.07	$\\
56048.9026	&	829616	&	CAHA/Omega$_{2000}$&$ J	$&$	60\times60	$&$	21.60	\pm	0.17	$\\
56049.9727	&	922069	&	CAHA/Omega$_{2000}$&$ J	$&$	30\times60	$&$	21.83	\pm	0.20	$\\
56052.2354	&	1117569	&	UKIRT/WFCAM		&$	J	$&$	4\times360	$&$	21.75	\pm	0.24	$\\
56054.2834	&	1294514	&	Gemini-N/NIRI	&$	J	$&$	1\times60	$&$	21.96	\pm	0.11	$\\
56065.2816	&	2244761	&	UKIRT/WFCAM		&$	J	$&$	6\times360	$&$	21.90	\pm	0.15	$\\
56049.1898	&	854426	&	P200/WIRC		&$	J	$&$	15\times240	$&$	21.68	\pm	0.16	$\\[2mm]

56039.3757	&	6493	&	UKIRT/WFCAM		&$	H	$&$	360			$&$	20.29	\pm	0.09	$\\
56039.3812	&	6963	&	UKIRT/WFCAM		&$	H	$&$	360			$&$	20.32	\pm	0.09	$\\
56039.3866	&	7432	&	UKIRT/WFCAM		&$	H	$&$	360			$&$	20.51	\pm	0.11	$\\
56039.3919	&	7894	&	UKIRT/WFCAM		&$	H	$&$	360			$&$	20.34	\pm	0.10	$\\
56040.3685	&	92266	&	UKIRT/WFCAM		&$	H	$&$	4\times360	$&$	21.65	\pm	0.42	$\\
56042.3348	&	262155	&	UKIRT/WFCAM		&$	H	$&$	4\times360	$&$	22.29	\pm	0.31	$\\[2mm]

56054.2376	&	1290555	&	Gemini-N/NIRI	&$	K	$&$	1\times60	$&$	21.46	\pm	0.14	$\\
\bottomrule
\end{tabular}
\hfill\center{Tab. \ref{tab:obs_log} --- continued}
\end{table}

\newpage
\newpage
\section{Late time observations}
\begin{table}
\caption{Summary of late-time observations}
\centering
\scriptsize
\begin{tabular}{l@{\hspace{1mm}}c@{\hspace{1mm}}c@{\hspace{1mm}}c@{\hspace{1mm}}c@{\hspace{1mm}}c}
\toprule
\multicolumn{1}{c}{MJD}		& \multicolumn{1}{c}{Epoch}	& \multicolumn{1}{c}{\multirow{2}{*}{Instrument}}	& \multicolumn{1}{c}{\multirow{2}{*}{Filter}}	& Exposure\\
\multicolumn{1}{c}{(days)}	& \multicolumn{1}{c}{(s)}&													& 												& time (s)\\
\midrule
 56205.1849 & 14332405 & CAHA/BUSCA & $u'$ &$  13\times45$\\
 56206.1974 & 14419882 & CAHA/BUSCA & $u'$ &$ 50\times45$\\
 56208.1930 & 14592304 & CAHA/BUSCA & $u'$ &$ 21\times45$\\
 56209.1754 & 14677188 & CAHA/BUSCA & $u'$ &$ 52\times45$\\
 56205.1849 & 14332405 & CAHA/BUSCA & $g'$ &$ 13\times45$\\
 56206.1974 & 14419882 & CAHA/BUSCA & $g'$ &$ 50\times45$\\
 56208.1930 & 14592304 & CAHA/BUSCA & $g'$ &$ 21\times45$\\
 56209.1754 & 14677188 & CAHA/BUSCA & $g'$ &$ 52\times45$\\
 56205.1849 & 14332405 & CAHA/BUSCA & $r'$ &$ 13\times45$\\
 56206.1974 & 14419882 & CAHA/BUSCA & $r'$ &$ 50\times45$\\
 56208.1930 & 14592304 & CAHA/BUSCA & $r'$ &$ 21\times45$\\
 56209.1754 & 14677188 & CAHA/BUSCA & $r'$ &$ 52\times45$\\
 56205.1849 & 14332405 & CAHA/BUSCA & $z'$ &$ 13\times45$\\
 56206.1974 & 14419882 & CAHA/BUSCA & $z'$ &$ 50\times45$\\
 56208.1930 & 14592304 & CAHA/BUSCA & $z'$ &$ 21\times45$\\
 56209.1754 & 14677188 & CAHA/BUSCA & $z'$ &$ 52\times45$\\
 56245.1818 & 17788140 &     LT/IO:O      & $r'$ &$  5\times100$\\
 56254.1677 & 18564517 &     LT/IO:O      & $r'$ &$  9\times100$\\
 56270.1558 & 19945888 &     LT/IO:O      & $r'$ &$  9\times100$\\
 56275.1809 & 20380060 &     LT/IO:O      & $r'$ &$  9\times100$\\
 56277.2116 & 20555512 &     LT/IO:O      & $r'$ &$  9\times100$\\
 56279.0852 & 20717394 &     LT/IO:O      & $r'$ &$  9\times100$\\
 56282.2041 & 20986865 &     LT/IO:O      & $r'$ &$  9\times100$\\
 56283.0932 & 21063685 &     LT/IO:O      & $r'$ &$  9\times100$\\
 56284.1360 & 21153781 &     LT/IO:O      & $r'$ &$  9\times100$\\
 56296.0438 & 22182616 &     LT/IO:O      & $r'$ &$  9\times100$\\
 56300.9702 & 22608259 &     LT/IO:O      & $r'$ &$  9\times100$\\
 56302.1352 & 22708908 &     LT/IO:O      & $r'$ &$  9\times100$\\
 56303.1084 & 22792994 &     LT/IO:O      & $r'$ &$  9\times100$\\
 56303.9884 & 22869026 &     LT/IO:O      & $r'$ &$  9\times100$\\
 56305.0037 & 22956752 &     LT/IO:O      & $r'$ &$  9\times100$\\
 56306.0300 & 23045418 &     LT/IO:O      & $r'$ &$  9\times100$\\
 56310.0862 & 23395876 &     LT/IO:O      & $r'$ &$ 15\times100$\\
 56310.9812 & 23473205 &     LT/IO:O      & $r'$ &$ 15\times100$\\
 56312.1042 & 23570230 &     LT/IO:O      & $r'$ &$ 15\times100$\\
 56360.9795 & 27793059 &     LT/IO:O      & $r'$ &$ 15\times100$\\
 56364.8835 & 28130364 &     LT/IO:O      & $r'$ &$ 15\times100$\\
 56365.9379 & 28221466 &     LT/IO:O      & $r'$ &$ 15\times100$\\
 56370.9299 & 28652774 &     LT/IO:O      & $r'$ &$ 15\times100$\\
 56309.4711 & 23342731 & Gemini-N/GMOS	  & $g'$ &$ 5\times100$\\
 56309.4704 & 23342671 & Gemini-N/GMOS	  & $r'$ &$ 5\times100$\\
 56309.4624 & 23341981 & Gemini-N/GMOS	  & $i'$ &$ 5\times100$\\
\bottomrule
\end{tabular}
\tablefoot{Column "Epoch" shows the logarithmic mean-time after the burst in the observer
	frame.
	}
\label{tab:obs_log_late_time}
\end{table}
\newpage
\newpage
\newpage
\newpage
\section{X-shooter spectra of the afterglow and host galaxy's nucleus}

\begin{figure*}
\centering
\includegraphics[angle=0, width=2\columnwidth]{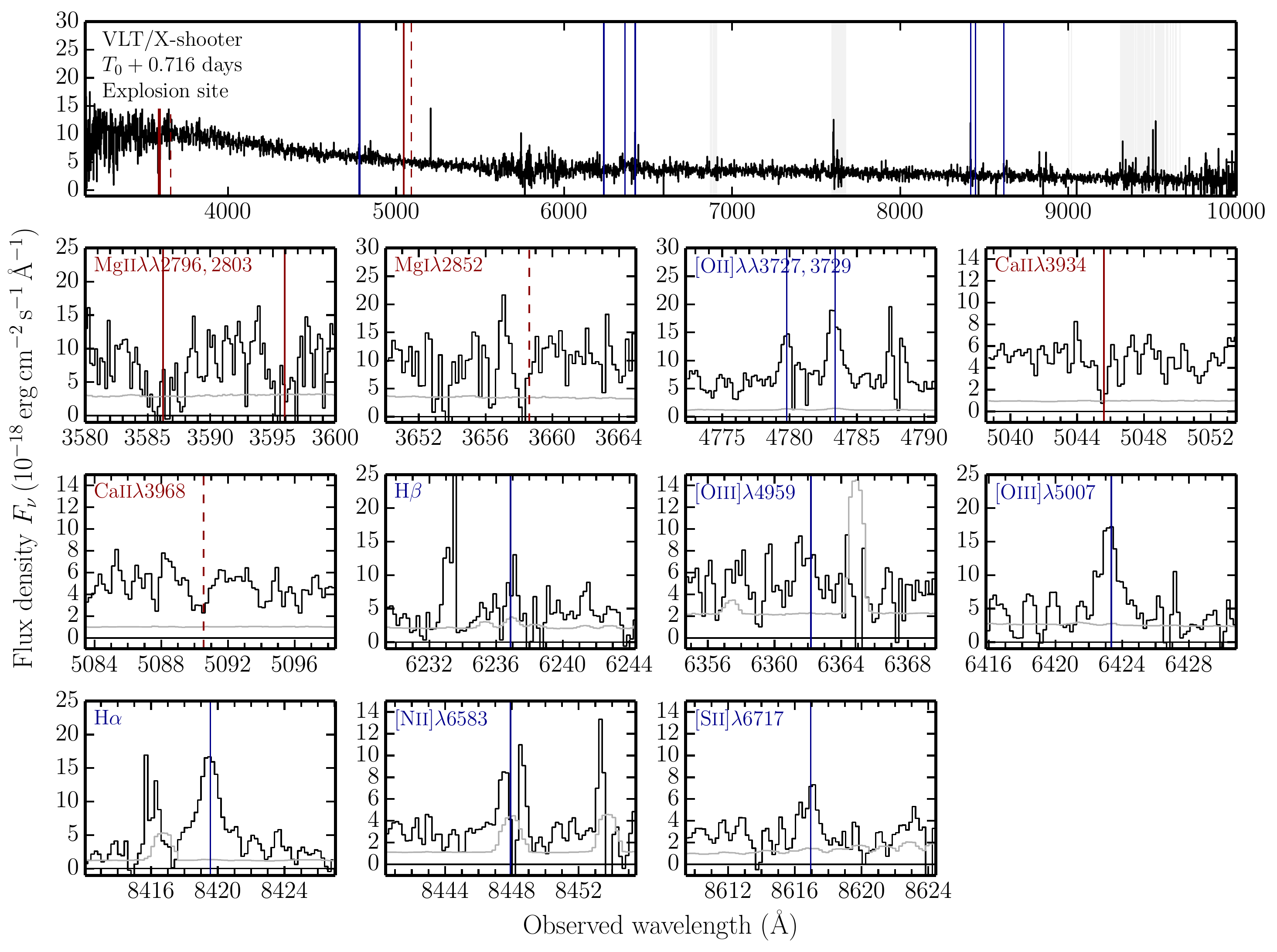}
\caption{X-shooter spectrum of GRB 120422A's afterglow obtained 0.716 days after the
burst. The top panel shows the combined UVB- and VIS-arm spectrum from 3150 to
10000~\AA. The absolute flux-calibrated spectrum is corrected for heliocentric
motion and Galactic reddening. The spectral data is shown in black, and the
corresponding noise level in grey. For illustrative purposes we rebinned the
spectrum to 2~\AA\ bins. The positions of absorption lines that are typically
associated with GRB absorbers are indicated by red lines (solid if detected and
dashed if a feature evaded detection). Nebular lines are shown in blue. The panels
below zoom in on each absorption and emission line (wavelength binning 0.15~\AA)
Table \ref{tab:GRB_abs_em_lines} summarises the fluxes and equivalent widths for
each line. Regions that are heavily affected by atmospheric absorption (transparency:
$<20\%$) are indicated by the grey shaded areas.
}
\label{fig:Spec_AG}
\end{figure*}

\begin{figure*}
\centering
\includegraphics[angle=0, width=2\columnwidth]{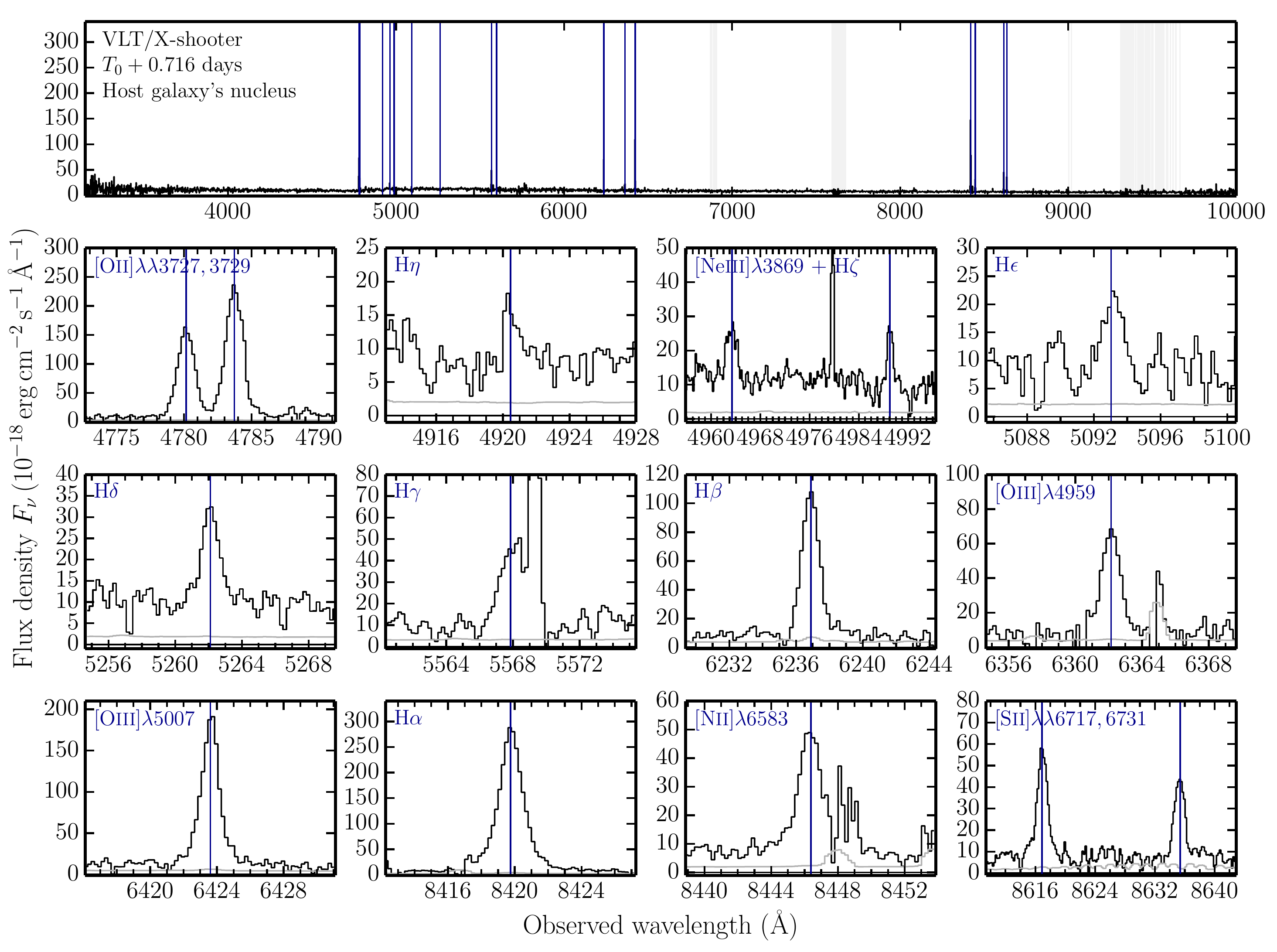}
\caption{Same as Fig. \ref{fig:Spec_AG} but for the host galaxy's nucleus. Absorption lines
are omitted since none was detected.
}
\label{fig:Spec_Nuc}
\end{figure*}

\end{document}